\documentclass[12pt]{article}
\RequirePackage{etex}
\usepackage[nodisplayskipstretch]{setspace}
\usepackage[utf8]{inputenc} 
\usepackage{booktabs}       
\usepackage{hyperref}       
\usepackage{url}            
\usepackage{amsfonts}       
\usepackage{nicefrac}       
\usepackage{microtype}      
\usepackage{lipsum}		    
\usepackage{graphicx}
\graphicspath{ {./images/} }
\usepackage{natbib}
\usepackage{doi}
\usepackage{enumerate}
\usepackage{bm}
\usepackage{bbm}
\usepackage{amssymb} 
\usepackage{amsmath}
\usepackage{amsthm}
\usepackage{float}
\usepackage{array}
\usepackage{makecell}
\usepackage{authblk}
\usepackage{caption}        
\usepackage{subcaption}     
\usepackage[normalem]{ulem}
\usepackage{enumitem}
\setlist{nosep}            
\newcommand{\markblankline}{\par\mbox{}\par}
\usepackage[linesnumbered,ruled,vlined]{algorithm2e}
\usepackage{xcolor}
\let\oldproof\proof
\renewcommand{\proof}{\color{blue}\oldproof}
\SetKwInput{KwInput}{Input}                
\SetKwInput{KwOutput}{Output}              

\newcommand{\mO}{\mathcal{O}}

\newcommand{\bx}{\bm{x}}
\newtheorem{definition}{Definition}[section]

\newtheorem{theorem}{Theorem}[section]
\newcommand{\blind}{1}

\usepackage{cleveref}
\usepackage{autonum}
\allowdisplaybreaks

\renewcommand{\sout}[1]{\unskip}

\date{}
\begin{document}

\def\spacingset#1{\renewcommand{\baselinestretch}%
{#1}\small\normalsize} \spacingset{1}

\if1\blind
{
  \title{\bf Operator-induced structural variable selection for identifying materials genes}
  \author[1]{Shengbin Ye \thanks{sy53@rice.edu}}
  \author[2]{Thomas P.~Senftle \thanks{tsenftle@rice.edu}}
  \author[1]{Meng Li \thanks{meng@rice.edu}}
  \affil[1]{Department of Statistics, Rice University, Houston, TX 77005}
  \affil[2]{Department of Chemical and Biomolecular Engineering, Rice University, Houston, TX 77005}
  \maketitle
} \fi

\if0\blind
{
  \bigskip
  \bigskip
  \bigskip
  \begin{center}
    {\LARGE\bf Operator-induced structural variable selection for identifying materials genes}
  \end{center}
  \markblankline
  \markblankline
 \markblankline
  \markblankline
  \markblankline
  \medskip
} \fi

\begin{abstract}
In the emerging field of materials informatics, a fundamental task is to identify physicochemically meaningful descriptors, or materials genes, which are engineered from primary features and a set of elementary algebraic operators through compositions. Standard practice directly analyzes the high-dimensional candidate predictor space in a linear model; statistical analyses are then substantially hampered by the daunting challenge posed by the astronomically large number of correlated predictors with limited sample size. We formulate this problem as variable selection with operator-induced structure (OIS) and propose a new method to achieve unconventional dimension reduction by utilizing the geometry embedded in OIS. Although the model remains linear, we iterate nonparametric variable selection for effective dimension reduction. This enables variable selection based on \textit{ab initio} primary features, leading to a method that is orders of magnitude faster than existing methods, with improved accuracy. To select the nonparametric module, we discuss a desired performance criterion that is uniquely induced by variable selection with OIS; in particular, we propose to employ a Bayesian Additive Regression Trees (BART)-based variable selection method. Numerical studies show superiority of the proposed method, which continues to exhibit robust performance when the input dimension is out of reach of existing methods. Our analysis of single-atom catalysis identifies physical descriptors that explain the binding energy of metal-support pairs with high explanatory power, leading to interpretable insights to guide the prevention of a notorious problem called sintering and aid catalysis design. 
\end{abstract}
  \markblankline
\vfill

\newpage
\doublespacing
\section{Introduction} \label{sec:intro}
The Materials Genome Initiative set up by the White House is a large-scale effort concerning the utilization of computational tools to accelerate the pace of discovery and deployment of advanced material systems. Since its inception in 2011, there has been a surge of interest in data-driven materials design and understanding \citep{Zhong2020, hart2021machine, keith2021combining, Lin2021, liu2021gaussian}. In this nascent area called materials informatics, computational methods that account for physical and chemical mechanisms of a material system play a central role in aiding, augmenting, or even replacing the time-consuming trial and error experimentation. A fundamental task is to identify physicochemically meaningful \textit{descriptors}, or \textit{materials genes} \citep{role_of_descriptor,foppa2021materials}. These descriptors, for example, are key to modeling single-atom catalysis and finding or developing more efficient catalytically active materials. In statistical terms, descriptors are high-dimensional predictors but with strong structure in that they are functional transformations of a set of primary features $\bm{X} = (\bm{x}_1,\ldots,\bm{x}_p)$. For instance, a simple example of descriptors is $f(\bm{X}) = \{\exp(\bm{x}_1) - \exp(\bm{x}_2)\}^2$, which can be constructed using exponential and squared functions in combination with subtraction.

Suppose the response vector $\bm{y}$ measures the material property of interest, and the primary features matrix $\bm{X} = (\bm{x}_1,\ldots,\bm{x}_p)$ collects physical or chemical properties of the materials such as atomic radii, ionization energies, etc. Then the space of engineered predictors (or descriptors) up to order $M$ is $\mathcal{O}^{(M)}(\bm{X})$, which consists of nonlinear predictors with explicit functional form resulting from $M$-order compositions of operators $\mathcal{O}$ on $\bm{X}$: 
\begin{equation} \label{eq:operator}
    \mathcal{O}^{(M)}(\bm{X}) = \mathcal{O} \circ \mathcal{O}^{(M-1)}(\bm{X}) = \underbrace{\mO \circ \cdots \circ \mO}_{M \text{ times}} (\bm{X}).
\end{equation}
For example, some commonly used operators in materials genome are 
\begin{equation}\label{eq:operator_set}
    \mathcal{O} = \{+, -, \times, /, |-|, I, \exp, \log, |\cdot|, \sqrt{}, ^{-1}, ^2, \sin(\pi\cdot), \cos(\pi\cdot)\},
\end{equation}
and the aforementioned descriptor $f(\bm{X}) = \{\exp(\bm{x}_1) - \exp(\bm{x}_2)\}^2$ belongs to $\mathcal{O}^{(3)}(\bm{X})$. We refer to this distinctive geometry encoded in $\mathcal{O}^{(M)}(\bm{X})$ as \textit{operator-induced structure} (OIS). The aforementioned descriptors in materials genome are thus the predictors in a linear model with OIS. Henceforth, we will use \textit{descriptor selection} and variable selection in the presence of OIS interchangeably. Note that the specification of $\mathcal{O}$ depends on domain knowledge, and we intentionally include the absolute difference operator $|-|$ in $\mathcal{O}$ because it is directly interpretable in materials science, and it often provides clear intuition on many physical phenomena, such as the metal-oxide binding energy \citep{Chunyen_domain_paper}. Treating it as a single operator reduces the required number of iterations to generate related descriptors. 

A common practice in materials genome~\citep{O'Connor2018, SISSO, MgO} is to employ modern statistical variable selection developed for linear models. However, the geometry of OIS defined by operators $\mathcal{O}$ and high-order compositions induces high correlation and ultra-high dimension to the feature space. As detailed in Section~\ref{sec:2_OIS_framework}, the dimension of $\mathcal{O}^{(M)}(\bm{X})$ increases \textit{double exponentially} with $M$ and the number of binary operators in $\mathcal{O}$. For example, with $p = 59$ in our real data application, enumerating $\mathcal{O}^{(3)}(\bm{X})$ gives $1.01 \times 10^{17}$ predictors while only a handful of them are associated with the response. Moreover, these predictors are highly correlated as a result of iteratively applying unary operators. This along with a small size such as $n = 91$ in our real data application substantially hurdles the performance of existing methods that rely on linear variable selection methods. Indeed, materials genomes are an analog concept to genomes, but the dimension of predictors and inherent strong correlation in materials genome-wide association studies, or \textit{materials GWAS}, pose unprecedented challenges to statistical analysis. 

In this article, we aim to develop a powerful method for materials GWAS in which we effectively identify materials genes that are associated with the response of interest. To achieve dimension reduction in materials GWAS, we consider an iterative approach by applying a small set of operators and immediately identifying the relevant descriptors, $\mathcal{D}$, before constructing more complex descriptors. This step is iterated in light of the composition structure in OIS, in striking contrast to existing literature that aims to exhaustively generate $\mathcal{O}^{(M)}(\bm{X})$. In each iteration, $\mathcal{O}(\mathcal{D})$ is typically substantially smaller than $\mathcal{O}^{(M)}(\bm{X})$, and such sparsity achieves dimension reduction and tackles the daunting computational challenges posed by materials GWAS.

Iterative dimension reduction, however, faces two intertwined challenges. First, the constructed descriptors in intermediate steps, unlike the astronomically large $\mathcal{O}^{(M)}(\bm{X})$, are not necessarily linearly associated with the response. To address this, we propose to use \textit{nonparametric variable selection} for dimension reduction to ensure selection accuracy under the geometry of OIS. That is, while the model is assumed to be linear, we employ nonparametric variable selection to achieve dimension reduction while maintaining high selection accuracy. We refer to this key novelty of our proposed method as ``parametrics assisted by nonparametrics", or PAN.

The second challenge pertains to the selection of the nonparametric module. Unlike traditional nonparametric variable selection, OIS variable selection calls for new performance criteria for the nonparametric module to ensure OIS selection accuracy (see Section~\ref{sec:iBART} for more details). We introduce a \textit{PAN criterion} to reassess nonparametric selection methods, which elucidates an asymmetric effect between false positives and false negatives and highlights a desired invariance property to unary transformations. In particular, we propose to use a Bayesian additive regression tree variable selection method, BART-G.SE \citep{bleich2014}, as the nonparametric module, which we show is well suited to satisfy the PAN criterion.

Coupling the PAN strategy with BART-G.SE, together with additional considerations to address the complexities of materials GWAS, leads to a new method for materials GWAS, which we call iterative BART, or iBART. The iterative framework of iBART reduces the size of the effective descriptor space significantly, mitigating collinearity in the process, and the use of nonparametric variable selection accounts for structural model misspecification in intermediate variable selection steps. Our extensive experiments show that iBART gives excellent performance with accuracy and scalability that are not seen in existing methods. Note that iBART is not a new BART variant for nonparametric regression, but rather an iterative use of BART within the PAN framework specifically tailored for materials GWAS.

The outline of the article is as follows. Section~\ref{sec:1.1_related_work} reviews related work in materials genome. In Section~\ref{sec:2_OIS_framework}, we introduce the OIS framework, describe the PAN selection procedure, and discuss how to choose the nonparametric module in PAN and some practical considerations regarding PAN. Section~\ref{sec:3_simulation} contains a simulation study that shows superior performance of iBART relative to existing methods. In Section~\ref{sec:4_real_data}, we apply iBART to a single-atom catalysis data set and it identifies physical descriptors that explain the binding energy of metal-support pairs with high explanatory power, leading to interpretable insights to guide the prevention of a notorious problem called sintering. We close in Section~\ref{sec:5_discussion} with a discussion. All proofs, details of variants of iBART, and additional simulation results and discussion are deferred to the Supplementary Material.

\subsection{Related work} \label{sec:1.1_related_work}
Descriptor selection has attracted growing attention in materials science. Recent methods often build on a \textit{one-shot} descriptor generation and selection scheme followed by modern statistical variable selection approaches~\citep{O'Connor2018, SISSO, MgO}. In particular, they first construct descriptors by applying operators iteratively $M$ times on the primary feature space $\bm{X}^{(0)} \equiv \bm{X} = (\bm{x}_1,\ldots,\bm{x}_p) \in \mathbb{R}^{n \times p}$ to construct an ultra-high dimensional descriptor space $\bm{X}^{(M)} \equiv \mathcal{O}^{(M)}(\bm{X})$ of $O(p^{2^M})$ descriptors, assuming binary operators are used in each iteration. Then variants of generic statistical methods are adopted to select variables from $\bm{X}^{(M)}$. Along this line, the method SISSO (Sure Independence Screening and Sparsifying Operator) proposed by \cite{SISSO} builds on Sure Independence Screening, or SIS \citep{SIS}, \citet{O'Connor2018} uses LASSO \citep{LASSO}, and \cite{MgO} adopts Bayesian variable selection methods. 

SISSO is widely perceived as one of the most popular methods for materials genome. It utilizes SIS to screen out $P$ descriptors, from which the single best descriptor is selected using an $\ell_0$-penalized regression. If a total of $k$ descriptors are desired, this process will be iterated for $k$ times yielding $k$ sets of SIS-selected descriptors, followed by an $\ell_0$-penalized regression to select the best $k$ descriptors from all the SIS-selected descriptors. Note that in each iteration, SIS is employed to screen out $P$ descriptors from the remaining descriptor set with an updated response vector given by its least squares residuals projected onto the space spanned by previously SIS-selected descriptors. Users must define the composition complexity of the descriptors through $M$, i.e., the order of compositions of operators. In a typical application of SISSO, the composition complexity $M$ is no greater than 3, the number of candidate descriptors in each SIS iteration is less than 100, and the number of descriptors $k$ is no larger than 5 \citep{SISSO}. Note that selecting 5 descriptors with 100 SIS-selected descriptors in each iteration amounts to fitting at most $\binom{500}{5} \approx 2.6 \times 10^{11}$ different regressions, which is computationally intensive.

A major drawback of these one-shot descriptor construction procedures is the introduction of a highly correlated and ultra-high dimensional descriptor space $\bm{X}^{(M)}$. High correlation often hampers the performance of a variable selection method, and the ultra-high dimensional descriptor space with large $p$, as common in modern applications, can make it computationally prohibitive for such methods. In practice, these methods often resort to ad hoc adaptation or sacrifice the complexity level of candidate descriptors. 

Another thread of work is the well-developed \textit{automatic feature engineering} in machine learning, aiming to generate complex features from given constructor functions adaptively~\citep{FICUS, Feurer2015, Khurana2018}. However, the overwhelming focus of this literature is on increasing the predictive power of the primary features $\bm{X}$. We instead focus on discovering the underlying functional relationship between the response and the predictors and revealing data-driven insights into the underlying physics of materials design. In addition, the sample size in materials genome is typically limited, hampering the use of machine learning methods that rely on large training data. 

In statistics, transformations have been commonly used to expand the predictor space, including polynomials, logarithmic, power transformations, and the previously noted interactions. The induced feature spaces from these elementary transformations are often overly simple to capture the intricate dynamics of the response in materials genome, particularly compared to high-order compositions of a larger operator set. There has been a rich literature on nonparametric variable selection, but descriptor selection relies on a linear model with feature engineering that favorably points to interpretable insights for domain experts as the functional forms of selected variables are explicitly given and the feature space could be composed using domain-related knowledge. The nonparametric module in PAN only serves as a dimension reduction tool, and the desired performance calls for new investigation under the context of OIS. Overall, materials GWAS may play an analogous role that GWAS have played in motivating new statistical methods and concepts, and to the best of our knowledge, the present article is the first statistical work on this topic.

\section{The OIS framework}\label{sec:2_OIS_framework}
\subsection{Operator-induced structural model} \label{sec:model}
We begin with a standard ultra-high dimensional linear regression model
\begin{equation} \label{eq:linear.model}
    \bm{y} = \beta_0 + \beta_1\bm{x}_1 + \cdots + \beta_{p^\star}\bm{x}_{p^\star} + \bm\varepsilon, 
\end{equation}
where $\bm\varepsilon \sim \mathcal{N}_n(\bm{0}, \sigma^2\bm{I})$ is a Gaussian noise vector, and the regression coefficients $\bm\beta$ are sparse. The dimension of predictors $p^\star$ is ultra-high, at the materials GWAS scale that typically exceeds the maximum size of matrices allowed by a modern personal computer, while the sample size $n$ is on the order of tens. In this paper, we assume that this seemingly ultra-dimensional descriptor space obeys an operator-induced structure (OIS). In particular, we assume that the predictors $\bm{x}_1,\ldots,\bm{x}_{p^\star}$ in \eqref{eq:linear.model}, or \textit{descriptors}, are generated by applying operators in $\mathcal{O}$ iteratively $M$ times on a primary feature space $\bm{X} = (\bm{x}_1,\ldots,\bm{x}_p) \in \mathbb{R}^{n \times p}$, 
\begin{equation} \label{eq:OIS.def} 
    (\bx_1,\ldots, \bx_{p^{\star}}) \equiv \bm{X}^{(M)} = \mathcal{O}^{(M)}(\bm{X}) = \mathcal{O} \circ \mathcal{O}^{(M-1)}(\bm{X}) = \underbrace{\mO \circ \cdots \circ \mO}_{M \text{ times}} (\bm{X}),
\end{equation}
where $\bm{X}^{(M)} = \mathcal{O}^{(M)}(\bm{X})$ denotes $M$-composition of $\mathcal{O}$ on $\bm{X}$ and can be defined iteratively as above. The operator set $\mathcal{O}$ is user-defined; for concreteness, we focus on the common example given in \eqref{eq:operator_set}, unless stated otherwise.

We adopt the following convention. Evaluation of operators on vectors is defined to be entry-wise, e.g., $\bm{x}^2_1 = (x_{1,1}^2,\ldots,x_{n,1}^2)^T$ and $\bm{x}_1 + \bm{x}_2 = (x_{1,1} + x_{1,2},\ldots,x_{n,1} + x_{n,2})^T$. Throughout this article, we assume that all descriptors in $\bm{X}^{(M)}$ are uniquely defined in terms of their numerical values. For instance, only one of the descriptors in $\{\bm{x}_1^2, \bm{x}_1\times \bm{x}_1\}$ will be kept in $\bm{X}^{(M)}$.  This can be easily achieved in practice by identifying and removing perfectly correlated descriptors.

We hereafter refer to the linear regression model in \eqref{eq:linear.model} along with the operator-induced structure (OIS) in \eqref{eq:OIS.def} as the OIS model. To facilitate a precise OIS model definition using predictors with nonzero coefficients, we define \emph{$M$-composition descriptor} as follows.
\begin{definition} [$M$-composition descriptor] \label{def:M-composition}
    We define $f^{(M)}(\bm{X})$ to be an $M$-composition descriptor if it is constructed via $M$ compositions of operators on some primary features $\bm{X}$:
        $f^{(M)}(\bm{X}) = o^{(M)}(\bm{X}) = o_{M} \circ f^{(M-1)}(\bm{X}) = o_{M} \circ o_{M-1} \circ \cdots \circ o_1 (\bm{X}),$
    where $o_m \in \mathcal{O}$ is the $m$-th composition operator(s) for $1 \leq m \leq M$, and $f^{(1)}(\bm{X}),\ldots,f^{(M-1)}(\bm{X})$ are the necessary intermediate descriptors for constructing the descriptor $f^{(M)}(\bm{X})$. 
\end{definition}
Note that if the $m$-th composition operator is a binary operator, there may exist two $(m-1)$-th composition operators but we suppress the notation in the definition above for simplicity. Furthermore, if an $M$-composition descriptor $f^{(M)}(\bm{X})$ only depends on a subset of primary features $\bm{X}_\mathcal{S}$, where $\mathcal{S} \subseteq [p] = \{1, \ldots, p\}$, we also write it as $f^{(M)}(\bm{X}_\mathcal{S})$ and call it an $(M, \mathcal{S})$-descriptor. 
\begin{definition} [$(M,\mathcal{S})$-composition OIS model] \label{def:MS_OIS_model}
    An $(M, \mathcal{S})$-OIS model assumes
    \begin{equation} \label{eq:MCSV}
        \bm{y} = \beta_0 + \sum_{k=1}^K f_k^{(M_k)}(\bm{X}_{\mathcal{S}_k})\beta_k + \bm\varepsilon,
    \end{equation}
    where $M = \max_{k=1,\ldots,K} M_k$ denotes the highest order of operator compositions, $K$ denotes the number of additive descriptors, $\bm{X}_{\mathcal{S}_k}$ is the set of primary features used in the $k$-th descriptors, and $\mathcal{S} = \bigcup_{k=1}^K \mathcal{S}_{k}$ is the set of all active primary feature indices.
\end{definition}

\begin{figure}[hbt!]
    \centering
    \includegraphics[height = 6.5cm]{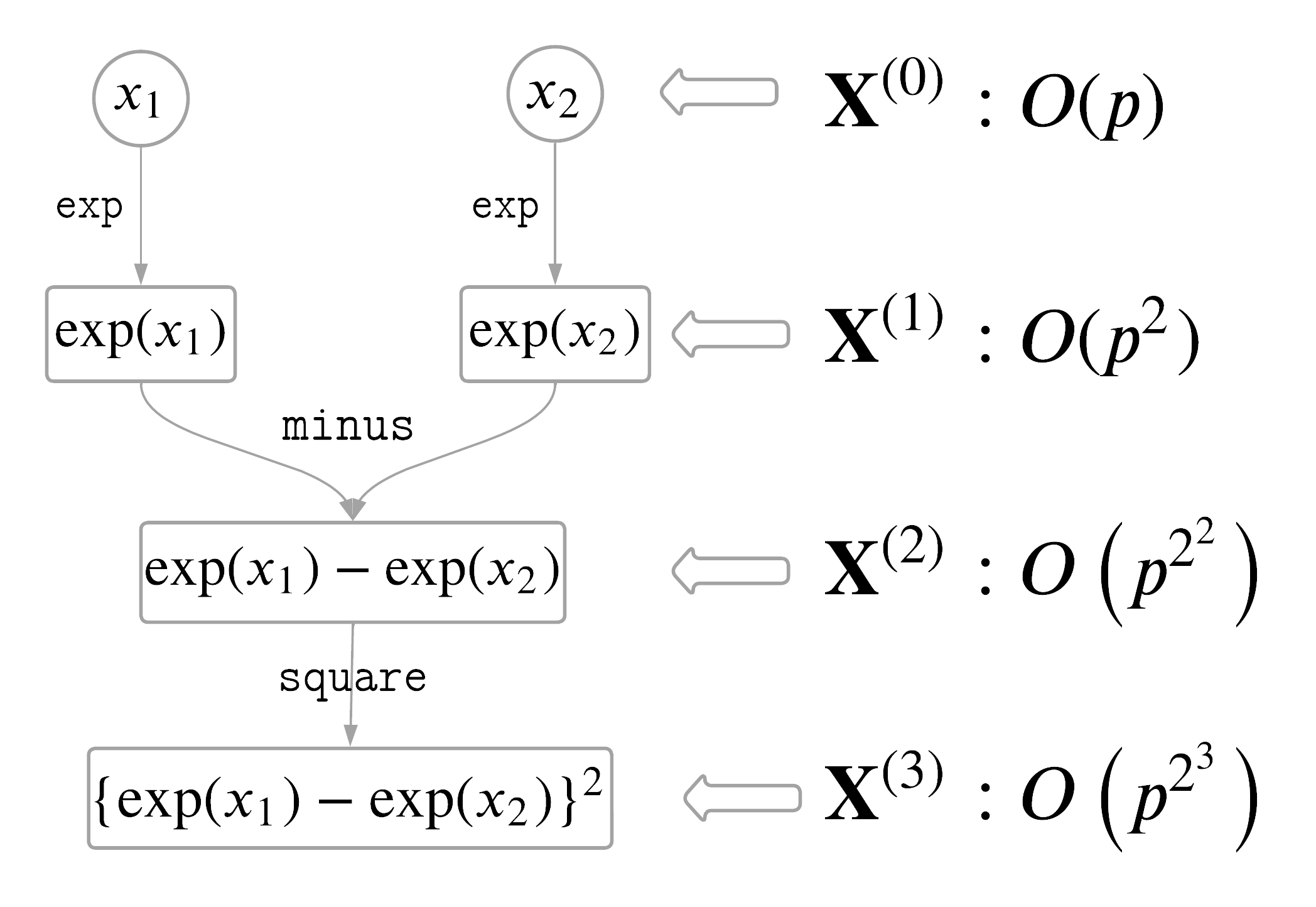}
    \caption{A tree diagram for generating $\{\exp(\bm{x}_1)-\exp(\bm{x}_2)\}^2$. The dimension of descriptor space increases double exponentially with the composition complexity.} 
    \label{fig:descriptor_diagram}
\end{figure}

Throughout the article, we assume the data follows an $(M, \mathcal{S})$-OIS model in \eqref{eq:MCSV}. We use $M_0$ for the oracle highest composition complexity and $\mathcal{S}_0$ for the oracle set of active primary feature indices. Descriptors in the $(M, \mathcal{S})$-OIS model and their intermediate descriptors are called \textit{active}. We next use a toy example to illustrate the introduced concepts in OIS and the challenges posed by descriptor selection. Suppose that the data-generating model is 
\begin{equation} 
\bm{y} = \beta_0 + \beta_1 f^{(M_1)}_1(\bm{X}_{\mathcal{S}_1}) + \beta_2 f^{(M_2)}_2(\bm{X}_{\mathcal{S}_2}) + \bm\varepsilon,
\end{equation} 
where $M_1 = 3, M_2 = 2, \mathcal{S}_1 = \{1, 2\}, \mathcal{S}_2 = \{3, 4\}, f^{(M_1)}_1(\bm{X}) = \{\exp(\bm{x}_1)-\exp(\bm{x}_2)\}^2,$ and $f^{(M_2)}_2(\bm{X}) = \sin(\pi\bm{x}_3\bm{x}_4).$ Here $\{\exp(\bm{x}_1)-\exp(\bm{x}_2)\}^2$ is a $3$-composition descriptor or a $(3, \{1, 2\})$-descriptor, and $\sin(\pi\bm{x}_3\bm{x}_4)$ is a $2$-composition descriptor or a $(2, \{3, 4\})$-descriptor. Both descriptors arise from applying $\mathcal{O}$ iteratively three times on the primary features: $\bm{X}^{(3)} = \mathcal{O} \circ \mathcal{O} \circ \mathcal{O}(\bm{X})$. The composition of operators resembles a tree-like structure for generating descriptors; Figure \ref{fig:descriptor_diagram} describes the tree-like workflow for generating $\{\exp(\bm{x}_1)-\exp(\bm{x}_2)\}^2$.

To see how the descriptor space dimension increases with the number of iterations, let $c_u$ and $c_b$ denote the number of unary and binary operators, respectively, and let $p_j$ denote the dimension of the $j$-th descriptor space $\bm{X}^{(j)}$. Note that the dimension of $\bm{X}^{(1)}$ is $p_1 = c_up + c_bp(p-1)/2$, which is on the order of $O(p^2)$; the dimension of $\bm{X}^{(2)}$ is $p_2 = c_up_1 + c_bp_1(p_1-1)/2 \approx O(p^{2 \cdot 2}) = O(p^{2^2})$; the dimension of $\bm{X}^{(3)}$ is $p_3 = c_up_2 + c_bp_2(p_2-1)/2 \approx O(p^{2 \cdot 2 \cdot 2}) = O(p^{2^3})$. The dimension of descriptor space will increase double exponentially with the number of binary operator compositions, e.g., with $M$ compositions of binary operators on $\bm{X}$ the resulting descriptor space has a dimension of order $O(p^{2^M})$. Similarly, the double exponential expansion applies to the number of binary operators $c_b$. Excluding redundant descriptors does not prevent this double exponential expansion. As shown in Section \ref{sec:simulation}, building $\bm{X}^{(2)}$ from $p = 59$ primary features results in an astronomical descriptor space containing over $5.5 \times 10^7 = O(59^{2^2})$ descriptors even after removing redundant and non-physical descriptors. In addition, the number of active (intermediate) descriptors will be non-increasing with $M$ as shown in Figure~\ref{fig:descriptor_diagram}: there are four active descriptors $\exp(\bm{x}_1), \exp(\bm{x}_2), \bm{x}_3, \bm{x}_4$ in $\bm{X}^{(1)}$, but only two active descriptors $\exp(\bm{x}_1) - \exp(\bm{x}_2)$ and $\bm{x}_3\bm{x}_4$ in $\bm{X}^{(2)}$, and two active descriptors $\{\exp(\bm{x}_1)-\exp(\bm{x}_2)\}^2$ and $\sin(\pi\bm{x}_3\bm{x}_4)$ in $\bm{X}^{(3)}$. 

\subsection{PAN descriptor selection for OIS model} \label{sec:method} 
We propose an iterative descriptor construction and selection procedure PAN for the OIS model, which generates descriptors by iteratively applying operators and selecting the potentially useful intermediate descriptors between each iteration of descriptors synthesis. The iterative descriptor selection procedure excludes irrelevant intermediate descriptors from the descriptor generating step, achieving a \textit{progressive} variable selection and enabling variable selection based on \textit{ab initio} primary features. This reduces the dimension of the subsequent descriptor space $\bm{X}^{(m)}$ and mitigates collinearity among the descriptors in comparison to the one-shot descriptor construction approaches.

To describe the method in its most general form, we allow \emph{different sets of operators} $\mathcal{O}_m \subseteq \mathcal{O}$ for each iteration $m = 1,\ldots,M$, leading to the descriptor space
\begin{equation} \label{eq:OIS.alt.def} 
    \mO_M \circ \mO_{M-1} \circ \cdots \circ \mO_2 \circ \mO_1(\bm{X}).
\end{equation}
The framework of our iterative descriptor selection procedure is as follows.

\noindent \textbf{PAN descriptor selection procedure:} 
\begin{enumerate}
    \item \textbf{Repeat} the following until at least one descriptor exhibits a strong linear association with the response variable $\bm{y}$ ($i = 0$):
    \begin{enumerate}
        \item Use a nonparametric variable selection procedure to perform descriptor selection on $\bm{X}^{(i)}$ and obtain the selected descriptors ${\bm{X}^{(i)}}'$;
        \item Apply the $i$-th operator set $\mathcal{O}_i$ on all of the previously selected descriptors, $\bigcup_m {\bm{X}^{(m)}}'$, yielding a new descriptor space, $\bm{X}^{(i+1)} = \mathcal{O}_i(\bigcup_m {\bm{X}^{(m)}}')$, where $\mathcal{O}_i$ can be different for each iteration $i$;
    \end{enumerate}
    \item \textbf{Once} there exist descriptor(s) that exhibit a strong linear association with response variable $\bm{y}$, use a linear parametric variable selection procedure to perform descriptor selection on $\bm{X}^{(i)}$, and obtain the selected descriptors, $\bm{X}^\star \subseteq \bm{X}^{(i)}$.
\end{enumerate}

We keep all the selected descriptors in the main loop to facilitate the creation of high-order complexity descriptors with the help of low-order complexity descriptors. For instance, constructing $\bm{x}_1^3$ using $\mathcal{O}$ defined in \eqref{eq:operator_set} requires us to keep $\bm{x}_1 \in {\bm{X}^{(0)}}'$ selected at the base iteration and $\bm{x}_1^2 \in {\bm{X}^{(1)}}'$ selected at the first iteration. 

To see how this iterative procedure helps reduce the dimension of descriptor space significantly, let $s_i = |{\bm{X}^{(i)}}'|$ be the number of descriptors selected in the $i$-th iteration and $p_i = |\bm{X}^{(i)}|$ be the dimension of the $i$-th descriptor space. Suppose that the number of selected descriptors is sparse in each iteration, i.e., $s_i \ll p_i$. Assuming binary operators were used, the dimension of the $(i+1)$-th descriptor space in PAN is on the order of $O(s_i^2) \ll O(p_i^2)$, and this holds for all iteration $i \geq 0$. If we further assume $s_i \approx O(p)$ for all $i \geq 0$, where $p = |\bm{X}|$ is the number of primary features, then the dimension of the $(i+1)$-th descriptor space for PAN is on the order of $O(p^2)$, compared to $O(p^{2^{i+1}})$ for the one-shot methods. Note these assumptions are reasonable according to the discussion in Section~\ref{sec:model}. In the simulation study in Section~\ref{sec:simulation} with $p = 10$ primary features, we observed that $s_i \approx O(10^1)$ and $p_i \approx O(10^2)$ for all iterations of the PAN procedure. On the contrary, a one-shot method, such as SISSO, generates a descriptor space containing $9.26 \times 10^9$ descriptors in the same setting. 

The use of a nonparametric variable selection procedure in Step 1(a) is necessary because the intermediate descriptors may not have a strong linear association with the response variable. Thus, a method that accounts for model misspecification, such as a nonparametric method, is more suitable for preliminary screening of the intermediate descriptors. In addition, a suitable nonparametric module for PAN needs to account for the geometry embedded in OIS and the unconventional goal of selecting operators along with variables; the next section discusses performance requirements for this step and introduces an implementation of PAN, iBART, that is particularly suitable for OIS. In Step 2, the oracle descriptors are linearly associated with the response. Hence, we employ linear parametric variable selection methods, such as LASSO \citep{LASSO}, to reduce false positives and select the final descriptors.

\subsection{Choosing nonparametric module in PAN and iBART} \label{sec:iBART}
In OIS, the ultimate goal is to recover or approximate the true functional relationship between the response and the primary features. This goal together with PAN entails new performance requirements for the nonparametric module in PAN. To illustrate such a need, let us consider a simple OIS model
\begin{equation}
\label{eq:property_model}
\bm{y} = f^{(M)}(\bm{X}) + \bm\varepsilon = \sqrt{\bm{x}_1 + \bm{x}_2} + \bm\varepsilon.
\end{equation}
In traditional nonparametric variable selection problems, the regression model only sees the primary features, $\bm{X}$, and the desired performance is successful identification of the active index set, $\mathcal{S}_0 = \{1,2\}$. The base iteration of PAN has a similar goal as we would want the selected index set $\hat{\mathcal{S}}$ to be a superset of $\mathcal{S}_0$. However, the intermediate descriptor space $\bm{X}^{(m)}$ in the $m$-th iteration consists of nonlinear transformations of the selected primary features $\bm{X}_{\hat{\mathcal{S}}}$, and the active index set is no longer well-defined. 

Due to the iterative structure of PAN, a good nonparametric selection method for PAN must be able to generate and identify the $m$-composition descriptors $f_k^{(m)}(\bm{X})$ at the $m$-th iteration (i.e., all active intermediate descriptors). To this end, a suitable nonparametric module should satisfy the following \textit{PAN criterion}:
\begin{center}
    It selects \emph{all} of the $m$-composition descriptors that are necessary for constructing the true $M$-composition descriptor, for all $0 \leq m \leq M$ iterations.
\end{center}
Taking model~\eqref{eq:property_model} as an example, failing to select either $f_1^{(0)} = \bm{x}_1$ or $f_2^{(0)} = \bm{x}_2$ in the base iteration will preclude the generation of $f^{(1)} = (\bm{x}_1 + \bm{x}_2)$ and $f^{(2)} = \sqrt{\bm{x}_1 + \bm{x}_2}$ in subsequent iterations. The PAN criterion thus favors a ``conservative'' nonparametric method, in which false positives during selection are allowed but false negatives \emph{must not} occur in any iteration. Here false positives are defined within each intermediate descriptor space $\bm{X}^{(m)}$.

The literature has provided a rich menu of nonparametric selection methods; however, the asymmetric effect of false positives and false negatives on descriptor selection illustrated by the PAN criterion motivates our choice of tree-based nonparametric approaches. To see this, suppose the true descriptor is $f(\bm{x}_1) = \log(\bm{x}_1)$ and the design matrix $\mathcal{O}(\bm{X})$ consists of 1-composition transformations of the primary features $\bm{X} = (\bm{x}_1,\ldots,\bm{x}_p)$. A typical nonparametric method aims to identify the oracle primary predictors ($\bm{x}_1$ in this case) that associate with $\bm{y}$ through an unknown function $f(\cdot)$, without considering transformations of $\bm{x}_1$ as possible candidate predictors. However, the goal in the presence of OIS is to identify the true primary predictors ($\bm{x}_1$) \emph{and} the correct operator composition ($\log(\cdot)$). This goal, in the presence of many non-signal but highly correlated unary transformations of $\bm{x}_1$, namely $\sqrt{\bm{x}_1}, |\bm{x}_1|$, etc., is shown to be difficult for many nonparametric methods; see Supplementary Material Section A.2.1 for detailed analyses. Tree-based methods are invariant to monotonic transformations and thus tend to be robust to related transformations that are often piecewise monotonic. Consequently, they may select unary transformations of $\bm{x}_1$ in addition to $f(\bm{x}_1) = \log(\bm{x}_1)$---such false positives, although increasing the candidate search space, are favorably compatible with the PAN criterion. We next provide a review of BART~\citep{chipman2010} and describe BART-G.SE~\citep{bleich2014}---the default nonparametric module in the PAN framework.

BART is a Bayesian nonparametric ensemble tree method for modeling $\bm{y} = f(\bm{X}) + \bm\varepsilon$, the unknown relationship between a response vector $\bm{y}$ and a set of predictors $\bm{x}_1,\ldots,\bm{x}_p$. More specifically, BART models the regression function $f$ by a sum of regression trees
\begin{equation}\label{eq:BART}
    \bm{y} = \sum_{i=1}^m g_i(\bm{x}_1,\ldots,\bm{x}_p; \mathcal{T}_i, \bm\mu_i) + \bm\varepsilon, \qquad\bm\varepsilon \sim \mathcal{N}_n(\bm{0}, \sigma^2\bm{I}_n).
\end{equation}
Each binary regression tree $g_i$ consists of a tree structure $\mathcal{T}_i$ partitioning observations into $B_i$ terminal nodes, and a set of terminal parameters $\bm\mu_i = \{\mu_{i1},\ldots,\mu_{iB_i}\}$ attached to these nodes. Observations within a given terminal node $b$ are constrained to have the same terminal parameter $\mu_{ib}$. The prior distributions for $(\mathcal{T}_i, \bm\mu_i)$ constrain each tree to be small so that each tree contributes to approximate $f$ in a small and distinct fashion. Readers are referred to \cite{chipman2010} for the full details of BART and posterior sampling.

The primary usage of BART is prediction, and the predicted values $\hat{\bm{y}}$ for $\bm{y}$ serve no purpose in variable selection. However, a variable inclusion rule can be defined based on the variable inclusion proportion $q_i$ of $\bm{x}_i$, which can be easily estimated from the posterior samples. To this end, we adopt the permutation-based selection threshold based on the permutation null distribution of $\bm{q} = (q_1,\ldots,q_p)$ proposed by \cite{bleich2014}.

Specifically, $B$ permutations of the response vector $\bm{y}_1^*,\ldots,\bm{y}_B^*$ are generated, and a BART model is fitted for each of the permuted response vectors with the same predictors $\bm{x}_1,\ldots,\bm{x}_p$. The variable inclusion proportions from the permuted BART models $\bm{q}_1^*,\ldots,\bm{q}_B^*$ are then used to create a permutation null distribution for the non-permuted variable inclusion proportion $\bm{q}$. The predictor $\bm{x}_i$ is selected if $q_i > m_i + C^* \cdot s_i$, where $m_i$ and $s_i$ are the mean and standard deviation of the permuted variable inclusion proportion $q_i^* = (q_{i,1}^*,\ldots,q_{i,B}^*)$, and $C^* = \inf_{C \in \mathbb{R}^+} \left\{ \forall i, \frac{1}{B}\sum_{b=1}^B \mathbb{I}(q_{i,b}^* \leq m_i + C \cdot s_i) > 1 - \alpha \right\}$ is the smallest global standard error multiplier (G.SE) that gives a simultaneous $1-\alpha$ coverage across the permutation null distributions of $q_i$ for all predictors. We refer to BART with the permutation-based selection procedure described above as BART-G.SE.
 
Various BART-related methods have been recently developed~\citep{DART, horiguchi2021assessing,ABC_forest}. They often incorporate sparsity-inducing priors into BART and prove to be highly effective in various tasks. However, it is unclear whether the excellent performance of them developed in traditional settings carries over to being compatible with the PAN criterion. Indeed, we have found that methods aiming at optimally choosing nonparametric variables in traditional settings may incur fewer false positives but have a higher chance to miss the true descriptors in intermediate iterations---such false negatives are devastating in the context of PAN and OIS variable selection. The PAN criterion provides a useful guide in choosing not only the regression method but also the selection rule, for which we recommend BART-G.SE. In our numerical experiments, we vary the nonparametric module in PAN by comparing several recent BART-related methods and other nonparametric selection methods, and find that the proposed iBART, PAN with BART-G.SE as the nonparametric module, is particularly well suited for OIS variable selection and tends to give the best overall performance; see the Supplementary Material for numerical results and a comprehensive discussion.

\subsection{Practical consideration and the algorithm} \label{sec:practical_consideration}
The operators in $\mathcal{O}$ in \eqref{eq:operator_set} can be classified into the unary operators $\mathcal{O}_u = \{I, \exp, \log, |\cdot|, \sqrt{}, ^{-1}, ^2, \sin(\pi\cdot), \cos(\pi\cdot)\}$ and the binary operators $\mathcal{O}_b = \{+,-,\times,/,|-|\}$, each posing different challenges to descriptor selection. The unary operators $\mathcal{O}_u$ induce strong collinearity among the engineered descriptors; for instance, $\text{Cor}(\bm{x}_1^2, |\bm{x}_1|) > 0.9$ when $\bm{x}_1 \sim \mathcal{N}_n(\bm{0},\bm{I})$ with $n = 200$. The binary operators $\mathcal{O}_b$ increase the descriptor space dimension double exponentially and generate complex nonlinear descriptors. These two issues are compounded when the two operator sets are used together. Therefore, we propose to decouple the two operator sets and alternate them, leading to two special cases of \eqref{eq:OIS.alt.def}:
\begin{align}
    \mathcal{O}_{A_u}^{(M)}(\bm{X}) &= \underbrace{\cdots \mO_b \circ \mO_u \circ \mO_b \circ \mO_u}_{M \text{ times}}(\bm{X}), \label{eq:unary_first}\\
    \mathcal{O}_{A_b}^{(M)}(\bm{X}) &= \underbrace{\cdots \mO_u \circ \mO_b \circ \mO_u \circ \mO_b}_{M \text{ times}}(\bm{X}) =\mathcal{O}_{A_u}^{(M - 1)} \circ \mO_b(\bm{X}).\label{eq:binary_first}
\end{align}
In addition to the binary operators identified earlier, we include an additional binary operator, $\pi_1: \mathbb{R}^2 \to \mathbb{R}$, defined by $\pi_1(a,b) = a$, which allows intermediate descriptors to be passed down unchanged. Note the two alternating descriptor spaces $\mathcal{O}_{A_u}^{(M)}(\bm{X})$ and $ \mathcal{O}_{A_b}^{(M)}(\bm{X})$ are not equivalent to the full descriptor space $\mathcal{O}^{(M)}(\bm{X})$ with the same composition complexity $M$. However, one can show that $\mathcal{O}_{A_u}^{(M_u)}(\bm{X})$ and $\mathcal{O}_{A_b}^{(M_b)}(\bm{X})$ can recover $\mathcal{O}^{(M)}(\bm{X})$ with some $M_u > M$ and $M_b > M$, respectively. This is formally described in the theorem below and its proof is available in Supplementary Material Section D.
\begin{theorem} \label{thm:alt_descriptor_space}
    Let $\bm{X} = (\bm{x}_1,\ldots,\bm{x}_p) \in \mathbb{R}^{n \times p}$ be a primary feature space and $\mathcal{O}$ be a set of operators such that it can be partitioned into a unary operator set $\mathcal{O}_u$ and a binary operator set $\mathcal{O}_b$. Suppose that $I \in \mathcal{O}_u$ and $\pi_1 \in \mathcal{O}_b$. Then for any $M \in \mathbb{N}$, there exists $M_u \geq M$ and $M_b \geq M$ such that $\mathcal{O}_{A_u}^{(M_u)}(\bm{X}) \supseteq \mathcal{O}^{(M)}(\bm{X})$ and $\mathcal{O}_{A_b}^{(M_b)}(\bm{X}) \supseteq \mathcal{O}^{(M)}(\bm{X})$, respectively.
\end{theorem}

For instance, the 2-composition descriptor space $\mathcal{O}^{(2)}(\bm{X})$ generated using \eqref{eq:OIS.def} contains descriptors such as $f_1 = (x_i^2 + x_j^2)$ and $f_2 = (x_i+x_j)^2$. These two descriptors can be generated using $\mathcal{O}_{A_b}^{(M)}(\bm{X})$ in \eqref{eq:binary_first}: $f_1 = \texttt{add} (\texttt{square} \circ \pi_1(x_i, x_j), \texttt{square} \circ \pi_1(x_j, x_i)) \in \mathcal{O}_{A_b}^{(3)}(\bm{X})$ and $f_2 = \texttt{square} \circ \texttt{add}(x_i,x_j) \in \mathcal{O}_{A_b}^{(2)}(\bm{X})$. Using \eqref{eq:unary_first}, $f_1$ and $f_2$ can also be generated as $f_1 = \texttt{add}(\texttt{square}(x_i), \texttt{square}(x_j)) \in \mathcal{O}_{A_u}^{(2)}(\bm{X})$ and $f_2 = \texttt{square} \circ \texttt{add}(I(x_i), I(x_j)) \in \mathcal{O}_{A_u}^{(3)}(\bm{X})$. As such, under the $M$-composition OIS model in \eqref{eq:MCSV}, one may consider using $\mathcal{O}^{(M)}$, $\mathcal{O}_{A_u}^{(M_u)}(\bm{X})$, or $\mathcal{O}_{A_b}^{(M_b)}(\bm{X})$ as long as they contain the true descriptors. In what follows we will focus on $\mathcal{O}_{A_u}^{(M)}(\bm{X})$ and $\mathcal{O}_{A_b}^{(M)}(\bm{X})$ instead of $\mathcal{O}^{(M)}$ for their aforementioned advantages. 

We adopt the descriptor generating strategies in \eqref{eq:unary_first} and \eqref{eq:binary_first} for different scenarios. In particular, if the primary features $\bm{X}$ are believed to generate the model through their intricate interactions that will be captured by binary operators and compositions of such binary operators, we recommend \eqref{eq:binary_first}. This is often the case in real-world applications, such as in Section~\ref{sec:4_real_data}, where the domain scientists have chosen a large set of potentially useful primary features, and unary transformations of these primary features are less interpretable and thus less desired. If such prior knowledge is not available and the relevant functional form of primary features is unknown, as in Section~\ref{sec:3_simulation}, we would recommend \eqref{eq:unary_first} to first identify such functional forms by selecting between unary descriptors.

The stopping criterion in Step 1 can be easily modified depending on practical needs. For example, we can specify the maximum composition complexity $M_{\max}$, like SISSO, or implement a data-driven criterion that terminates Step 1 when there exists a descriptor such that its absolute correlation with response variable $\bm{y}$ exceeds a pre-specified threshold $\rho_{\max}$ that is close to 1. We note that iBART allows larger $M_{\max}$ than SISSO as iBART does not rely on one-short feature engineering, and the use of $\rho_{\max}$ allows early termination.

It is also common in practice that one may only want to select $k$ descriptors for easy interpretation, such as $k \leq 5$. If the cardinality of the selected descriptors $\bm{X}^\star$ is greater than $k$, then an $\ell_0$-penalized regression may be used to choose the best $k$ descriptors. 

We allow user-specified options for these considerations when implementing iBART, summarized in Algorithm~\ref{alg:ibart}; an \texttt{R} package for iBART is available at \url{https://github.com/mattsheng/iBART}. The following default settings will be used in our experiments. We use the BART-G.SE thresholding variable selection procedure implemented in the \texttt{R} package \texttt{bartMachine} to perform nonparametric variable selection for Step 1(a) in PAN and use LASSO implemented in the \texttt{R} package \texttt{glmnet} to perform parametric variable selection for Step 2 in PAN. For BART-G.SE, we set the number of trees to 20 that is recommended by~\cite{bleich2014}, the number of burn-in samples to 10,000, the number of posterior samples to 5,000, the number of permutations of the response to 50, and the rest of the parameters to the default values. With these values, our Markov chain Monte Carlo (MCMC) runs appear to have reached a sufficient number of iterations in our experiments. We choose the penalty term $\lambda$ in LASSO by minimizing the mean squared error loss through a 10-fold cross-validation procedure and set the other parameters to the default values. Unreported results using real data indicate that other tuning methods to debias LASSO yield similar performance. The algorithm terminates if the composition complexity $M$ reaches $M_{\max} = 4$ or the maximum absolute correlation $|\rho|$ reaches $\rho_{\max} = 0.95$, whichever occurs first. Setting $M_{\max}$ to 4 appears sufficient for the considered materials GWAS application as descriptors with more complexity become challenging to interpret.

\begin{algorithm}[h!]
\SetAlgoLined
\KwInput{
    $\rho_{\max} \in (0,1) =$ maximum absolute correlation with the response variable; \\ 
    $M_{\max} =$ maximum composition complexity;\\ 
    \texttt{Lzero} = whether to perform $\ell_0$-penalized regression;\\
    $k =$ number of selected descriptors by $\ell_0$-penalized regression. Only required when \hangindent=0.7cm \texttt{Lzero == TRUE}.
}
\KwOutput{$\bm{X}_k^\star$: selected descriptors}
\KwData{$\bm{X}$: primary features; $\mathcal{O}_u$: set of unary operators; $\mathcal{O}_b$: set of binary operators; $\bm{y}$: response vector}
$M = 0$\\
$\rho = \max_{\bm{x} \in \bm{X}^{(M)}} \text{cor}(\bm{x},\bm{y})$\\
\While{$M \leq M_{\max}$ or $|\rho| \leq \rho_{\max}$}
{
    ${\bm{X}^{(M)}}' \leftarrow$ BART-G.SE selected descriptors on $\bm{X}^{(M)}$\\
    $\bm{X}^{(M+1)} \leftarrow \mathcal{O}_{M+1} \left(\cup_i {\bm{X}^{(i)}}'\right)$\\
	 $M \leftarrow M + 1$\\
	 $\rho \leftarrow \max_{\bm{x} \in \bm{X}^{(M)}} \text{cor}(\bm{x},\bm{y})$
}
$\bm{X}^\star \leftarrow$ LASSO selected descriptors on $\bm{X}^{(M)}$\\
\If{Lzero == TRUE and $|\bm{X}^\star| > k$}
{
	$\bm{X}_k^\star \leftarrow$ best $k$ descriptors from $\ell_0$-penalized regression
}
\Else
{
	$\bm{X}_k^\star \leftarrow \bm{X}^\star$
}
\caption{iBART} \label{alg:ibart}
\end{algorithm}

\section{Simulation} \label{sec:3_simulation}
\subsection{Outline}
In Sections \ref{sec:unary} and \ref{sec:binary} we demonstrate that the employed BART-G.SE tends to satisfy the PAN criterion when selecting unary and binary operators, respectively, and evaluate its false positives. Section \ref{sec:simulation} assesses iBART relative to several existing descriptor selection methods in view of the PAN criterion and OIS variable selection accuracy using a complex simulation setting. Section \ref{sec:p_study} showcases the robustness of iBART in the initial input dimension $p$. Section \ref{sec:mis-model} examines the performance of iBART when the operator set $\mathcal{O}$ can not generate the ground truth model. We use Algorithm~\ref{alg:ibart} and the default settings described in Section~\ref{sec:practical_consideration} when implementing iBART, unless stated otherwise. Each simulation is replicated 100 times. We also compare variants of iBART by varying the nonparametric module; see the Supplementary Material for details.

\subsection{Unary operators} \label{sec:unary}
We consider all unary transformations of five primary features $\bm{X} = (\bm{x}_1, \ldots, \bm{x}_5)$, i.e., the descriptor space is $\mathcal{O}_u(\bm{X}) = \bigcup_{i=1}^5 \left\{\bm{x}_i, \bm{x}_i^{-1}, \bm{x}_i^2, \sqrt{\bm{x}_i}, \log(\bm{x}_i), \exp(\bm{x}_i), |\bm{x}_i|, \sin(\pi\bm{x}_i), \cos(\pi\bm{x}_i)\right\}$. For each unary operator $u_j \in \mathcal{O}_u$, we generate the response vector by
\begin{equation} \label{eq:sim.model1}
    \bm{y} = 10u_j(\bm{x}_1) + \bm\varepsilon, \qquad\bm\varepsilon \sim \mathcal{N}_n(\bm{0}, \bm{I}),
\end{equation}
with sample size $n = 200$, yielding nine independent models in total. We draw the primary features $\bm{X}$ independently from the standard normal distribution if the domain of $u_j$ is $\mathbb{R}$, and the \text{Lognormal}(2, 0.5) distribution if the domain is $\mathbb{R}_{\geq 0}$.

The left plot in Figure \ref{fig:TP_all} shows the number of true positives (TP) of BART-G.SE. For each of the nine models in \eqref{eq:sim.model1}, the true descriptor is selected 100/100 times. This suggests that BART-G.SE is capable of identifying the true descriptor with high probability when the descriptor space is populated with unary operators $\mathcal{O}_u$. Other nonparametric methods did not select all TP 100/100 times for all nine scenarios, failing the PAN criterion; see the Supplementary Material for further results and discussion.

The left plot in Figure \ref{fig:FP_all} shows the number of FP of BART-G.SE for each simulation setting. Although BART-G.SE is capable of capturing the true descriptor with high probability, it also selects some non-signal descriptors in some cases. The number of FP is especially high when the true descriptor is $\bm{x}_1$, $|\bm{x}_1|, \exp(\bm{x}_1), \log(\bm{x}_1), \bm{x}_1^2$, or $\sqrt{\bm{x}_1}$. This is due to high collinearity among these six descriptors. In particular, the empirical Pearson correlation between $\log(\bm{x}_1)$ and $\sqrt{\bm{x}_1}$ is over 0.99 under simulation setting \eqref{eq:sim.model1} and thus some false positives are expected. Selecting inactive descriptors in one iteration is less of a concern in variable selection with OIS as they do not constitute misspecified models, and can be further screened out during Step 2 of PAN.

\subsection{Binary operators} \label{sec:binary}
In this simulation study, we apply binary operators $\mathcal{O}_b = \{+, -, \times, /, |-|, \pi_1\}$ on the five primary features $\bm{X} = (\bm{x}_1,\ldots,\bm{x}_5)$. The descriptor space $\mathcal{O}_b(\bm{X}) \in \mathbb{R}^{200 \times 55}$ contains all binary transformations of all possible pairs of the five primary features. For each binary operator $b_j \in \mathcal{O}_b,$ we generate the response vector by
\begin{equation} \label{eq:binary_sim}
    \bm{y} = 10b_j(\bm{x}_1,\bm{x}_2) + \bm\varepsilon, \qquad\bm\varepsilon \sim \mathcal{N}_n(\bm{0}, \bm{I}),
\end{equation}
with sample size $n = 200$, yielding a total of six independent models. We generate the primary features $\bm{X}$ following $\bm{x}_1,\ldots,\bm{x}_5 \overset{\text{i.i.d.}}\sim \mathcal{N}_n(\bm{0},\bm{I}).$

\begin{figure}[hbt!]
    \centering
    \begin{subfigure}[b]{0.8\textwidth}
        \centering
        \includegraphics[width=\textwidth]{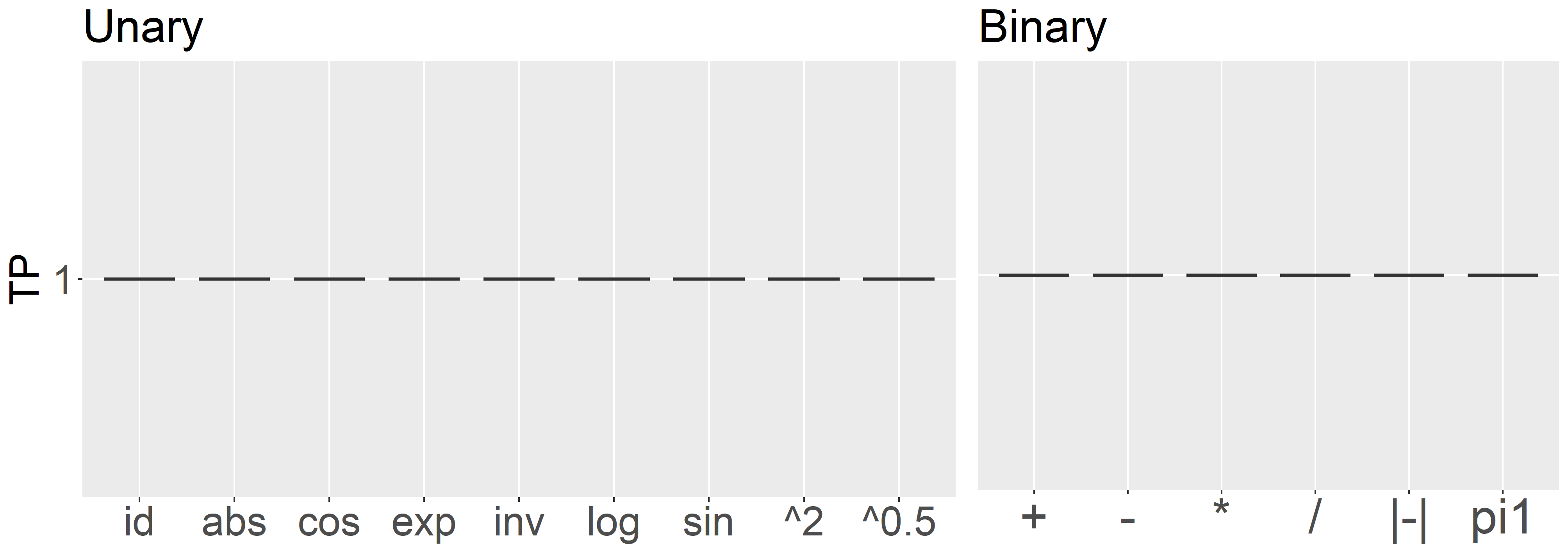}
        \caption{TP}
        \label{fig:TP_all}
    \end{subfigure}
    \begin{subfigure}[b]{0.8\textwidth}
        \centering
        \includegraphics[width=\textwidth]{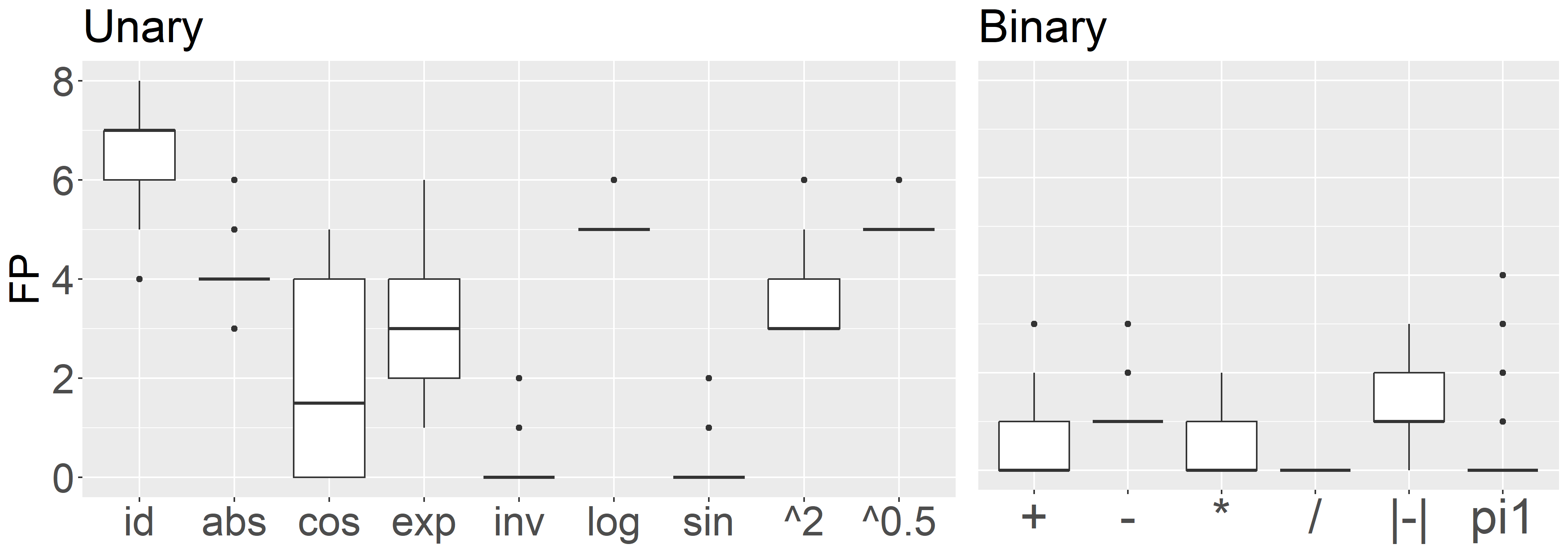}
        \caption{FP}
        \label{fig:FP_all}
    \end{subfigure}
    \caption{Boxplots of TP and FP over 100 simulation replicates under \eqref{eq:sim.model1} and \eqref{eq:binary_sim}.}
\end{figure}

The right plot in Figure \ref{fig:TP_all} shows the number of TP for each of the six models in \eqref{eq:binary_sim}. BART-G.SE is able to identify the true binary descriptor 100/100 times in all six settings, similar to earlier observations in Section~\ref{sec:unary}. The right plot in Figure \ref{fig:FP_all} shows that BART-G.SE may also select some irrelevant descriptors but considerably fewer than in Section \ref{sec:unary}. In particular, when the true descriptor is $(\bm{x}_1 - \bm{x}_2)$, BART-G.SE correctly selects $(\bm{x}_1 - \bm{x}_2)$ but also selects $|\bm{x}_1 - \bm{x}_2|$ 100/100 times. Unlike in Section \ref{sec:unary}, the inclusion of $|\bm{x}_1 - \bm{x}_2|$ is not due to high correlation between $|\bm{x}_1 - \bm{x}_2|$ and $(\bm{x}_1 - \bm{x}_2)$. In fact, their empirical Pearson correlation is merely 0.07. We suspect that $|\bm{x}_1 - \bm{x}_2|$ was selected due to its piecewise monotonicity in $(\bm{x}_1 - \bm{x}_2)$ that tends to be invariant to tree-based methods. Similarly, BART-G.SE selects both $(\bm{x}_1 - \bm{x}_2)$ and $|\bm{x}_1 - \bm{x}_2|$ with high probability when the true descriptor is $|\bm{x}_1 - \bm{x}_2|$. When the true descriptor is $(\bm{x}_1 + \bm{x}_2)$, $\bm{x}_1 \times \bm{x}_2$, $\bm{x}_1/\bm{x}_2$, or $\pi_1(\bm{x}_1, \bm{x}_2)$, the FP of BART-G.SE is nearly zero. This shows that BART-G.SE is not too conservative when high collinearity does not exist.

Our investigation using unary and binary descriptors suggests that the permutation threshold of BART-G.SE tends to satisfy the PAN criterion without being too conservative, making it a good candidate for PAN and OIS variable selection. We next use a more complex simulation to assess iBART in view of the PAN criterion and OIS selection accuracy.

\subsection{Complex descriptors with high-order compositions} \label{sec:simulation} 
In this section, we compare iBART with existing approaches under a complex model and demonstrate superior performance of iBART. We use the following model 
\begin{equation} \label{eq:sim_4}
    \bm{y} = 15\{\exp(\bm{x}_1)-\exp(\bm{x}_2)\}^2 + 20\sin(\pi\bm{x}_3\bm{x}_4) + \bm\varepsilon, \qquad\bm\varepsilon\sim \mathcal{N}_n(\bm{0},\sigma^2\bm{I}),
\end{equation}
with $n = 250$, $p = 10$, and $\sigma = 0.5$. Here the number of primary features $p$ is set to a relatively small number since competing one-shot methods cannot complete the simulation when $p \geq 20$ due to the ultra-high dimension of their descriptor spaces. In Section \ref{sec:p_study}, we demonstrate that iBART scales well in $p$ and gives a robust performance. We use the operator set $\mathcal{O}$ defined in~\eqref{eq:operator_set} with $\pi_1$, and the primary feature vectors $\bm{x}_i$ are drawn independently from a uniform distribution, namely, $\bm{x}_1,\ldots,\bm{x}_p \overset{\text{i.i.d.}}\sim \text{U}_n(-1,1).$ Section B of Supplementary Material demonstrates the effect of dependent primary features on iBART and other methods using the same functional relationship in \eqref{eq:sim_4}.

iBART is implemented in \texttt{R} based on Algorithm 1 in Section~\ref{sec:practical_consideration} using the default settings. We chose the descriptor generating process \eqref{eq:unary_first} in Section~\ref{sec:3_simulation} since the i.i.d. primary features do not show strong collinearity. Two versions of iBART are considered: iBART without and with $\ell_0$-penalized regression, labeled as ``iBART" and ``iBART$+\ell_0$", respectively. The $\ell_0$-penalization finds the best subset of $k$ variables from the set of selected variables using the Akaike information criterion (AIC) with $k \in \{1,2,3,4\}$. We compare the performance of iBART and iBART$+\ell_0$ with SISSO and LASSO. SISSO is implemented using the Fortran 90 program published by \cite{SISSO} on GitHub with the following settings: the descriptor magnitude allowed in the descriptor space is set to $[1 \times 10^{-6}, 1 \times 10^5]$; the size of the SIS-selected subspace is set to 20; the operator composition complexity $M$ is set to 3; the maximum number of operators in a descriptor is set to 6; and the number of selected descriptors $k \in \{1,2,3,4\}$ is tuned by AIC. The \texttt{R} package \texttt{glmnet} is used to implement LASSO and the penalty parameter $\lambda$ is tuned via 10-fold cross-validation to minimize the mean squared error loss. Since the size of $\mathcal{O}^{(3)}(\bm{X})$ exceeds the limit of an \texttt{R} matrix, we give LASSO an advantage by reducing the descriptor space to $\mathcal{O}_u \circ \mathcal{O} \circ \mathcal{O}(\bm{X})$ that aligns with the true data-generating process. We also utilize the $\ell_0$-penalized regression step as in SISSO and iBART+$\ell_0$ for LASSO, leading to LASSO$+\ell_0$. 

For each method, we calculate the number of TP selections, FP selections, and false negative (FN) selections. We use the $F_1$ score as an overall metric to quantify the performance of each method: $F_1 = 2\cdot\frac{\text{Precision} \cdot \text{Recall}}{\text{Precision} + \text{Recall}},$ where $\text{Precision} = \frac{\text{TP}}{\text{TP} + \text{FP}}$ and $\text{Recall} = \frac{\text{TP}}{\text{TP} + \text{FN}}$. The value $F_1 = 1$ means correct identification of the true model without having any FP and FN. In this simulation, the two TPs are $f_1(\bm{X}) = \{\exp(\bm{x}_1)-\exp(\bm{x}_2)\}^2$ and $f_2(\bm{X}) = \sin(\pi\bm{x}_3\bm{x}_4)$.

\begin{figure}[hbt!]
    \centering
    \includegraphics[width = 0.9\linewidth]{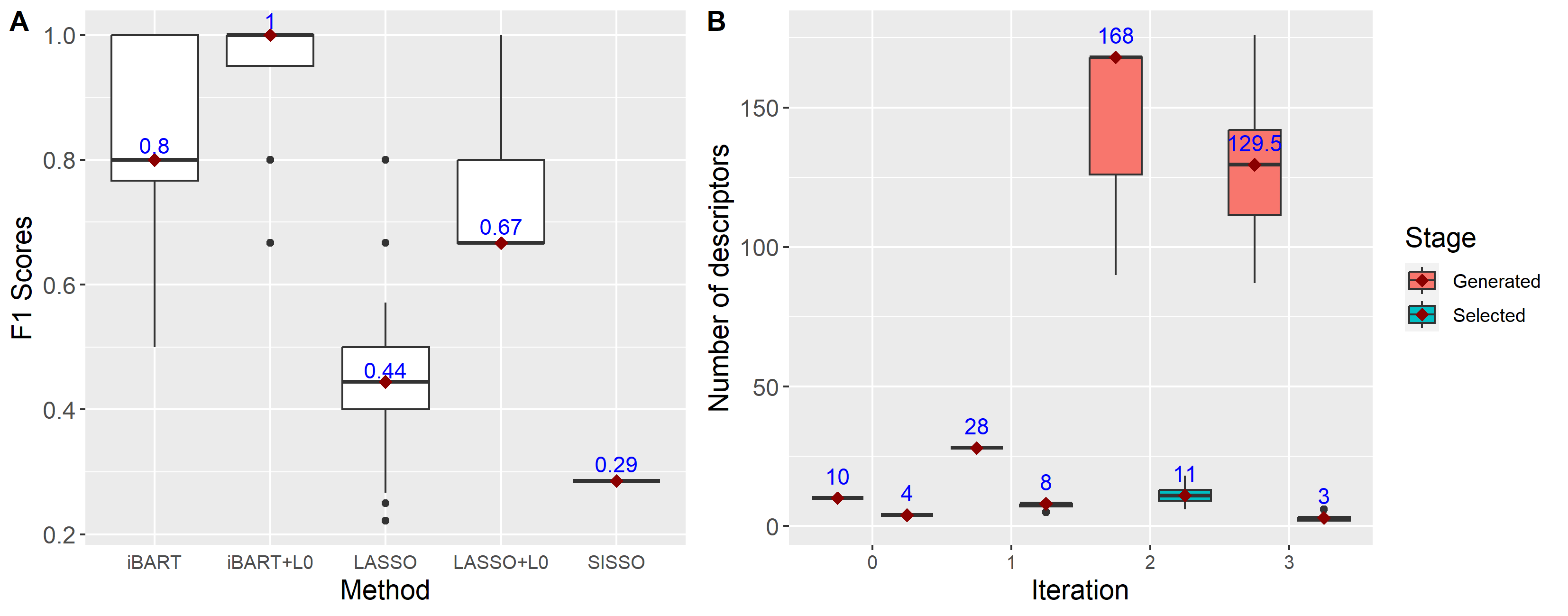}
    \caption{Left: Boxplots of $F_1$ scores over 100 simulations for different methods under Model \eqref{eq:sim_4}. Right: Boxplots of iBART generated and selected descriptors in each iteration.}
    \label{fig:Sec3.4}
\end{figure}

As shown in Figure \ref{fig:Sec3.4}A, both iBART-based methods achieve very high $F_1$ scores, with a median $F_1$ of 0.8 and 1 for iBART and iBART$+\ell_0$, respectively. In particular, both iBART and iBART$+\ell_0$ have 2 TPs in all simulations while incurring an average FP of 0.93 and 0.3, respectively. This demonstrates that iBART satisfies the PAN criterion in all iterations as it is able to identify the 2 TPs 100/100 times in simulations. Furthermore, iBART$+\ell_0$ gives a very low FP, scoring a perfect $F_1$ score 75/100 times. LASSO-based methods and SISSO have lower $F_1$ scores than iBART and iBART$+\ell_0$ but for different reasons. LASSO selects 37.65 descriptors on average, and the 2 TPs are always selected. This means that LASSO also enjoys the PAN criterion but its $F_1$ score is hindered by the large number of FP. With the help of the $\ell_0$-penalized regression, LASSO$+\ell_0$ reduces the average number of FP from 35.65 to 2 while maintaining 2 TPs 100/100 times, substantially increasing the $F_1$ score to 0.67, the third-highest score but still considerably lower than iBART and iBART$+\ell_0$. SISSO, on the other hand, has only 1 TP but 3 FPs in all simulations, which unfortunately does not satisfy the PAN criterion. In particular, SISSO can identify $\sin(\pi\bm{x}_3\bm{x}_4)$ 100/100 times but always selects a false signal, $|\exp(\bm{x}_1) - \exp(\bm{x}_2)|$, a descriptor that has an absolute correlation over 0.9 with the TP, $\{\exp(\bm{x}_1)-\exp(\bm{x}_2)\}^2$. In summary, both iBART-based and LASSO-based methods satisfy the PAN criterion but LASSO-based methods incur more FPs. SISSO, however, fails to identify one of the two descriptors 100/100 times but selects a highly correlated counterpart, indicating its relatively weakened ability to distinguish the true descriptor when there are descriptors highly correlated with the TP. Results not reported here show that replacing BART-G.SE with LASSO in the PAN framework missed TPs with high probability even at the base iteration, indicating the importance of nonparametric variable selection in PAN.

To gain insight into the scalability of iBART, Figure \ref{fig:Sec3.4}B shows the boxplots of the number of iBART generated and selected descriptors in each iteration. Throughout the 100 simulation replicates, iBART generates no more than $168 \leq 2p^2$ descriptors, which is significantly less than the number of descriptors generated by SISSO ($9.26 \times 10^9$) and LASSO ($1.2 \times 10^6$). Such dimension reduction not only reduces runtime and memory usage but also enables iBART to tackle data with much larger $p$, as we show in Sections \ref{sec:p_study} and \ref{sec:4_real_data}.

\subsection{Large \texorpdfstring{$p$}{p}} \label{sec:p_study}
In this section, we demonstrate that the performance of iBART is robust to increase in input dimension $p$ while one-shot descriptor selection methods are not. Under Model~\eqref{eq:sim_4} with $p = 20$, SISSO with the same parameter settings in the preceding section generates more than $3.8 \times 10^{11}$ descriptors and failed to complete one simulation within 24 hours on a server with 1.3TB of memory and 40 CPU cores available to SISSO. LASSO, on the other hand, failed at $p = 20$ since the \texttt{glmnet} function in the \texttt{R} package \texttt{glmnet} cannot handle a matrix of size $250 \times (1.57 \times 10^{7})$. The proposed method iBART instead scales well in the input dimension $p$ partly because of the efficient dimension reduction via the PAN strategy. In particular, when $p = 20$, iBART took an average of 497 seconds to finish, and its memory usage peaks out at 10.85 GB in 100 simulation replicates.

\begin{figure}[hbt!]
    \centering
    \includegraphics[width = 0.6\linewidth]{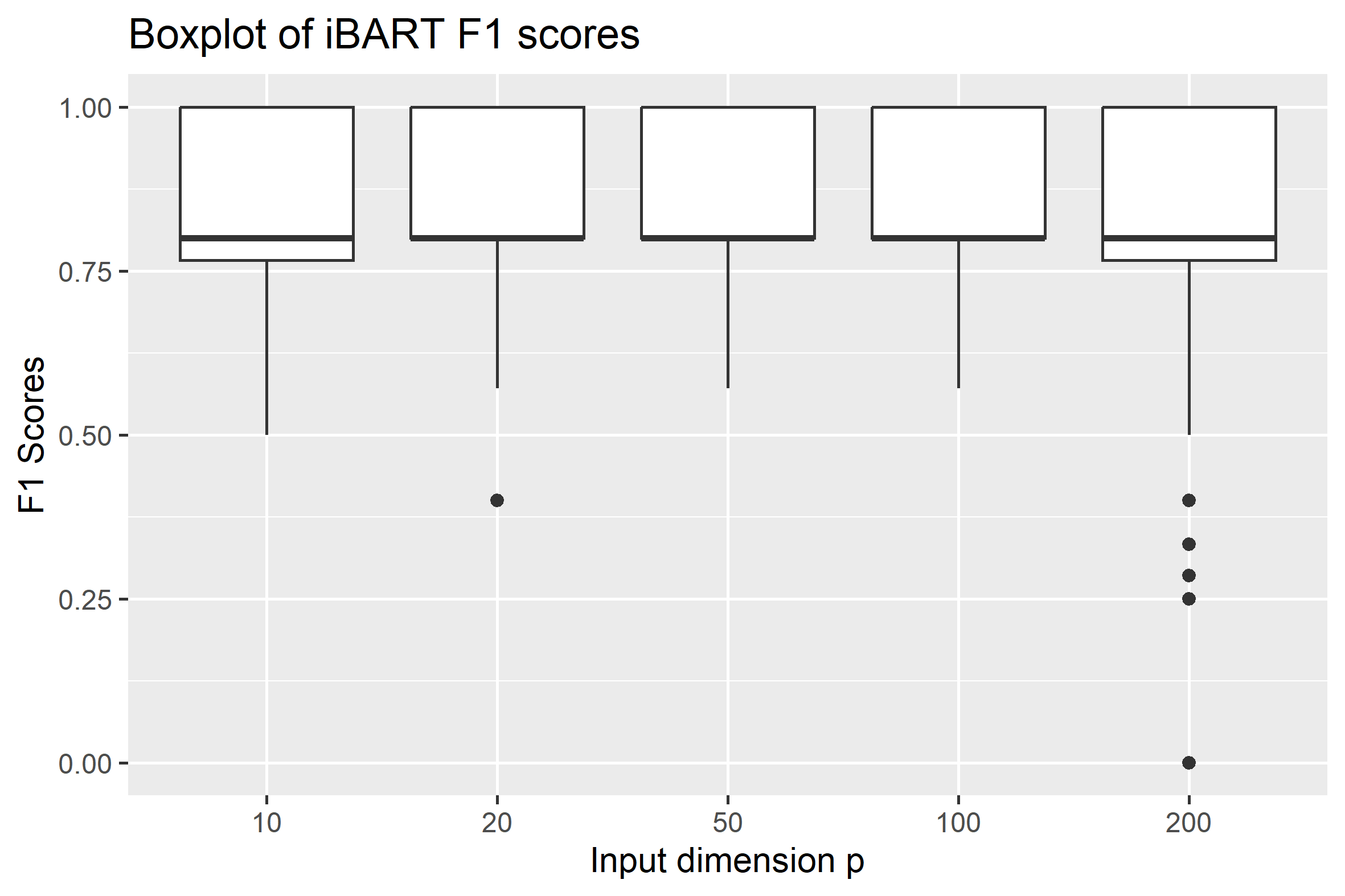}
    \caption{Boxplot of $F_1$ scores for iBART under Model \eqref{eq:sim_4} with $p \in \{10, 20, 50, 100, 200\}$.}
    \label{fig:iBART_diff_p}
\end{figure}
We implement iBART for $p \in \{10, 20, 50, 100, 200\}$ using the same simulation settings in Section~\ref{sec:simulation}. Figure \ref{fig:iBART_diff_p} shows that iBART's $F_1$ score is robust to the increase in $p$. In fact, benefiting from the \textit{ab initio} mechanism through the PAN strategy, the performance of iBART would be identical for varying $p$ as long as it selects $\bm{x}_1, \bm{x}_2, \bm{x}_3, \bm{x}_4$ in the first iteration, which is often the case based on Figure \ref{fig:iBART_diff_p} at least under the current simulation setting. We acknowledge that the stability of the $F_1$ scores across different values of $p$ may be attributed, in part, to the relatively small noise standard deviation. In contrast, one-shot descriptor selection methods suffer significantly from just a small increase in the input dimension $p$ since the descriptor space increases double exponentially in $p$. One way for one-shot description selection methods to circumvent the ultra-high dimension issue is to reduce the maximum composition complexity $M_\text{max}$. However, this undesirably rules out descriptors with the correct composition complexity. For example, the descriptor in \eqref{eq:sim_4} $f_1(\bm{x}) = \{\exp(\bm{x}_1)-\exp(\bm{x}_2)\}^2$ requires at least 3 compositions of operators and reducing $M_\text{max}$ to 2 means that $f_1(\bm{x})$ would never be generated and hence would never be selected. Thus, in a complex scenario, one-shot descriptor selection methods cannot generate and select the complex descriptors unless $p$ is small. We considered a small $p = 10$ in the preceding section to accommodate such limitation of one-shot methods. 

\subsection{Model misspecification} \label{sec:mis-model}
In this section, we examine the behavior of iBART when the operator set $\mathcal{O}$ is not sufficient to generate the data-generating function. We generate data from the model
$\bm{y} = \bm{x}_1^{1.7} + \bm\varepsilon$ 
with $\bm\varepsilon \sim \mathcal{N}_n(\bm{0}, \bm{I})$, $n = 250$, $p = 10$, use the same operator set $\mathcal{O}$ as in Section~\ref{sec:simulation}, and draw the primary features from a uniform distribution, i.e., $\bm{x}_1,\ldots,\bm{x}_p \overset{\text{i.i.d.}}\sim \text{U}_n(0, 3).$ In this simulation, iBART iterates once after screening the primary features. Since iBART only selects $\bm{x}_1$ in the first iteration, 100/100 times, it does not apply the binary operators in $\mathcal{O}$.

\begin{figure}[hbt!] 
    \centering
    \includegraphics[width = 0.9\linewidth]{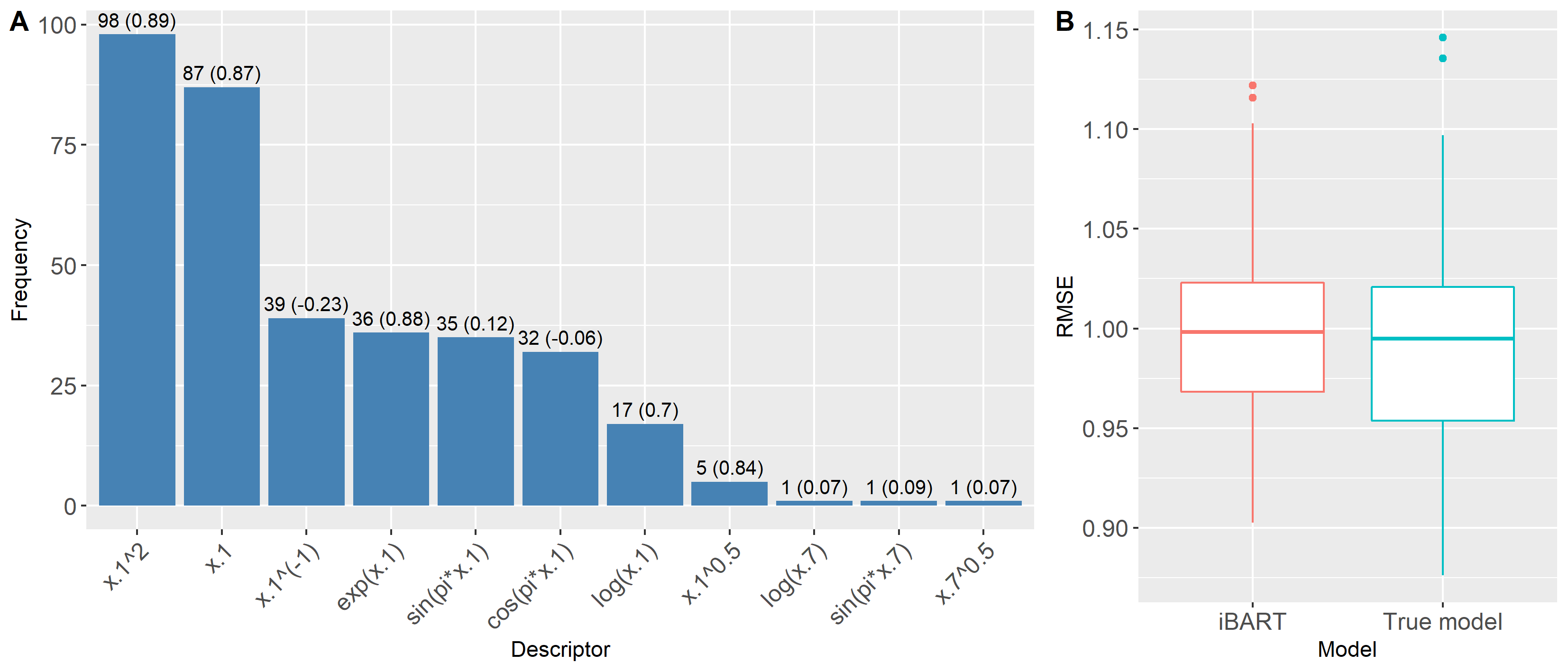}
    \caption{Left: frequency (average Pearson correlation with $\bm{x}_1^{1.7}$) of iBART selected descriptors. Right: root mean squares error (RMSE) of iBART models and the true models.}
    \label{fig:iBART_misspec}
\end{figure}

In contrast to the preceding sections, the computation of true and false positives, as well as the $F_1$ scores, is not feasible here due to the model specification by design. To evaluate the performance of our method, Figure~\ref{fig:iBART_misspec}A shows the frequency of unique descriptors selected by iBART and the average Pearson correlation between the selected descriptor and $\bm{x}_1^{1.7}$ in parentheses. Note that multiple descriptors might be selected in one replication, and therefore the sum of frequencies exceeds 100. We can see that, although the true descriptor $\bm{x}_1^{1.7}$ is not in the candidate descriptor space, iBART selects highly correlated descriptors from the candidate space, including $\bm{x}_1^2$ for 98\% of the time and $\bm{x}_1$ for 87\% of the time. Figure~\ref{fig:iBART_misspec}B shows that the RMSE of the iBART selected models does not deviate much from the RMSE of the true model, indicating a similar explanatory power. These observations suggest that when the employed operator space is not sufficient to generate the ground truth, iBART is capable of selecting a model that closely resembles the true model, at least in the considered simulation setting. This is reassuring as in many applications of OIS, such as the one in Section~\ref{sec:4_real_data}, the goal is precisely to find an accurate but interpretable model that well approximates complex real-world physical systems.

\section{Application to single-atom catalysis} \label{sec:4_real_data}
We apply the proposed method to analyze a single-atom catalysis dataset~\citep{O'Connor2018} in which the goal is to identify physical descriptors that are associated with the binding energy of metal-support pairs calculated by density functional theory (DFT). Single-atom catalysts are popular in modern materials science and chemistry as they offer high reactivity and selectivity while maximizing utilization of the expensive active metal component \citep{Yang2013, O'Connor2018, Wang2018}. However, single-atom catalysts suffer from a lack of stability caused by the tendency for single metal atoms to agglomerate in a process called sintering. To prevent sintering in single-atom catalysts, one can tune the binding strength between single metal atoms and oxide supports. While first principle simulations can calculate the binding energy for given metal-support pairs, modeling their association requires explicit statistical modeling and is key to aid the design of single-atom catalysts that are robust against sintering. Feature engineering leads to physical descriptors constructed using mathematical operators and physical properties of the supported metal and the support, and gains popularity in materials informatics as the obtained descriptors are interpretable and provide insights into the underlying physical relationship. The key challenge is to select the most relevant physical properties among large-scale candidate predictors that have explanatory and predictive power to the binding energy; often the sample size is small as first principle simulations are computationally intensive. 

The data comprise binding energy of $n = 91$ metal-support pairs and $p = 59$ physical properties of these metal-support pairs, in which the binding energy serves as the response $\bm{y}$ while the 59 physical properties constitute the primary features $\bm{X} = (\bm{x}_1,\ldots,\bm{x}_{59})$. We use the operator set given in \eqref{eq:operator_set} but exclude $\sin(\pi\cdot)$ and $\cos(\pi\cdot)$ because they are not physically meaningful in this data application. 

In this analysis, we compare the best $k$ descriptors ($k=1,\ldots,5)$ of iBART$+\ell_0$ (referred to as iBART for brevity), SISSO, and the method proposed in \cite{O'Connor2018}. The method proposed in \cite{O'Connor2018} combines LASSO with extensive domain knowledge; in particular, they rule out a substantial proportion of less promising descriptors using expert knowledge of single-atom catalysis. Such expert-guided, tailored dimension reduction is not needed for SISSO or iBART, but is necessary for LASSO-type methods because the astronomical size of $\mathcal{O}^{(2)}(\bm{X})$ in this application far exceeds the maximum matrix size allowed in many modern programming languages. We refer to \cite{O'Connor2018} as LASSO$^\star$ to distinguish it from LASSO$+\ell_0$ implemented in Section~\ref{sec:simulation}. To enhance interpretability, we eliminate non-physical descriptors, such as $\textit{volumn} + \textit{speed}$, by automatically comparing the units of the constructed descriptors. Results without this constraint are reported in the Supplementary Material, showing that this constraint does not alter our conclusions when comparing methods but substantially improves the interpretability of the identified model.

For iBART, we follow the settings described in Section~\ref{sec:practical_consideration} and use the descriptor generating process \eqref{eq:binary_first}. The SISSO parameters are set as followed. The maximum number of descriptors allowed in a model is set to 5; the descriptor magnitude allowed in the descriptor space is set to $[1 \times 10^{-8}, 1 \times 10^8]$; the size of the SIS-selected subspace is set to 40; the composition complexity $M$ is cap at 2 because $\mathcal{O}^{(3)}(\bm{X})$ and higher complexity space exceed the maximum number of elements allowed in a Fortran 90 array. The LASSO$^\star$ procedure is implemented using the MATLAB code published by \cite{O'Connor2018} with all parameters left as default.

Each method's performance is assessed using out-of-sample RMSE, runtime, and the number of generated descriptors. To calculate out-of-sample RMSE, we randomly partition the $n = 91$ observations into a 90\% training set (82 samples) and a 10\% testing set (9 samples) and repeat this process 50 times. For brevity, we herein refer to out-of-sample RMSE as RMSE.

\begin{figure}[ht!]
    \centering
    \includegraphics[width = 0.8\linewidth]{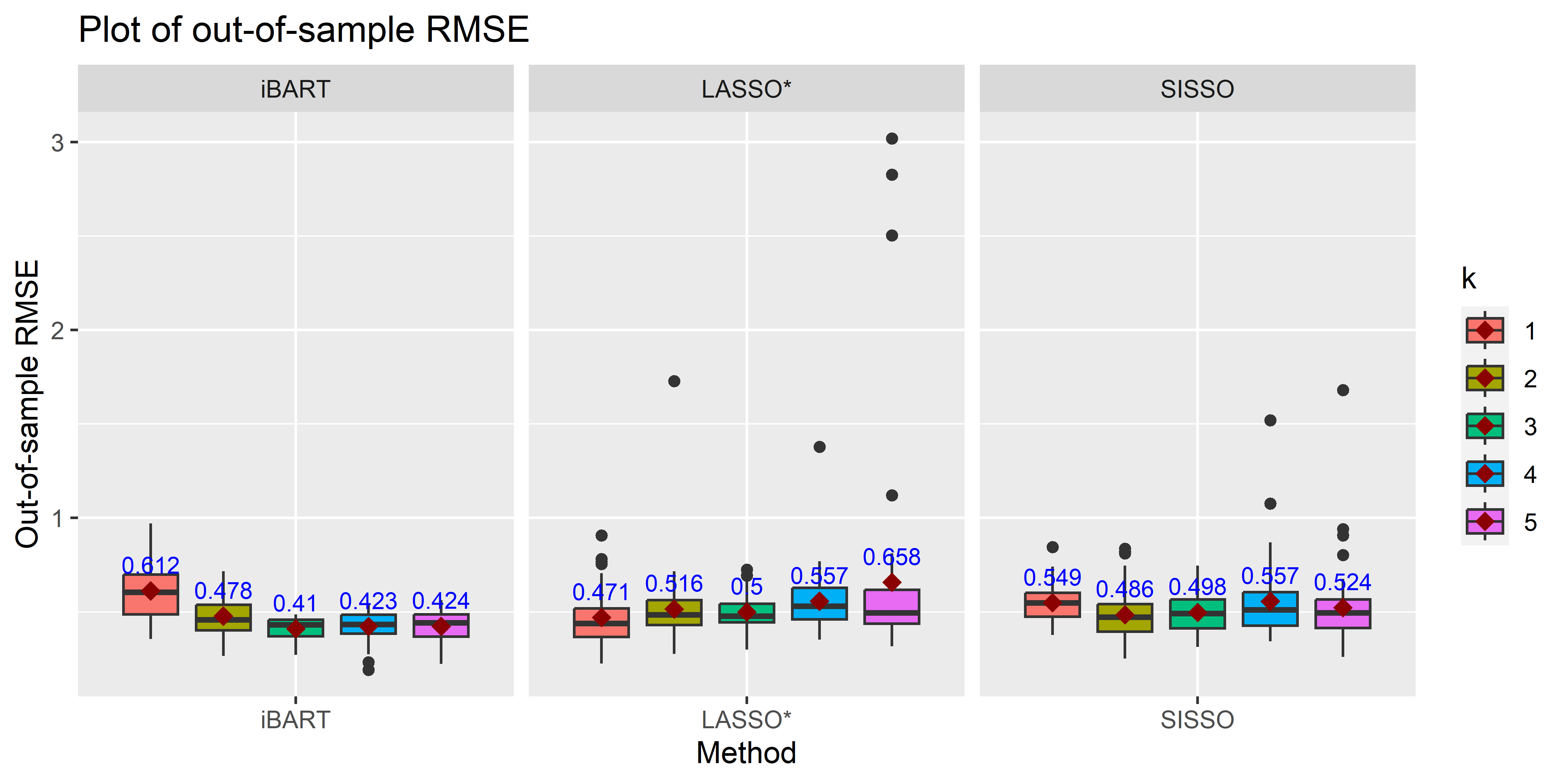}
    \caption{Boxplot of the out-of-sample RMSE for each method across 50 random partitions with $1 \leq k \leq 5$ (left to right in each plot). The blue numbers and the red rhombuses indicate the average out-of-sample RMSE.}
    \label{fig:Connor_RMSE_out}
\end{figure}

From Figure \ref{fig:Connor_RMSE_out}, the smallest average RMSE is attained at $k = 3$ for iBART (0.41), $k = 1$ for LASSO$^\star$ (0.471), and $k = 2$ for SISSO (0.486), that is, iBART reduces the RMSE by 13\% relative to LASSO$^\star$ and 16\% relative to SISSO; also see the first row of Table~\ref{tab:runtime}. iBART outperforms LASSO$^\star$  and SISSO with smaller average RMSE and reduced variability for $k \geq 2$. The larger average RMSE of iBART at $k = 1$ may be partially due to its smaller descriptor space as a result of its alternating descriptor generation process, and our implementation details that give advantages to LASSO$^\star$. Multiple descriptors are often needed to approximate complex physical systems, and the predictive performance of iBART reassuringly improves with more descriptors in the model. On the contrary, the RMSE of LASSO$^\star$ increases with more descriptors in the model, indicating overfitting. For all $k$, we observe iBART does not report large deviations from the average performance, suggesting robustness to training sets compared to LASSO$^\star$ and SISSO. 

In addition to the performance gain in RMSE, iBART also leads to a substantial reduction in computing time and memory usage. Table~\ref{tab:runtime} shows that iBART leads to over 30-fold ($6943 / 225 = 30.86$) speedup compared to SISSO, and over 480-fold ($5511 \times 20 / 225 = 489$) speedup compared to LASSO$^\star$, tested on an Intel Xeon Gold 6230 CPU @ 2.10 GHz using either 1 or 20 CPU cores. We use the published code by authors for competing methods, and the runtime reported here does not isolate the effect of various programming languages (R for iBART, MATLAB for LASSO$^\star$, and Fortran 90 for SISSO). We did not obtain an accurate single-core runtime of LASSO$^\star$ because of its poor scalability, and instead multiplied its 20-core runtime by 20 for comparison (i.e., $5511 \times 20 \times 50 / 3600 = 1530$ hours); the exact speedup of iBART compared to the single-core runtime of LASSO$^\star$ may vary due to the reduced communication cost among multiple cores and other factors. We remark that the LASSO$^\star$ method implemented here is given an advantage with an additional dimension reduction step, and an exploration of higher complexity descriptor space as in iBART is computationally prohibitive for LASSO$^\star$. The excellent scalability of iBART transforms into memory efficiency as the descriptor space in iBART is orders of magnitude smaller than that of competing methods. Table~\ref{tab:runtime} shows that iBART generates a descriptor space of size $627 < O(p^2)$ in the last iteration on average. Thanks to this significantly smaller descriptor space, we were able to run iBART on a laptop with only 16GB of memory; in contrast, LASSO$^\star$ and SISSO failed at the descriptor generation step owing to the enormous descriptor space they try to generate, and require server-grade computing facilities.   

\begin{table}[ht]
    \caption{\label{tab:runtime}Performance comparison of three methods: out-of-sample RMSE, runtime, and the number of generated descriptors, averaged over 50 cross validations.}
    \centering
    \fbox{
        \begin{tabular}[t]{lccc}
        & iBART & LASSO$^\star$ & SISSO \\
        \hline
        RMSE & 0.41 & 0.471 & 0.486 \\
        Runtime & 225 sec & 5511 sec $\times \,20$ & 6943 sec \\
        Number of generated descriptors & 627 & $3.3 \times 10^5$ & $5.5 \times 10^7$\\
        \end{tabular}}
\end{table}

\begin{table}[ht]
    \caption{\label{tab:k_model_table}Selected linear models by iBART for $k \in \{1,2,3,4,5\}$.}
    \renewcommand\arraystretch{2}
    \centering
    \fbox{
        \begin{tabular}[t]{ll}
        $k$ & Selected descriptors\\
        \hline
        1 & $\left|\frac{\Delta H_{\text{sub}} - \Delta H_{\text{f,ox,bulk}}}{\Delta E_\text{vac}}\right|$ \\ 
        2 & $\frac{EA^s \cdot (\Delta H_{\text{sub}} - \Delta H_{\text{f,ox,bulk}})}{N_{\text{val}}^m/CN_{\text{bulk}}^m}, \left|\frac{\Delta H_{\text{sub}} - \Delta H_{\text{f,ox,bulk}}}{\Delta E_\text{vac}}\right|$ \\
        3 & $EA^s \cdot \Delta H_{\text{f,ox,bulk}}, \left|\frac{\Delta H_{\text{sub}} - \Delta H_{\text{f,ox,bulk}}}{\Delta E_{\text{vac}}}\right|, \left|\frac{(\eta^{1/3})^m}{N_{\text{val}}^m \cdot \Delta E_{\text{vac}}}\right|$ \\
        4 & $EA^s \cdot \Delta H_{\text{f,ox,bulk}}, \frac{(\eta^{1/3})^m}{\phi^s},  \frac{(\eta^{1/3})^m \cdot IE_4^s}{(N_{\text{val}}^m)^2}, \left|\frac{EA^s \cdot (\Delta H_{\text{sub}} - \Delta H_{\text{f,ox,bulk}})}{\Delta E_{\text{vac}}}\right|$ \\
        5 & $\frac{(\eta^{1/3})^m}{\phi^s}, \frac{\Delta H_{\text{f,ox,bulk}}}{\Delta E_{\text{vac}}}, \frac{EA^s \cdot (\Delta H_{\text{sub}} - \Delta H_{\text{f,ox,bulk}})}{N_{\text{val}}^m/CN_{\text{bulk}}^m}, \frac{(\eta^{1/3})^m \cdot IE_4^s}{(N_{\text{val}}^m)^2}, \left|\frac{\Delta H_{\text{sub}} - \Delta H_{\text{f,ox,bulk}}}{\Delta E_\text{vac}}\right|$\\
        \end{tabular}}
\end{table}

\begin{table}[ht]
    \caption{\label{tab:primary_feature_names}Descriptions of the selected primary features by iBART.}
    \centering
    \fbox{
        \begin{tabular}{*{2}{l}}
        Primary feature               & Physical meaning\\
        \hline
        $\Delta H_{\text{sub}}$       & Heat of sublimation\\
        $\Delta H_{\text{f,ox,bulk}}$ & Oxidation energy of the bulk metal\\
        $\Delta E_\text{vac}$         & Oxygen vacancy energy\\
        $EA^s$                        & Electron affinity of support\\
        $N_{\text{val}}^m$            & Number of valence electrons in metal adatom\\
        $CN_{\text{bulk}}^m$          & {Coordination number of the surface metal atom in the bulk phase}\\
        $(\eta^{1/3})^m$              & Discontinuity in electron density of metal adatom\\
        $\phi^s$                      & Chemical potential of the electrons in support\\
        $IE_4^s$                      & \makecell[l]{4th ionization energy of support with the bulk metal in\\ the $4^+$ oxidation state}\\
        \end{tabular}}
\end{table}

\begin{figure}[hbt!]
    \centering
    \includegraphics[width = 0.8\linewidth]{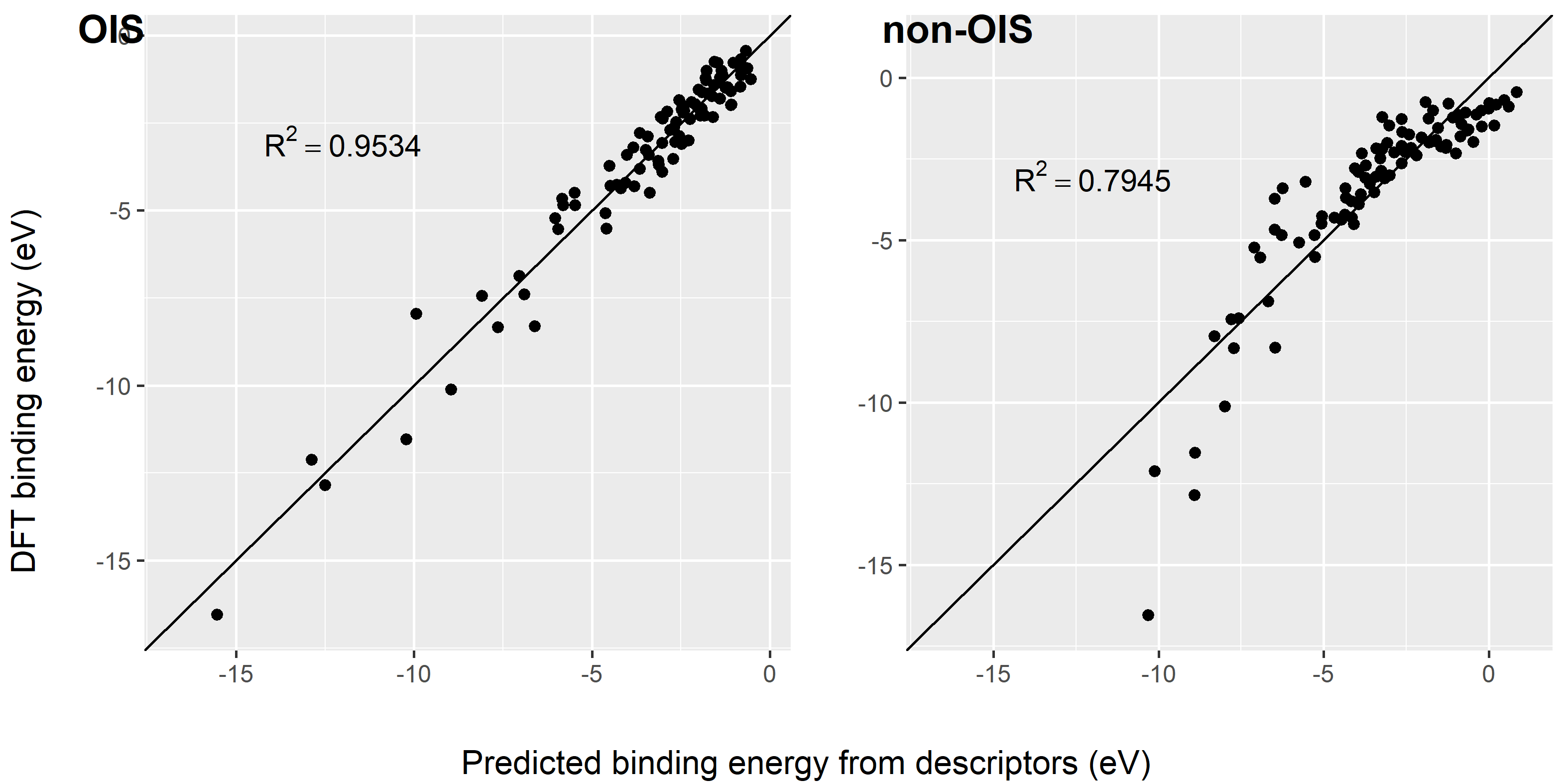}
    \caption{DFT binding energies versus predicted values using linear models with OIS and without OIS. The black line is $y = x$. Each model has $k=3$ descriptors.}
    \label{fig:in_sample_plot}
\end{figure}

Figure \ref{fig:in_sample_plot} demonstrates a clear advantage of the OIS model over the non-OIS model. The OIS model is the iBART model with $k=3$, the optimum model suggested by RMSE. The non-OIS model is a simple least squares model with $\bm{X}$ as the design matrix and no feature engineering step. For ease of comparison, the non-OIS model also has $k=3$ predictors determined by best subset selection. The OIS model yields an $R^2$ of 0.9534, indicating high explanatory power. In contrast, the non-OIS model gives an $R^2 = 0.7945$ and shows poor fitting performance as the scatter plot deviates from the diagonal line $y = x$ considerably. Both the OIS and the non-OIS models have analytical forms, but the OIS model gains more explanatory power and gives an insightful description of the response through nonlinearity of its predictors. In particular, the non-OIS model is $\hat{y}_{\text{non-OIS}} = -4.7 -0.4 \times \Delta H_{\text{f,ox}} + 0.3 \times N_{\text{val}}^s + 1.0 \times \Delta E_{\text{vac}}$, while the OIS model is
\begin{equation} 
    \hat{y}_{\text{OIS}} = - 0.01 + 0.4 \times (EA^s \cdot \Delta H_{\text{f,ox,bulk}}) - 0.6 \times \left|\frac{\Delta H_{\text{sub}} - \Delta H_{\text{f,ox,bulk}}}{\Delta E_{\text{vac}}}\right| - 19.6 \times \left|\frac{(\eta^{1/3})^m}{N_{\text{val}}^m \cdot \Delta E_{\text{vac}}}\right|.
\end{equation}
This OIS model pinpoints targeted descriptors to guide further investigation into the underlying physical discovery.

\section{Discussion} \label{sec:5_discussion}
In this article, we study variable selection in the presence of feature engineering that is widely applicable in many scientific fields to provide interpretable models. Unlike in classical variable selection, candidate predictors are engineered from primary features and composite operators. While this problem has become increasingly important in science, such as the emerging field of materials informatics, the induced new geometry has not been studied in the statistical literature. We propose a new strategy ``parametrics assisted by nonparametrics", or PAN, to efficiently explore the descriptor space and achieve nonparametric dimension reduction for linear models. Using BART-G.SE as the nonparametric module, the proposed method iBART iteratively constructs and selects complex descriptors. Compared to one-shot descriptor construction approaches, iBART does not operate on an ultra-high dimensional descriptor space and thus substantially mitigates the ``curse of dimensionality" and high correlations among descriptors. We introduce the OIS framework, define a PAN criterion, and assess iBART through the lens of this criterion. This sets a foundation for future research in interpretable model selection with feature engineering.

Other than the methodological contributions, we use extensive experiments to demonstrate appealing empirical features of iBART that may be crucial for practitioners. iBART automates the feature generating step; one-shot methods such as SISSO often require user intervention as otherwise they are not scalable to handle the overwhelmingly large descriptor space---this descriptor space increases double exponentially in the number of primary features and binary operators as a result of compositions. iBART has excellent scalability and leads to robust performance when the dimension of primary features increases to a level that renders existing methods computationally infeasible. Compared to SISSO, which is widely perceived as state-of-the-art in the field of materials genomes, our data application shows that iBART reduces the out-of-sample RMSE by 16\% with an over 30-fold speedup and a fraction of memory demand. Overall, the proposed method accomplishes traditionally ``server-required" tasks using a regular laptop or desktop with improved accuracy. Beyond materials genomes illustrated in Section \ref{sec:4_real_data}, iBART can be applied to a vast domain where interpretable modeling is of interest. 

There are interesting next directions building on our iBART approach. First, OIS provides a useful perspective for structured variable selection by introducing operators and compositions. Certain operator sets may be particularly useful depending on the application; for example, the composite multiply operator leads to high-order interactions. It is interesting to examine the performance of iBART with a diverse choice of operators. Also, we have focused on finite sample performance when assessing selection accuracy through simulations partly in view of the limited sample size typically available in the motivating example. It is nevertheless interesting to theoretically investigate the necessary conditions for the PAN criterion to hold for selected nonparametric variable selection methods such as BART when the sample size diverges.

\section*{Acknowledgments}
We thank the editor, associate editor, and reviewers for constructive comments that helped to improve the paper. \if0\blind{This work was partially supported by the grant [Blinded].}\fi 
\if1\blind{This research was partially supported by the Big-Data Private-Cloud Research Cyberinfrastructure MRI-award funded by NSF under grant CNS-1338099 and by Rice University's Center for Research Computing. Shengbin Ye's research was supported by NIH grant T32CA096520. Meng Li's research was partially supported by NSF grants DMS-2015569 and DMS/NIGMS-2153704. Thomas P. Senftle's research was supported by NSF grant CBET-2143941.}\fi 

\noindent \textbf{Disclosure Statement:}
The authors report there are no competing interests to declare.

\bibliographystyle{jasa_acs}
\bibliography{ref}

\end{document}


\def\spacingset#1{\renewcommand{\baselinestretch}%
{#1}\small\normalsize} \spacingset{1}

\newcommand{\titletext}{Supplementary Material for ``Operator-induced structural variable selection for identifying materials genes''}
\if1\blind
{
  \title{\bf \titletext}
  \author[1]{Shengbin Ye \thanks{sy53@rice.edu}}
  \author[2]{Thomas P.~Senftle \thanks{tsenftle@rice.edu}}
  \author[1]{Meng Li \thanks{meng@rice.edu}}
  \affil[1]{Department of Statistics, Rice University, Houston, TX 77005}
  \affil[2]{Department of Chemical and Biomolecular Engineering, Rice University, Houston, TX 77005}
  \maketitle
} \fi

\if0\blind
{
  \bigskip
  \bigskip
  \bigskip
  \begin{center}
    {\LARGE\bf \titletext}
  \end{center}
  \markblankline
  \markblankline
  \markblankline
  \markblankline
  \markblankline
  \medskip
} \fi
\doublespacing

\renewcommand\thesection{\Alph{section}} 
\setcounter{table}{0}
\renewcommand{\thetable}{S\arabic{table}}
\setcounter{figure}{0} 
\renewcommand{\thefigure}{S\arabic{figure}}    

This Supplementary Material contains a detailed comparison of variants of iBART by varying the nonparametric module with numerical experiments and discussion (Section~\ref{sec:iBART_variant}), additional simulation results using correlated primary feature (Section~\ref{sec:corr}), {additional real data application results without enforcing unit consistency (Section~\ref{sec:without_unit_info})}, and the proof of Theorem 2.1 (Section~\ref{sec:proof}).

\section{Variants of iBART} \label{sec:iBART_variant}

\subsection{Nonparametric variable selection in PAN}
We consider three nonparametric variable selection methods in addition to BART-G.SE \citep{bleich2014} in the PAN framework, including ABC Bayesian Forests \citep{ABC_forest}, DART \citep{DART}, and Iterative Nonparametric Independence Screening (INIS) \citep{INIS}. While there are many other nonparametric methods, we choose these three because of their empirical and theoretical performances and their readily available implementation. These three methods represent nonparametric variable selection in both Bayesian and frequentist regimes with ABC Bayesian Forests and DART being Bayesian while INIS being frequentist. We now give a brief overview of each method.

ABC Bayesian Forests is a BART-based method for variable selection. It utilizes the spike-and-forest priors \citep{BART_post_concentration} to induce sparsity on the regression tree by limiting the set of variables $\mathcal{S} \subseteq[p]$ that a regression tree can grow on, and performs variable selection by evaluating the marginal inclusion probabilities. \citet{ABC_forest} proposes to use Approximate Bayesian Computation (ABC) based on data-splitting for posterior sampling. The data-splitting strategy with a default of 50-50 proposal-acceptance/rejection split has been found to improve the acceptance rate of ABC in practice, but it can negatively impact the performance of ABC Bayesian Forests in cases of small sample size, as in often seen in applications of OIS variable selection.

DART is also based on BART but it induces sparsity differently from ABC Bayesian Forests. DART penalizes the splitting proportion of covariates through the use of Dirichlet prior. This means that all covariates are available for splitting but some covariates have a lower chance of being split on. Since the Dirichlet prior is conjugate to the BART priors, the existing backfitting MCMC algorithm for BART can be adapted to DART with minor modifications. This enables efficient sampling from the DART posterior. 

INIS is a frequentist method for nonparametric variable selection. It circumvents the high dimensional issue by considering $p$ marginal nonparametric regression problems by regressing the response on each $X_j$ for $1 \leq j \leq p$. Variable selection is achieved by thresholding the utilities of these $p$ marginal nonparametric regression models. 

In the next two sections, we compare the empirical performances of BART-G.SE, ABC Bayesian Forests, DART, and INIS in the PAN framework using both simulated and real data. We also compare their performances with LASSO and SISSO.

\subsection{Performance in simulated data}
In this section, we replicate the simulation study in Sections 3.2--3.4 of the main paper to comprehensively study the effect of different nonparametric variable selection methods on the PAN framework. ABC Bayesian Forests is implemented using an R package provided by the authors of \citet{ABC_forest}. DART is implemented using an R package written by the author of \citet{DART} (\url{https://github.com/theodds/dartMachine}). INIS is implemented using R source code written by the authors of \citet{INIS} (\url{https://github.com/yangfengstat/NIS}).

\subsubsection{Unary operators} \label{sec:unary_supp}
Following Section 3.2 of the main paper, we consider all unary transformations $\mathcal{O}_u = \{I, \cdot^{-1}, \cdot^2, \sqrt{\cdot}, \log, \exp, \text{abs}, \sin(\pi\cdot), \cos(\pi\cdot)\}$ of the five primary features $\bm{X} = (\bm{x}_1,\ldots,\bm{x}_5)$. For each unary operator $u_j \in \mathcal{O}_u$, we generate the corresponding response vector by
\begin{equation}
    \bm{y} = 10u_j(\bm{x}_1) + \bm\varepsilon, \qquad\bm\varepsilon \sim \mathcal{N}_n(\bm{0}, \bm{I}),
\end{equation}
with a sample size of $n = 200$, yielding nine independent models in total. That is, the correct descriptors in all nine models are unary transformations of $\bm{x}_1$, and $\bm{x}_2,\ldots,\bm{x}_5$ are irrelevant. For details on the simulation setup and BART-G.SE tuning parameters, please refer to the main paper. Below we outline the tuning parameters used for ABC Bayesian Forests, DART, and INIS.

ABC Bayesian Forests is trained with $B = 2,000$ ABC samples (default is $B = 1,000$), where only a fraction of ABC samples (top 20\%, default is top 10\%) are kept in the reference table for marginal inclusion probability (MIP) calculation. For each ABC sample, a BART with 20 trees is trained and we draw the last 2 BART samples after 200 burn-in MCMC iterations. We note the posterior iteration and burn-in iteration were the default settings in the R package. We adopt the Median Probability Model (MPM) to select descriptors. That is, descriptors with MIP $\geq 0.5$ were deemed as significant.

DART is trained with 50 trees (the default) and the posterior is summarized with 5,000 MCMC iterations after 5,000 burn-in iterations. All other tuning parameters are left as default. The variable selection criterion is the same as ABC Bayesian Forests.

INIS is trained with the INIS-penGAM algorithm outlined in Section 4.1 of \citet{INIS}. The algorithm iterates until convergence or after 50 iterations, whichever occurs first. All other tuning parameters are set to their default values.

\begin{figure}[h]
    \centering
    \includegraphics[width=0.96\textwidth]{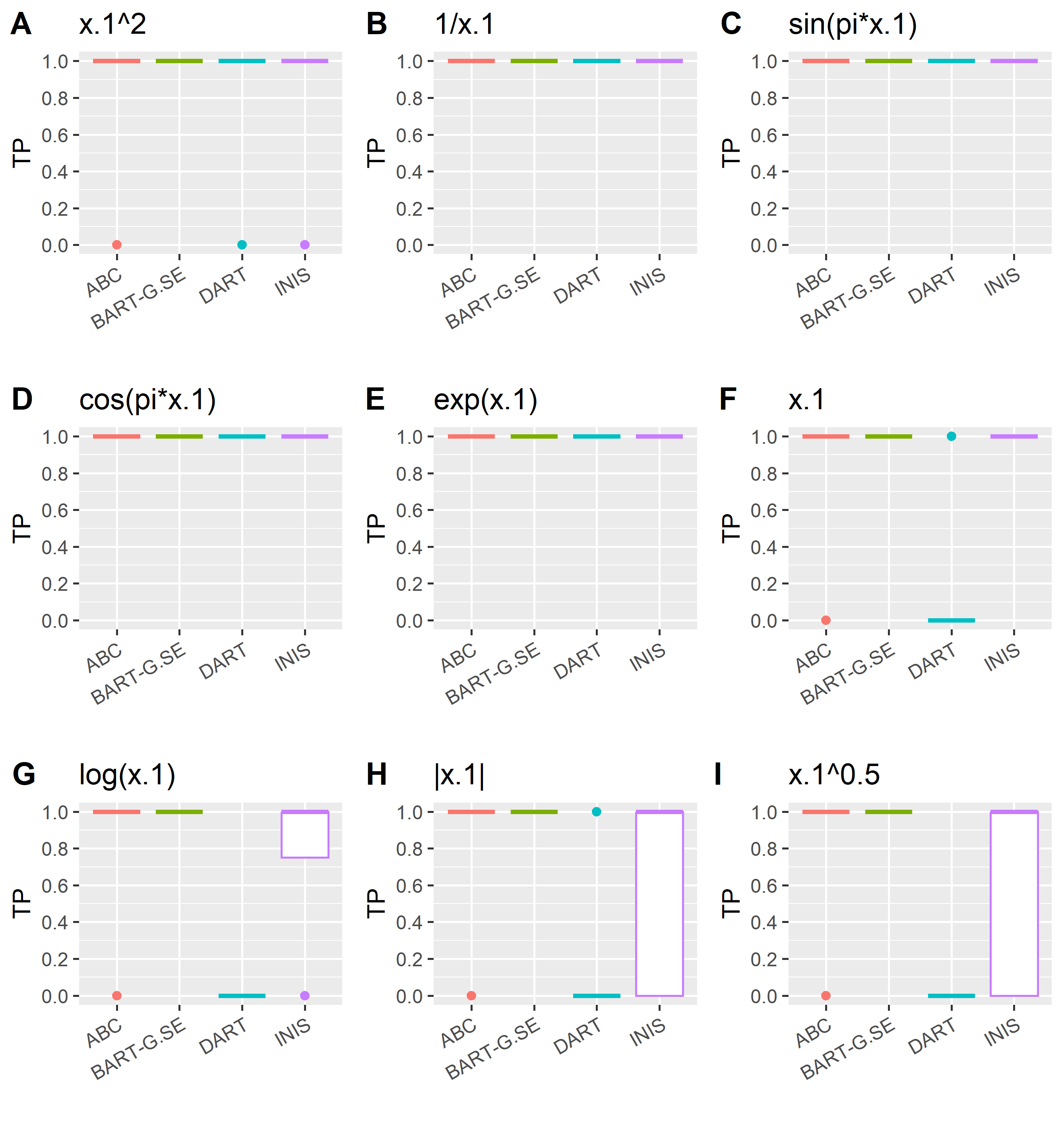}
    \caption{Boxplots of TP. \textbf{A}, The true descriptor is $\bm{x}_1^2$. \textbf{B}, The true descriptor is $\bm{x}_1^{-1}$. \textbf{C}, The true descriptor is $\sin(\pi\bm{x}_1)$. \textbf{D}, The true descriptor is $\cos(\pi\bm{x}_1)$. \textbf{E}, The true descriptor is $\exp(\bm{x}_1)$. \textbf{F}, The true descriptor is $\bm{x}_1$. \textbf{G}, The true descriptor is $\log(\bm{x}_1)$. \textbf{H}, The true descriptor is $|\bm{x}_1|$. \textbf{I}, The true descriptor is $\sqrt{\bm{x}_1}$.}
    \label{fig:s3.2_TP}
\end{figure}

Figure \ref{fig:s3.2_TP} shows the boxplots of TP for all four methods in nine different scenarios. Figures \ref{fig:s3.2_TP}A--E show that all methods have a median TP of 1 in the first five scenarios, i.e., the true descriptor is either $\bm{x}_1^2$, $\bm{x}_1^{-1}$, $\sin(\pi\bm{x}_1)$, $\cos(\pi\bm{x}_1)$, or $\exp(\bm{x}_1)$. This implies that these five unary operators can be easily identified by all four methods. Nonetheless, Figure \ref{fig:s3.2_TP}A shows that ABC Bayesian Forests, DART, and INIS did not select $\bm{x}_1^2$ in some simulation replications. Specifically, ABC Bayesian Forests, DART, and INIS did not select the true descriptor $11/100$, $1/100$, and $3/100$ of the time, respectively.

In the Scenarios F--I, not all methods were able to identify the true descriptor consistently. Both ABC Bayesian Forests and BART-G.SE have comparable TP performance but only BART-G.SE was able to identify the TP 100/100 of the time in Scenarios F--I. TP rates in Table~\ref{tab:unary_TP} indicate that INIS can identify the correct descriptor with decent probability and DART can hardly select the correct descriptor in Scenarios F--I.

\begin{table}[ht]
    \caption{\label{tab:unary_TP}TP rate over 100 simulations of four methods in Scenarios F--I.}
    \centering
    \fbox{
        \begin{tabular}[t]{lcccc}
        & ABC & BART-G.SE & DART & INIS \\
        \hline
        $\bm{x}_1$        & 99\% & 100\% & 1\% & 100\% \\
        $\log(\bm{x}_1)$  & 87\% & 100\% & 0\% & 75\% \\
        $|\bm{x}_1|$      & 92\% & 100\% & 1\% & 72\% \\
        $\sqrt{\bm{x}_1}$ & 81\% & 100\% & 0\% & 74\%
        \end{tabular}}
\end{table}

\begin{figure}[h]
    \centering
    \includegraphics[height=0.85\textheight]{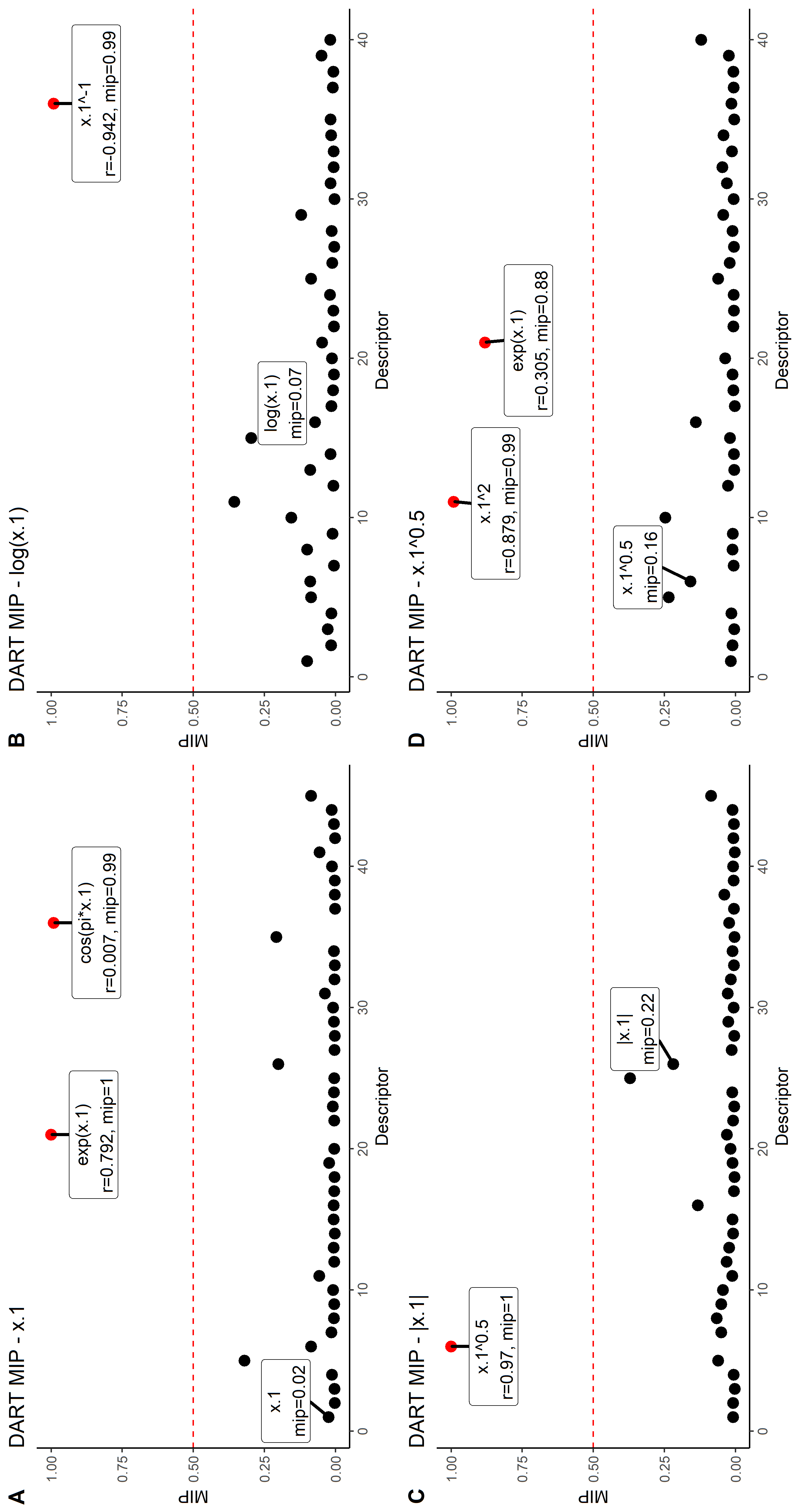}
    \caption{Marginal inclusion probability plot of DART. Red dots denote the selected descriptor by DART; Red dash line denotes the MPM threshold; $r$ denotes the average Pearson correlation with the correct descriptor over 100 simulations; MIP denotes the marginal inclusion probability of the descriptor over 100 simulations. \textbf{A}, The true descriptor is $\bm{x}_1$. \textbf{B}, The true descriptor is $\log(\bm{x}_1)$. \textbf{C}, The true descriptor is $|\bm{x}_1|$. \textbf{D}, The true descriptor is $\sqrt{\bm{x}_1}$.}
    \label{fig:s3.2_DART_mip}
\end{figure}

To see why DART had such a low TP rate in Scenarios F--I, we plot the MIP of each descriptor in Figure~\ref{fig:s3.2_DART_mip}. Note that MIP of the true descriptor is minimal while those of the selected non-signal descriptors are nearly 1. Figure~\ref{fig:s3.2_DART_mip} also shows that the MIPs of most descriptors are near 0, indicating a strong penalization of the Dirichlet prior. Hence, the presence of many transformations of active primary variables poses challenges to methods developed in traditional nonparametric selection settings, and they need to be reassessed by the PAN criterion and OIS variable selection. 

\begin{figure}[h]
    \centering
    \includegraphics[width=\textwidth]{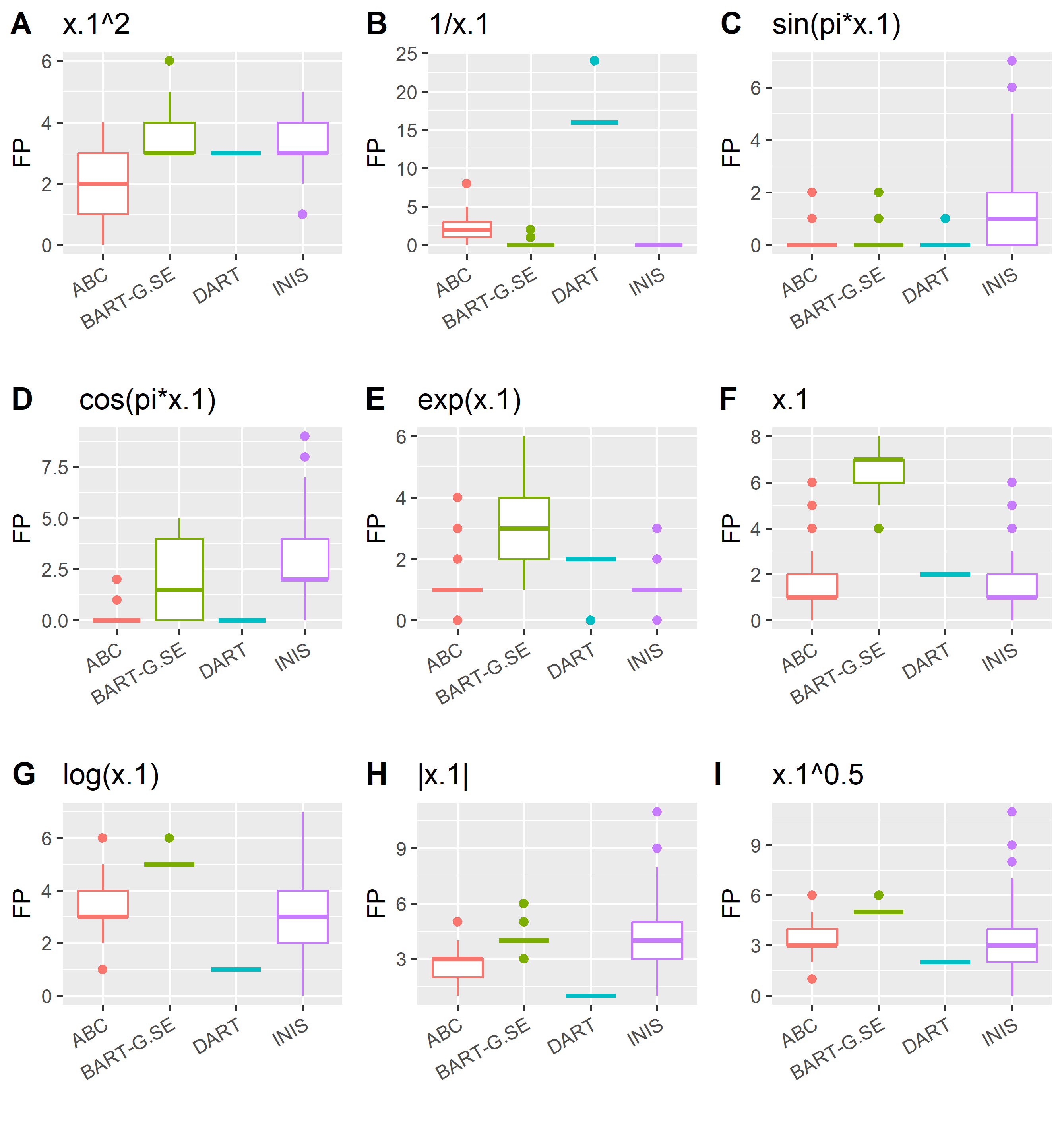}
    \caption{Boxplots of FP. \textbf{A}, The true descriptor is $\bm{x}_1^2$. \textbf{B}, The true descriptor is $\bm{x}_1^{-1}$. \textbf{C}, The true descriptor is $\sin(\pi\bm{x}_1)$. \textbf{D}, The true descriptor is $\cos(\pi\bm{x}_1)$. \textbf{E}, The true descriptor is $\exp(\bm{x}_1)$. \textbf{F}, The true descriptor is $\bm{x}_1$. \textbf{G}, The true descriptor is $\log(\bm{x}_1)$. \textbf{H}, The true descriptor is $|\bm{x}_1|$. \textbf{I}, The true descriptor is $\sqrt{\bm{x}_1}$.}
    \label{fig:s3.2_FP}
\end{figure}

Figure~\ref{fig:s3.2_FP} shows the FP distribution for each method in Scenarios A--I. The average median FP numbers across all scenarios are 1.7 for ABC Bayesian Forests, 2 for INIS, 3 for DART, and 3.2 for BART-G.SE. 

In summary, while BART-G.SE often selects slightly more false positives than competing methods, it is the \emph{only} method tested that achieves a 100\% TP rate in all nine unary scenarios. In view of the asymmetric effect of false positives and false negatives on descriptor selection, BART-G.SE satisfies the PAN criterion in all nine unary scenarios under the simulation settings and thus appears to be well suited for the PAN framework and descriptor selection. 

\subsubsection{Binary operators}
We consider all binary transformations $\mathcal{O}_b = \{+, -, \times, /, |-|, \pi_1\}$ on any pairs of five primary features $\bm{X} = (\bm{x}_1,\ldots,\bm{x}_5)$. For each binary operator $b_j \in \mathcal{O}_b$, we generate the response vector by
\begin{equation}
    \bm{y} = 10b_j(\bm{x}_1, \bm{x}_2) + \bm\varepsilon, \qquad\bm\varepsilon \sim \mathcal{N}_n(\bm{0}, \bm{I})
\end{equation}
with sample size $n = 200$, yielding six independent models. That is, the correct descriptors in all six models are binary transformations of $(\bm{x}_1, \bm{x}_2)$ and $\bm{x}_3,\ldots,\bm{x}_5$ are irrelevant. Tuning parameters for each method are identical to their settings in Section~\ref{sec:unary_supp}.

\begin{figure}[h]
    \centering
    \includegraphics[width=\textwidth]{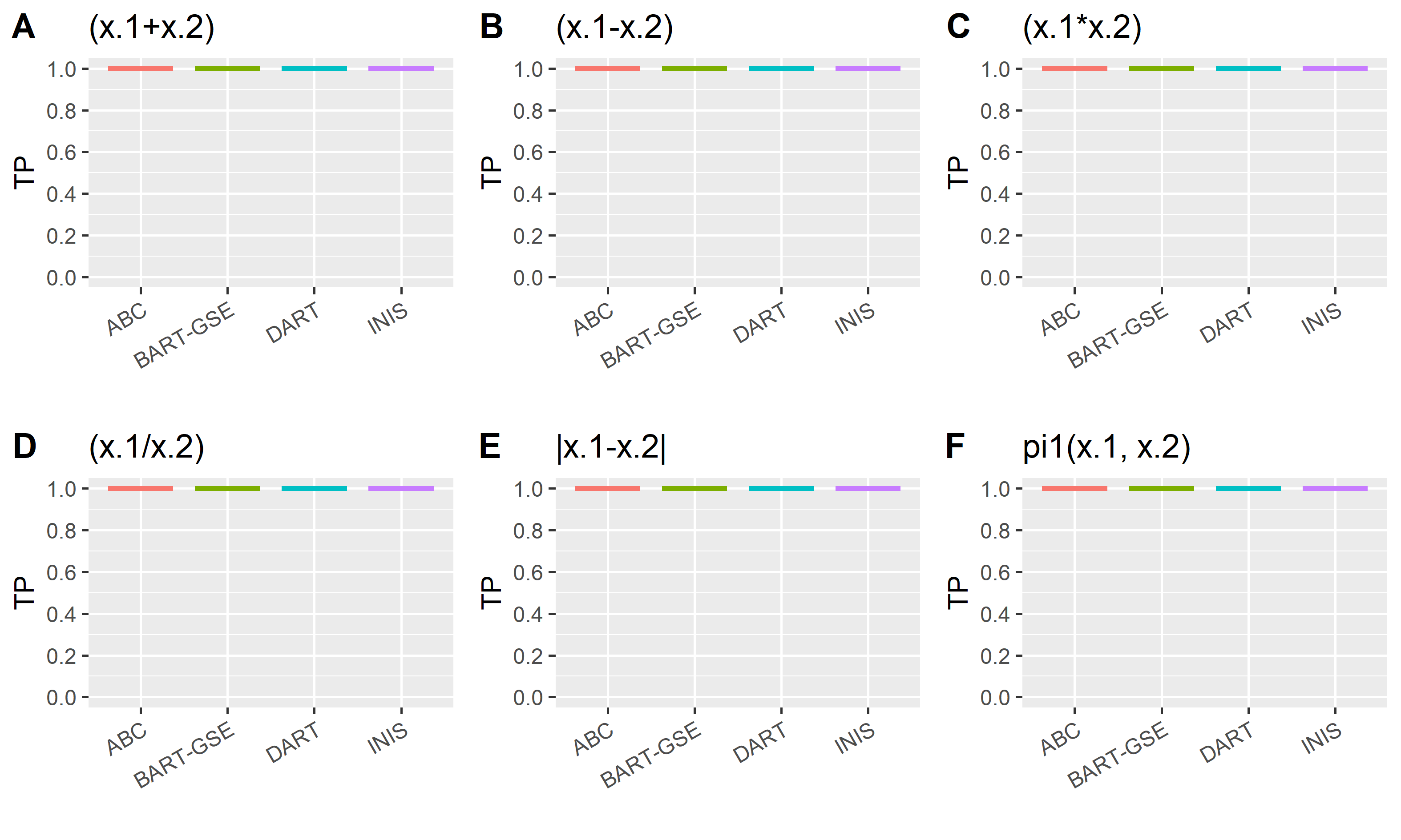}
    \caption{Boxplots of TP. \textbf{A}, The true descriptor is $(\bm{x}_1 + \bm{x}_2)$. \textbf{B}, The true descriptor is $(\bm{x}_1 - \bm{x}_2)$. \textbf{C}, The true descriptor is $(\bm{x}_1 \times \bm{x}_2)$. \textbf{D}, The true descriptor is $(\bm{x}_1 / \bm{x}_2)$. \textbf{E}, The true descriptor is $|\bm{x}_1 - \bm{x}_2|$. \textbf{F}, The true descriptor is $\pi_1(\bm{x}_1, \bm{x}_2)$.}
    \label{fig:s3.3_TP}
\end{figure}

Similar to the finding in the main paper, the selection of binary descriptors is a seemingly easy task as illustrated in Figure~\ref{fig:s3.3_TP}. All methods score a 100\% TP rate in all six binary scenarios. The FP selection in binary scenarios is also noticeably lower than in unary scenarios. The average median FP numbers across the six scenarios are 0 for ABC Bayesian Forests, 0.6 for BART-G.SE, and 1.2 for INIS. DART, on the other hand, exhibits an inconsistent FP performance. With $(\bm{x}_1/\bm{x}_2)$ being the true descriptor, DART was able to select the true descriptor, but it also selected an additional 10 FPs 99/100 times and 15 FPs 1/100 time. Recall in Section~\ref{sec:unary_supp} that DART selects significantly many FPs when the true descriptor is $\bm{x}_1^{-1}$ and this behavior extends to the $(\bm{x}_1/\bm{x}_2)$ case. These two descriptors are actually similar after rewriting $(\bm{x}_1/\bm{x}_2)$ as $(\bm{x}_1 \times \bm{x}_2^{-1})$. This seems to suggest that DART has a difficulty in distinguishing the inverse operator from the other unary operators, resulting in many FPs. 

\begin{figure}[h]
    \centering
    \includegraphics[width=\textwidth]{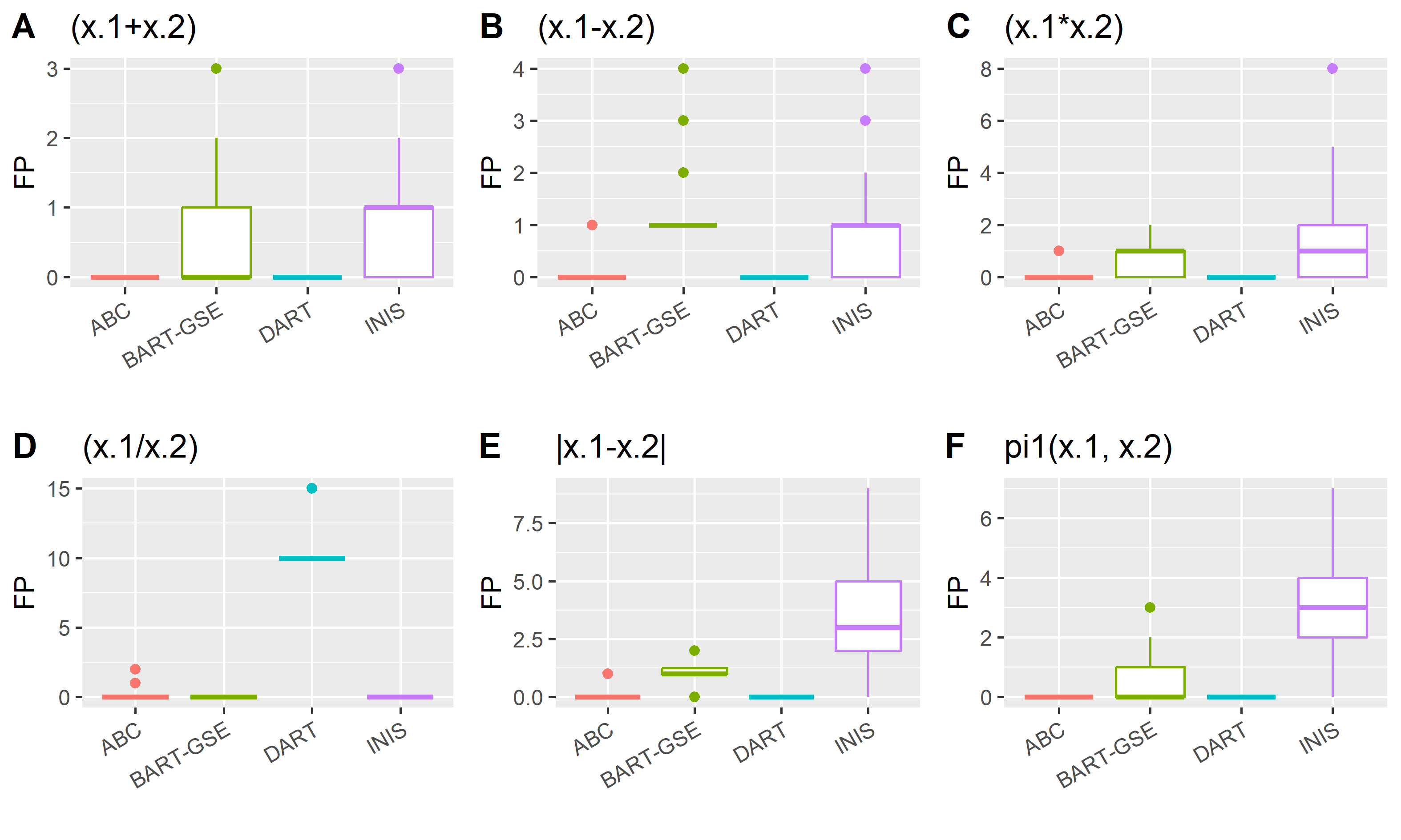}
    \caption{Boxplots of FP. \textbf{A}, The true descriptor is $(\bm{x}_1 + \bm{x}_2)$. \textbf{B}, The true descriptor is $(\bm{x}_1 - \bm{x}_2)$. \textbf{C}, The true descriptor is $(\bm{x}_1 \times \bm{x}_2)$. \textbf{D}, The true descriptor is $(\bm{x}_1 / \bm{x}_2)$. \textbf{E}, The true descriptor is $|\bm{x}_1 - \bm{x}_2|$. \textbf{F}, The true descriptor is $\pi_1(\bm{x}_1, \bm{x}_2)$.}
    \label{fig:s3.3_FP}
\end{figure}

\begin{figure}[h]
    \centering
    \includegraphics[width=\linewidth]{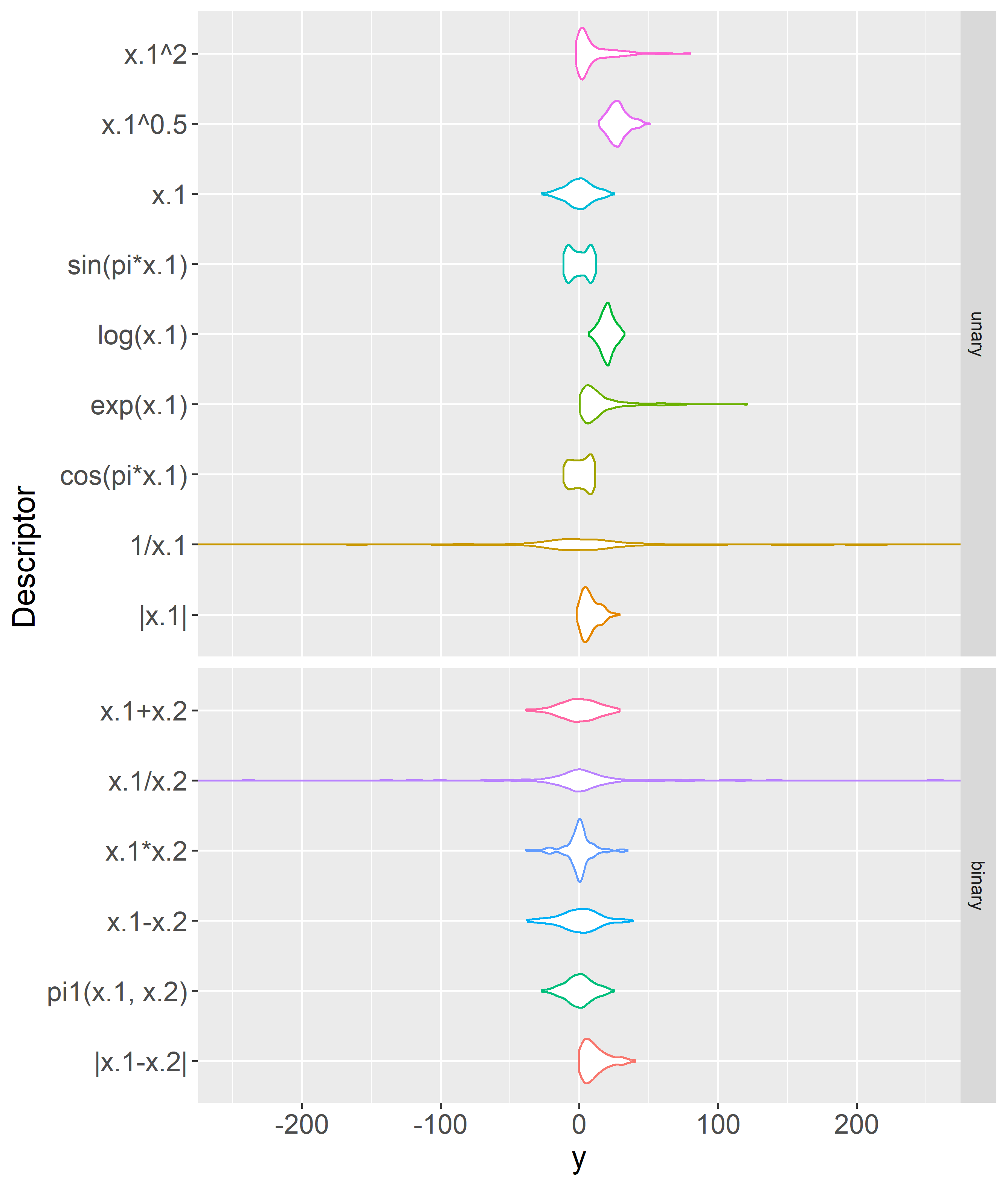}
    \caption{Violin plots of $\bm{y}$ under different models. Range of $\bm{y}$ is limited to $(-250, 250)$ for visualization. Range of $\bm{y} = 10\bm{x}_1^{-1} + \bm\varepsilon$ is $(-1762, 2145)$. Range of $\bm{y} = 10(\bm{x}_1/\bm{x}_2) + \bm\varepsilon$ is $(-573, 663)$.}
    \label{fig:s3.3_y_dist}
\end{figure}

To dissect this phenomenon, we visualize the distribution of the response vector $\bm{y}$ in each unary and binary scenario in Figure~\ref{fig:s3.3_y_dist}. The major difference between the inverse operator and the other operators lies in the tail behavior, where the inverse operator has much longer tails than the other operators. This suggests that the long tail characteristics of the inverse operator hamper the performance of DART, whereby the tree acceptance rate for the inverse operator case is much lower than that for the other operators. As shown in Figure~\ref{fig:DART_accept_inv}, the tree acceptance rate for the inverse operator case is near 0\% after the burn-in period, whereas those for the other cases are about 20\%. This indicates that DART encounters difficulties in efficiently exploring the full posterior space after the burn-in period, which may account for its poor performance in the inverse operator case.

\begin{figure}[h]
    \centering
    \begin{subfigure}[b]{0.48\textwidth}
        \centering
        \includegraphics[width = \textwidth]{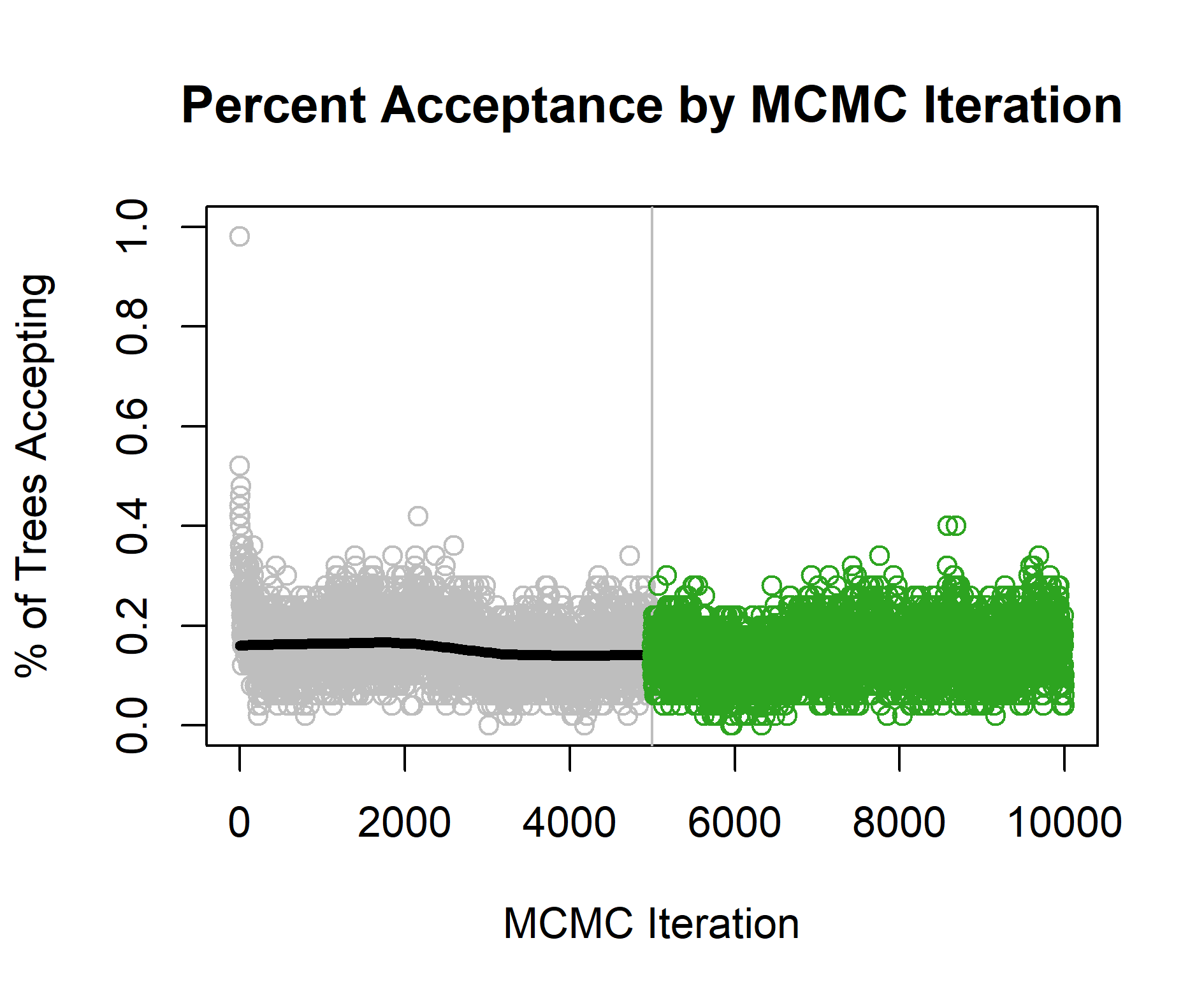}
        \caption{$\bm{y} = 10\bm{x}_1^2 + \bm\varepsilon$}
        \label{fig:DART_accept_sqre}
    \end{subfigure}
    \hfill
    \begin{subfigure}[b]{0.48\textwidth}
        \centering
        \includegraphics[width = \textwidth]{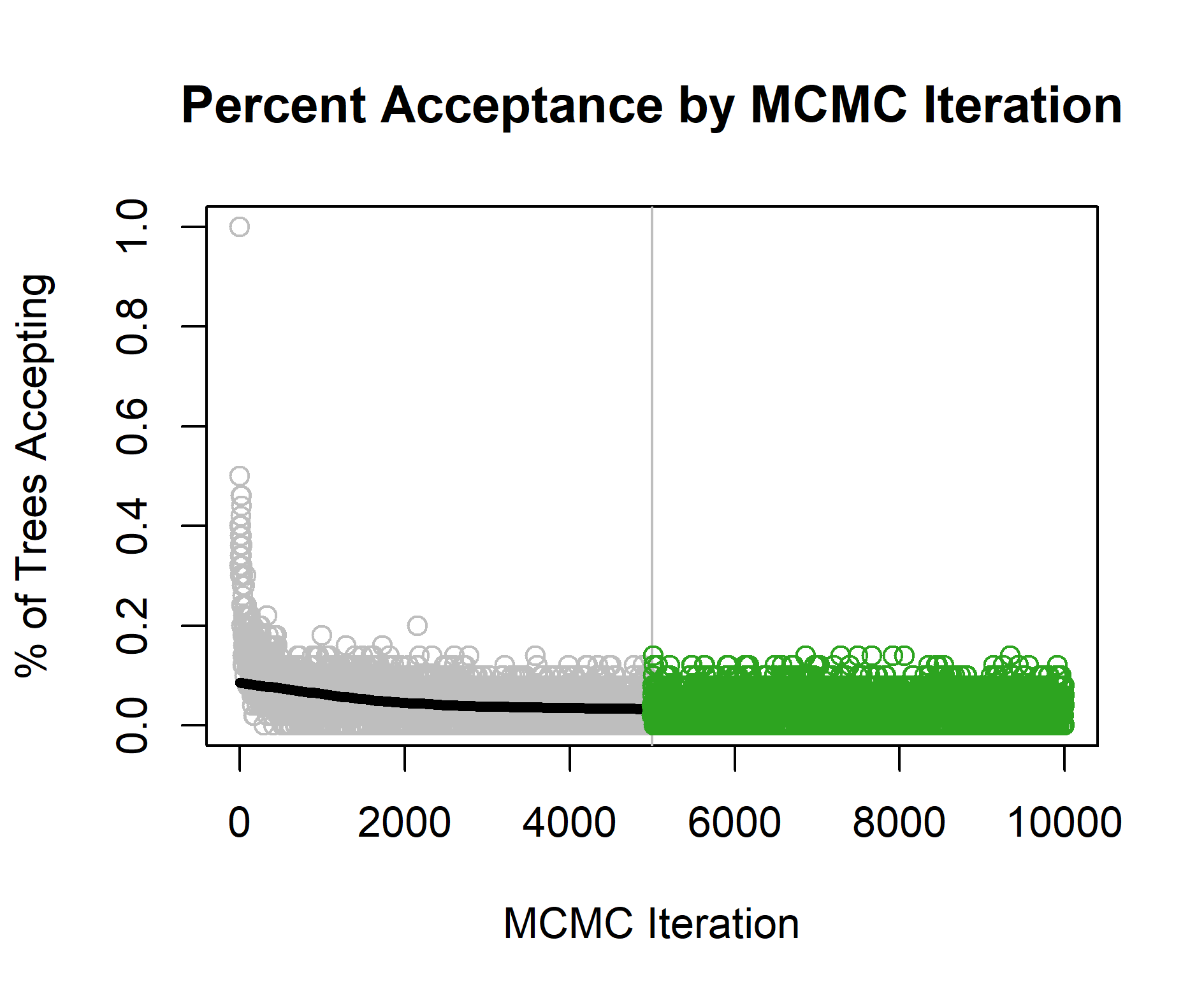}
        \caption{$\bm{y} = 10\bm{x}_1^{-1} + \bm\varepsilon$}
        \label{fig:DART_accept_inv}
    \end{subfigure}
    \begin{subfigure}[b]{0.48\textwidth}
        \centering
        \includegraphics[width = \textwidth]{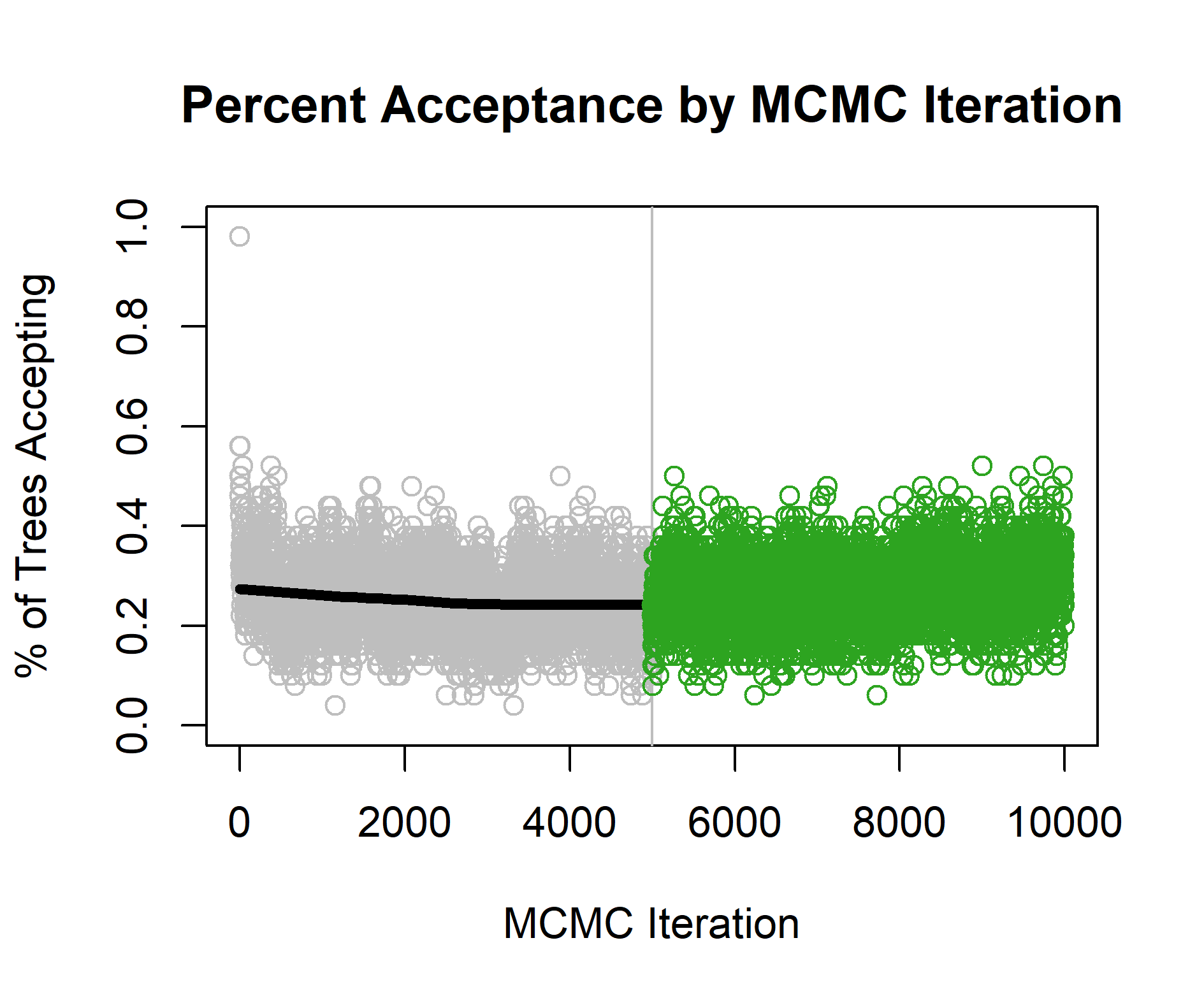}
        \caption{$\bm{y} = 10\sin(\bm{x}_1) + \bm\varepsilon$}
        \label{fig:DART_accept_sin}
    \end{subfigure}
    \hfill
    \begin{subfigure}[b]{0.48\textwidth}
        \centering
        \includegraphics[width = \textwidth]{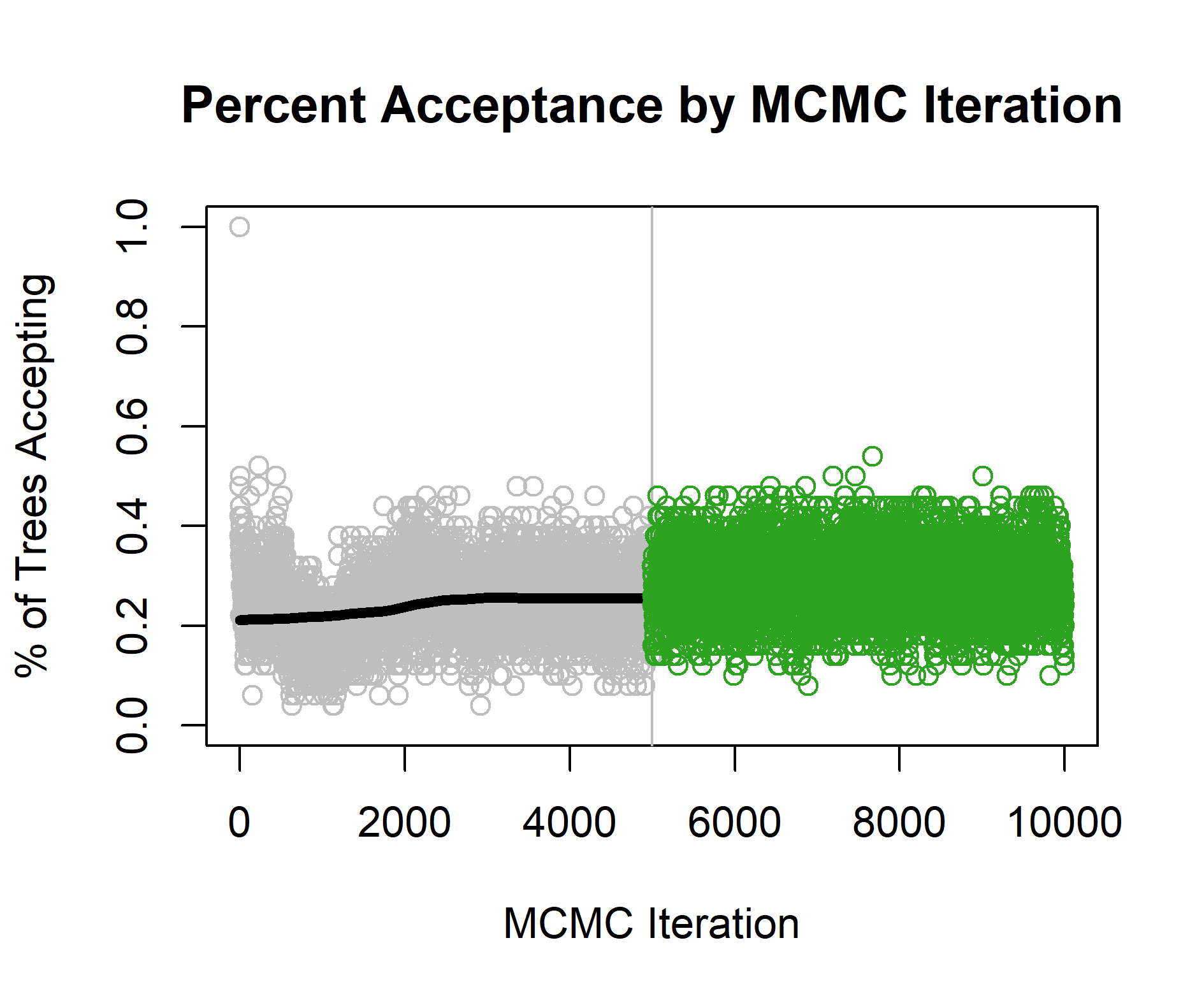}
        \caption{$\bm{y} = 10\cos(\bm{x}_1) + \bm\varepsilon$}
        \label{fig:DART_accept_cos}
    \end{subfigure}
    \begin{subfigure}[b]{0.48\textwidth}
        \centering
        \includegraphics[width = \textwidth]{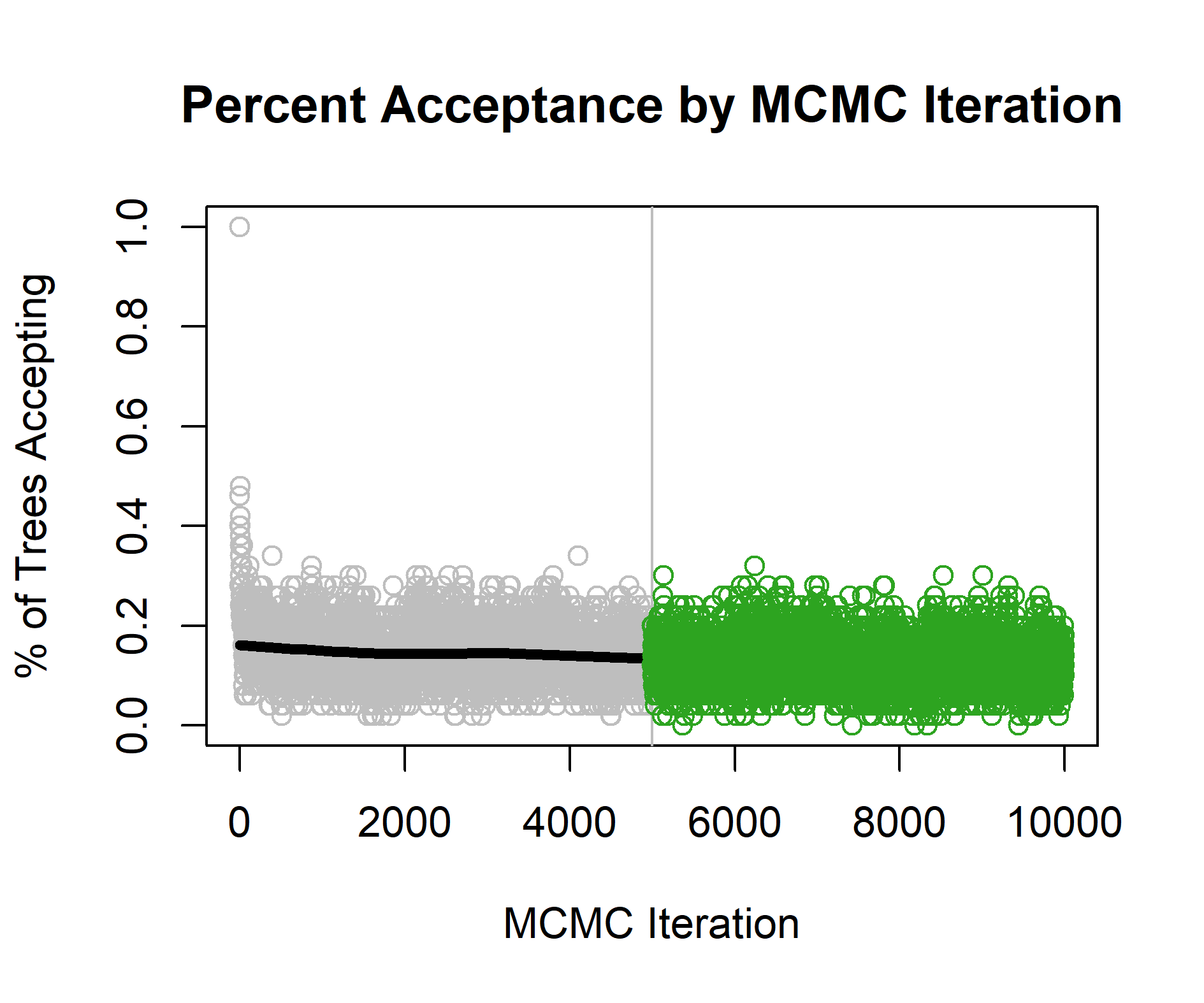}
        \caption{$\bm{y} = 10\exp(\bm{x}_1) + \bm\varepsilon$}
        \label{fig:DART_accept_exp}
    \end{subfigure}
    \hfill
    \begin{subfigure}[b]{0.48\textwidth}
        \centering
        \includegraphics[width = \textwidth]{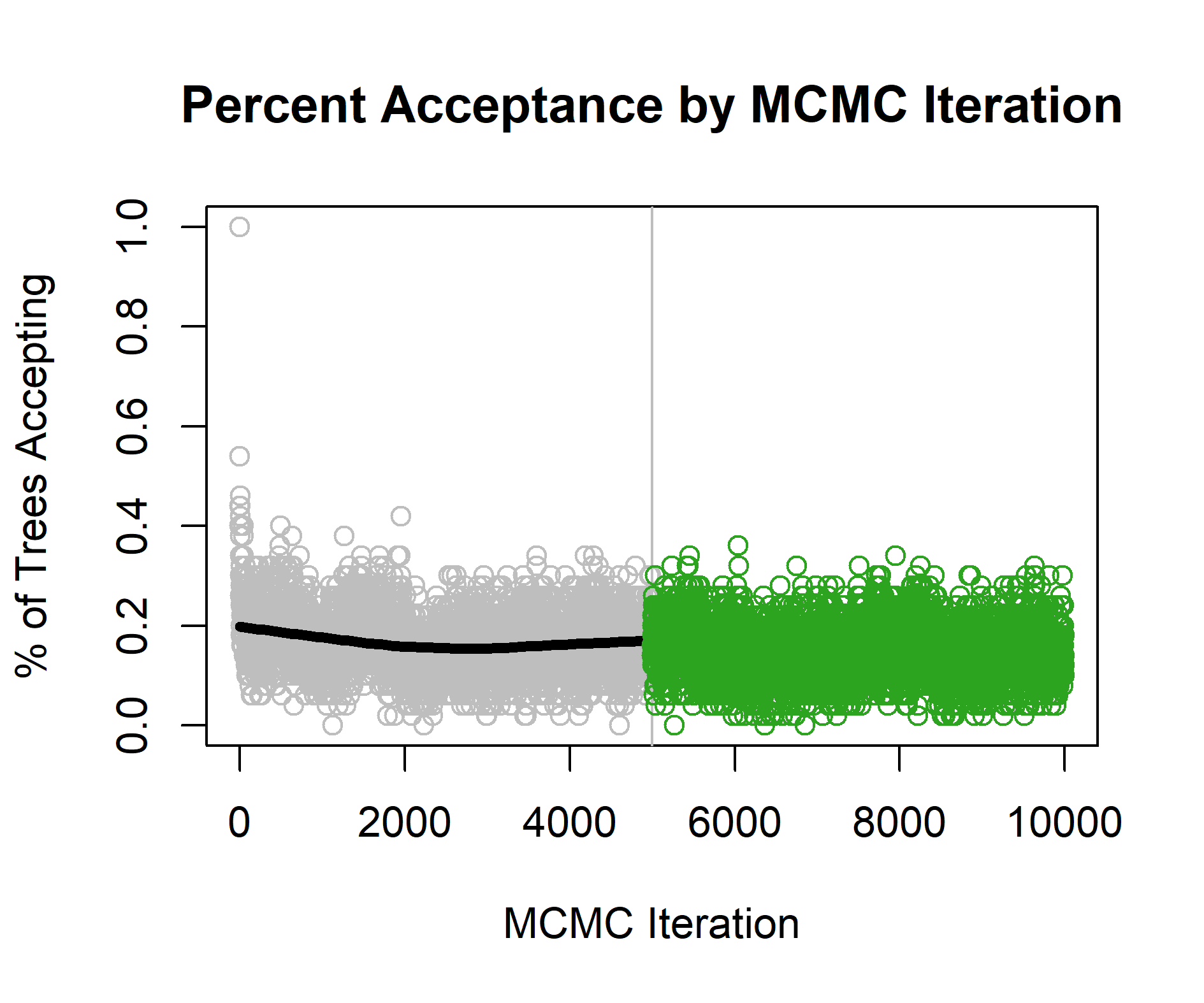}
        \caption{$\bm{y} = 10\bm{x}_1 + \bm\varepsilon$}
        \label{fig:DART_accept_id}
    \end{subfigure}
    \caption{Acceptance rate of DART in each unary scenarios. Gray open-circles represent acceptance rate in burn-in period. Green open-circles represent acceptance rate post burn-in period. Black solid lines represent average acceptance rate over MCMC iterations.}
    \label{fig:DART_accept}
\end{figure}

\begin{figure}[h]\ContinuedFloat
    \begin{subfigure}[b]{0.48\textwidth}
        \centering
        \includegraphics[width = \textwidth]{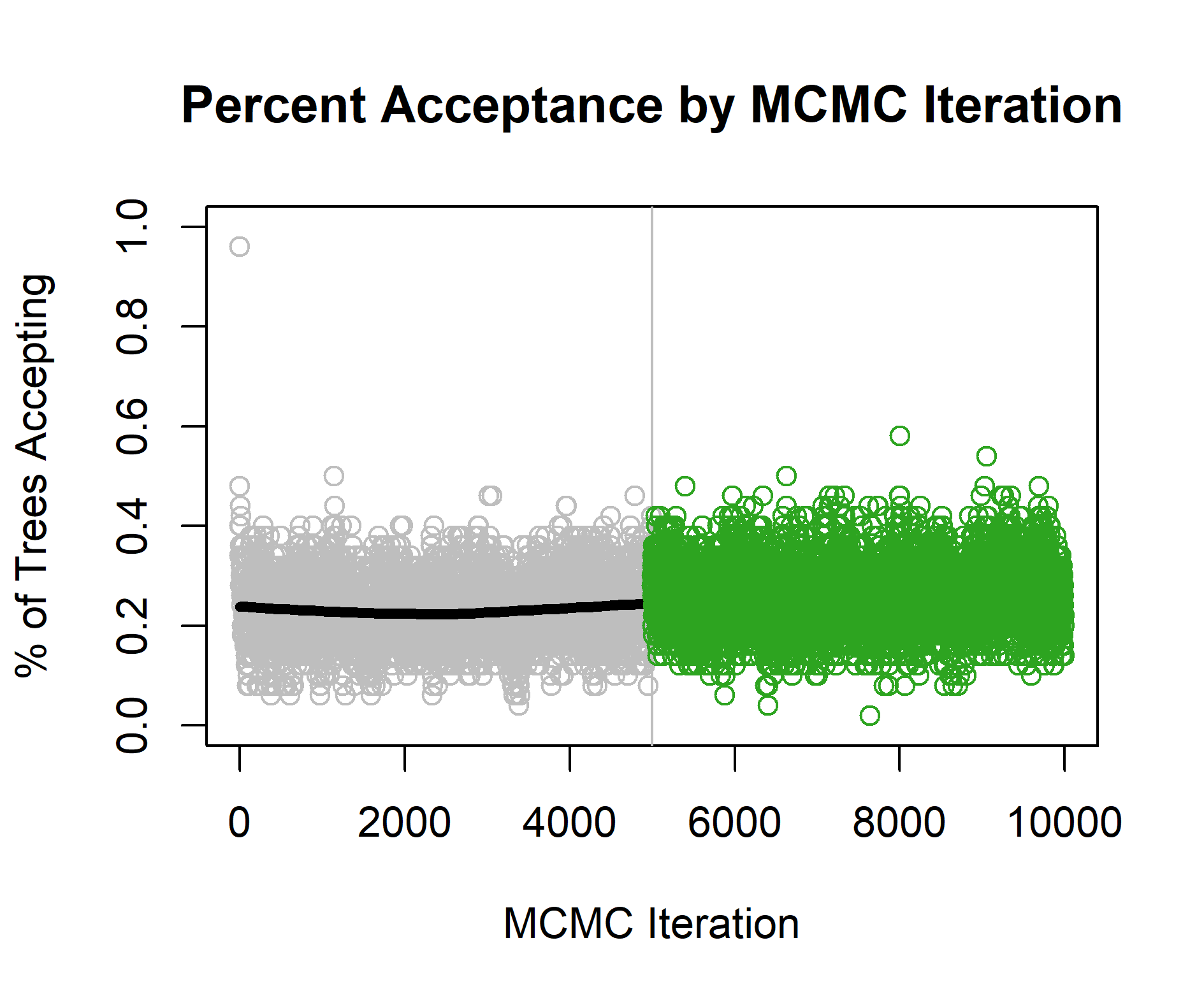}
        \caption{$\bm{y} = 10\log(\bm{x}_1) + \bm\varepsilon$}
        \label{fig:DART_accept_log}
    \end{subfigure}
    \hfill
    \begin{subfigure}[b]{0.48\textwidth}
        \centering
        \includegraphics[width = \textwidth]{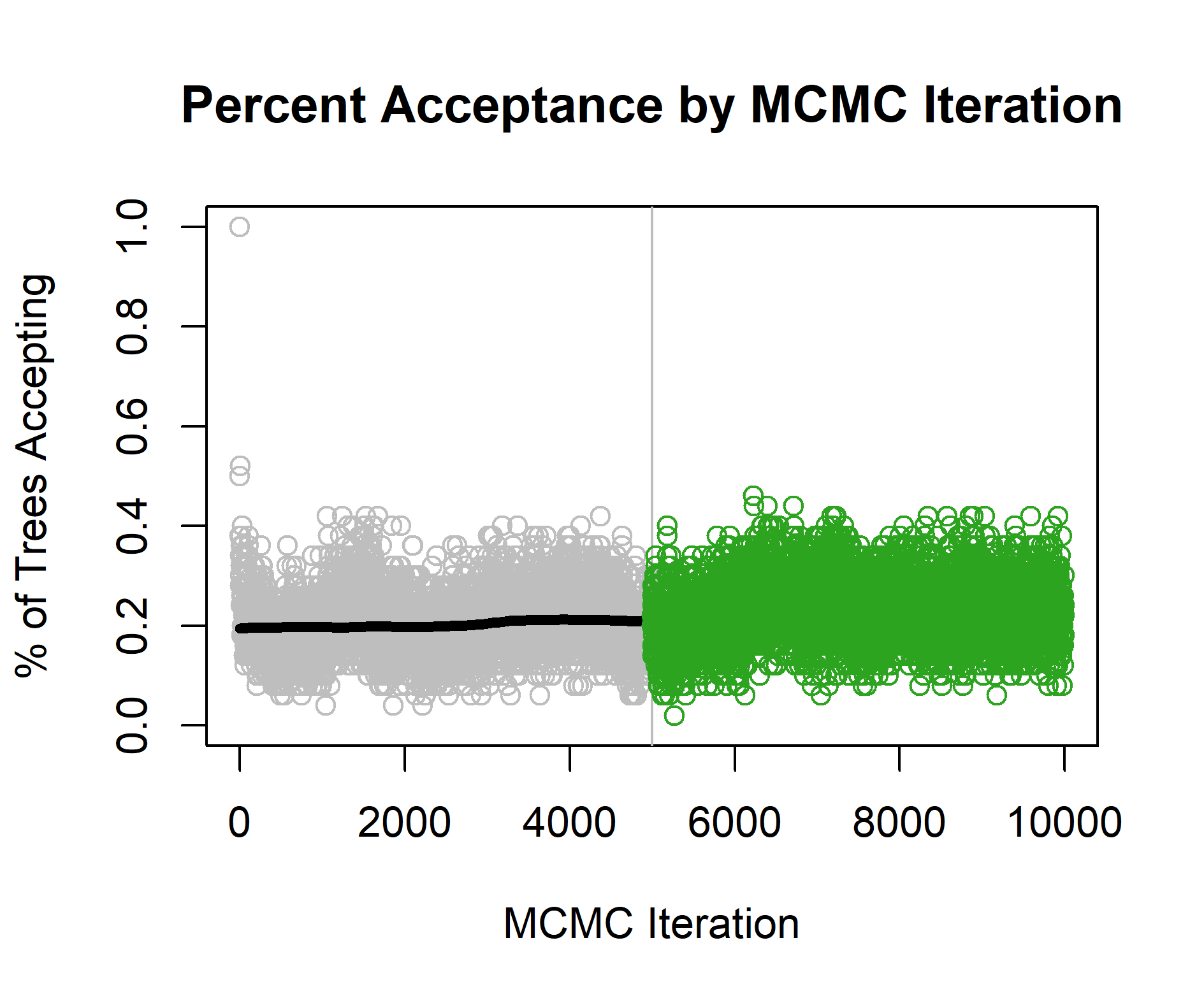}
        \caption{$\bm{y} = 10|\bm{x}_1| + \bm\varepsilon$}
        \label{fig:DART_accept_abs}
    \end{subfigure}
    \begin{subfigure}[b]{0.48\textwidth}
        \centering
        \includegraphics[width = \textwidth]{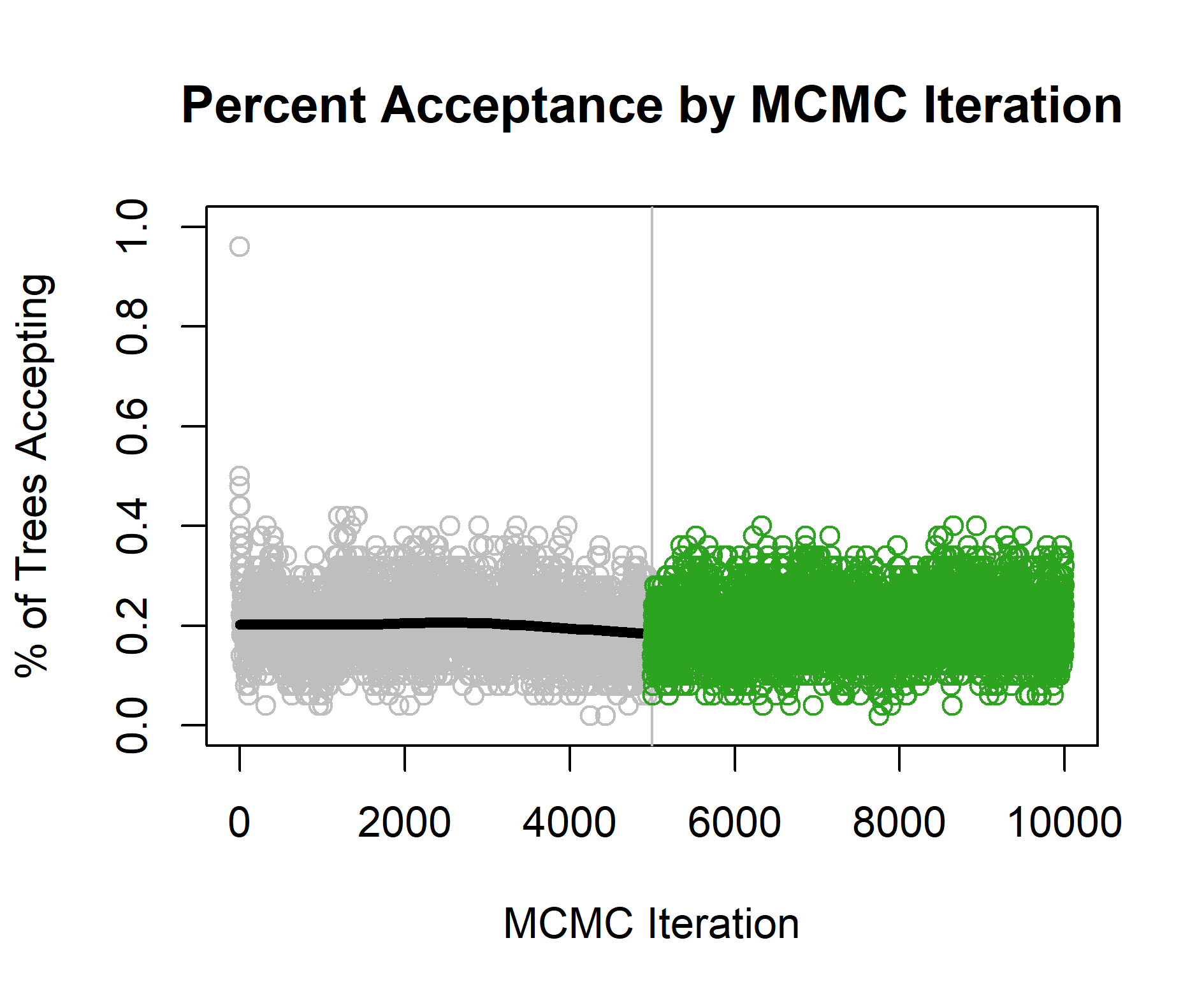}
        \caption{$\bm{y} = 10\sqrt{\bm{x}_1} + \bm\varepsilon$}
        \label{fig:DART_accept_sqrt}
    \end{subfigure}
    \caption{Acceptance rate of DART in each unary scenarios. Gray open-circles represent acceptance rate in burn-in period. Green open-circles represent acceptance rate post burn-in period. Black solid lines represent average acceptance rate over MCMC iterations.}
    \label{fig:DART_accept_2}
\end{figure}

\subsubsection{Complex descriptors with high-order compositions} \label{sec:supp_complex}
In this section, we compare the OIS selection accuracy of iBART using four different nonparametric variable selection methods. To facilitate a complex OIS model, we use model (15) in Section 3.4 of the main paper
\begin{equation} \label{eq:complex_eq}
    \bm{y} = 15\{\exp(\bm{x}_1) - \exp(\bm{x}_2)\}^2 + 20\sin(\pi\bm{x}_3\bm{x}_4) + \bm\varepsilon, \qquad \bm\varepsilon \sim \mathcal{N}_n(\bm{0}, \sigma^2\bm{I}),
\end{equation}
with $n \in \{100, 250\}$, $p = 10$, and $\sigma = 0.5$. The small sample size case $n = 100$ resembles real-world data size. The primary feature vectors $\bm{x}_i$ are drawn independently from a Uniform distribution, namely, $\bm{x}_1,\ldots,\bm{x}_p \overset{\text{i.i.d.}}{\sim} \text{U}_n(-1,1)$. Tuning parameters for BART-G.SE are unchanged from Section 3.4 of the main paper, and those for INIS are identical to their settings in Section~\ref{sec:unary_supp}.

ABC Bayesian Forests is trained with $B = 2,000$ ABC samples, where only a fraction of ABC samples (top 10\%) are kept in the reference table for calculating MIP. For each ABC sample, we increase the number of burn-in iterations from 200 in Section~\ref{sec:unary_supp} to 1,000 iterations due to a more complex simulation design. The numbers of trees and posterior samples remain identical to their settings in Section~\ref{sec:unary_supp}, which are 20 and 2, respectively. In addition to the MPM variable selection criterion described in Section~\ref{sec:unary_supp}, we adopt the permutation test from BART-G.SE to facilitate variable selection for ABC Bayesian Forests due to the poor performance of MPM in the small sample size $n=100$ setting. 

DART is trained with 20 trees and the posterior is summarized with 5,000 MCMC iterations after 10,000 burn-in iterations. All other tuning parameters are left as default. We adopt the MPM model for variable selection in DART.

We herein refer to PAN with BART-G.SE, ABC Bayesian Forests, DART, and INIS as iBART, iABC, iDART, and iNIS, respectively. Each of these methods also has an $\ell_0$-penalized version, leading to iBART$+\ell_0$, iABC$+\ell_0$, iDART$+\ell_0$, and iNIS$+\ell_0$. The $\ell_0$-penalization finds the best subset of $k$ variables from the set of selected variables using the Akaike information criterion (AIC) with $k \in \{1,2,3,4\}$.

\begin{figure}[h]
    \centering
    \includegraphics[width=0.94\textwidth]{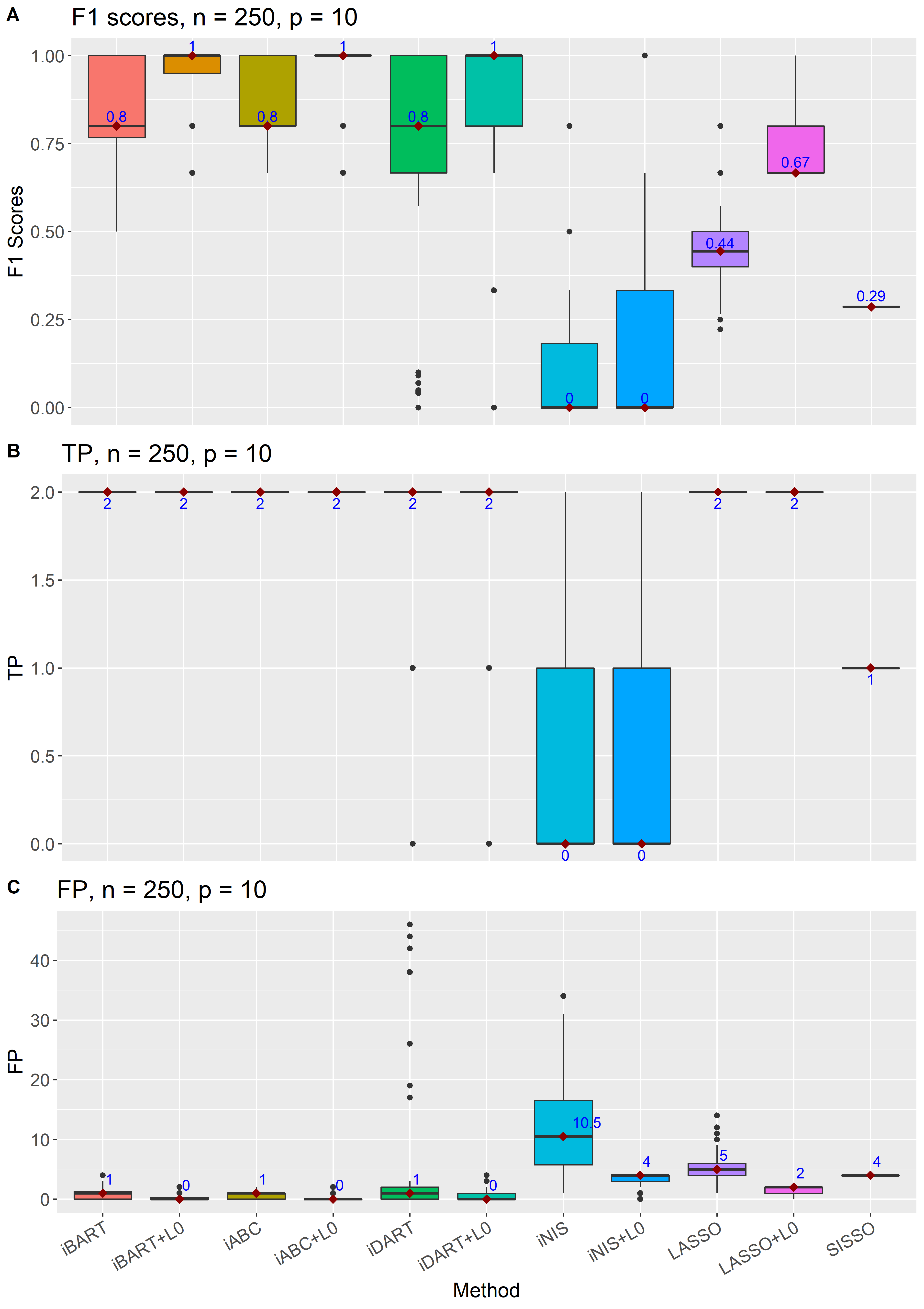}
    \caption{Selection accuracy when $n=250$ and $p=10$. Boxplots for iNIS and iNIS$+\ell_0$ are drawn using 42/100 simulations where the algorithm did not throw error. \textbf{A}, Boxplots of $F_1$ score. \textbf{B}, Boxplots of TP. \textbf{C}, Boxplots of FP.}
    \label{fig:supp3.4_n250}
\end{figure}

Figure~\ref{fig:supp3.4_n250} depicts the OIS selection accuracy of each method through $F_1$ score, TP, and FP. The three BART-based methods, iBART, iABC, iDART, and their $\ell_0$-penalized variants have similar accuracy in all three metrics, demonstrating superior OIS selection accuracy in this simulation setting. Despite having similar performance, iBART and iABC have fewer outliers than iDART as shown in Figure~\ref{fig:supp3.4_n250}, suggesting better stability. Both iBART and iABC score a 100\% TP rate in all 100 replications but iDART misses one descriptor 1 time and misses two descriptors 2 times. iNIS, on the contrary, only completed 42/100 simulation replications. In the other 58 replications, INIS gave numerical errors during QR decomposition of the descriptor space. Within those completed replications, both versions of iNIS have a median TP of 0 which leads to a median $F_1$ score of 0. Both versions of iNIS also select more FP than other methods, on average. 

\begin{figure}[h]
    \centering
    \includegraphics[width=0.94\textwidth]{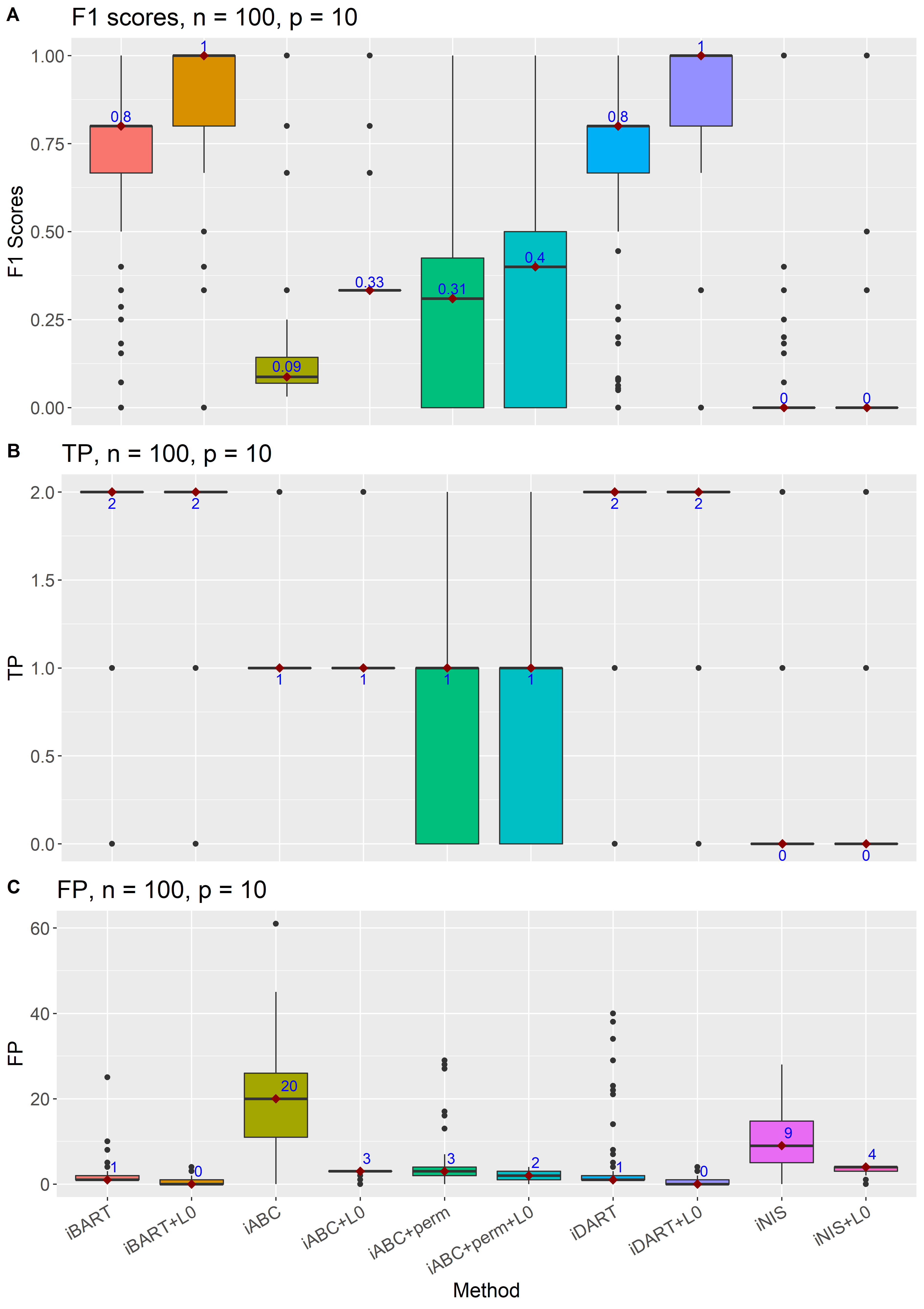}
    \caption{Selection accuracy when $n=100$ and $p=10$. Boxplots for iNIS and iNIS$+\ell_0$ are drawn using 50/100 simulations where the algorithm did not throw error. \textbf{A}, Boxplots of $F_1$ score. \textbf{B}, Boxplots of TP. \textbf{C}, Boxplots of FP.}
    \label{fig:supp3.4_n100}
\end{figure}

We now repeat the same analysis with a smaller sample size of $n=100$, which closely resembles the real data application section where we had $n=91$ samples. Both iBART and iDART are able to maintain similar selection accuracy with a reduction of sample size as shown in Figure~\ref{fig:supp3.4_n100}; their respective medians of $F_1$, TP, and FP are identical in both $n=100$ and $n=250$ cases. Such robust performance is not seen in iABC, which suffers significantly from the reduced sample size. Median $F_1$ scores of iABC and iABC$+\ell_0$ drop from 0.8 and 1 to 0.09 and 0.33, respectively. This may be caused by the data-splitting strategy in the ABC sampler which does not efficiently use all available data for a challenging high-dimensional variable selection problem. 

As another analysis to rule out the effect of MPM in iABC, we adopt the permutation test method of BART-G.SE in place of MPM for variable selection in ABC Bayesian Forests and denote it as iABC$+$perm. We rerun the simulation and report results in Figure~\ref{fig:supp3.4_n100}. Since both iABC and iABC$+$perm have the same posterior samples in each simulation replicate, differences in selection accuracy, if any, are solely due to the variable selection criteria. 

As shown in Figure~\ref{fig:supp3.4_n100}, the permutation test reduces the median FP of non-$\ell_0$-penalized iABC from 20 to 3 but the median TP rate remains at 50\%. This observation highlights two points: 1) the permutation test effectively filters out the false signals and 2) the permutation test coupled with iABC posterior samples (i.e., iABC$+$perm) is not able to detect one of the true descriptors, $f_1(\bm{x}) = \{\exp(\bm{x}_1) - \exp(\bm{x}_2)\}^2$, 91/100 times. There is still a substantial performance gap between iABC$+$perm and iBART after assigning the same selection rule to iABC, suggesting that the decline in iABC's sensitivity may be attributed to the inefficiency of the ABC sampler rather than the MPM selection criterion.

\subsection{Performance in real data}
Here we replicate the analysis in Section 4 with an addition of three methods, iABC, iDART, and iNIS. iABC is trained with $B = 4,000$ ABC samples, where only the top 10\% are kept in the reference table for MIP calculation. For each ABC sample, we increase the number of burn-in iterations from 1,000 in Section~\ref{sec:supp_complex} to 5,000. The numbers of trees and posterior samples remain identical to their respective setting in Section~\ref{sec:supp_complex}. The tuning parameter settings for iDART and iNIS are identical to their respective settings in Section~\ref{sec:supp_complex}. To facilitate a fair playground for the PAN variants, we restrict iABC, iBART, iDART, and iNIS to stop at iteration 3, i.e., they only construct descriptors up to composition complexity of $M = 3$. 

\begin{figure}[h]
    \centering
    \includegraphics[width=0.86\textwidth]{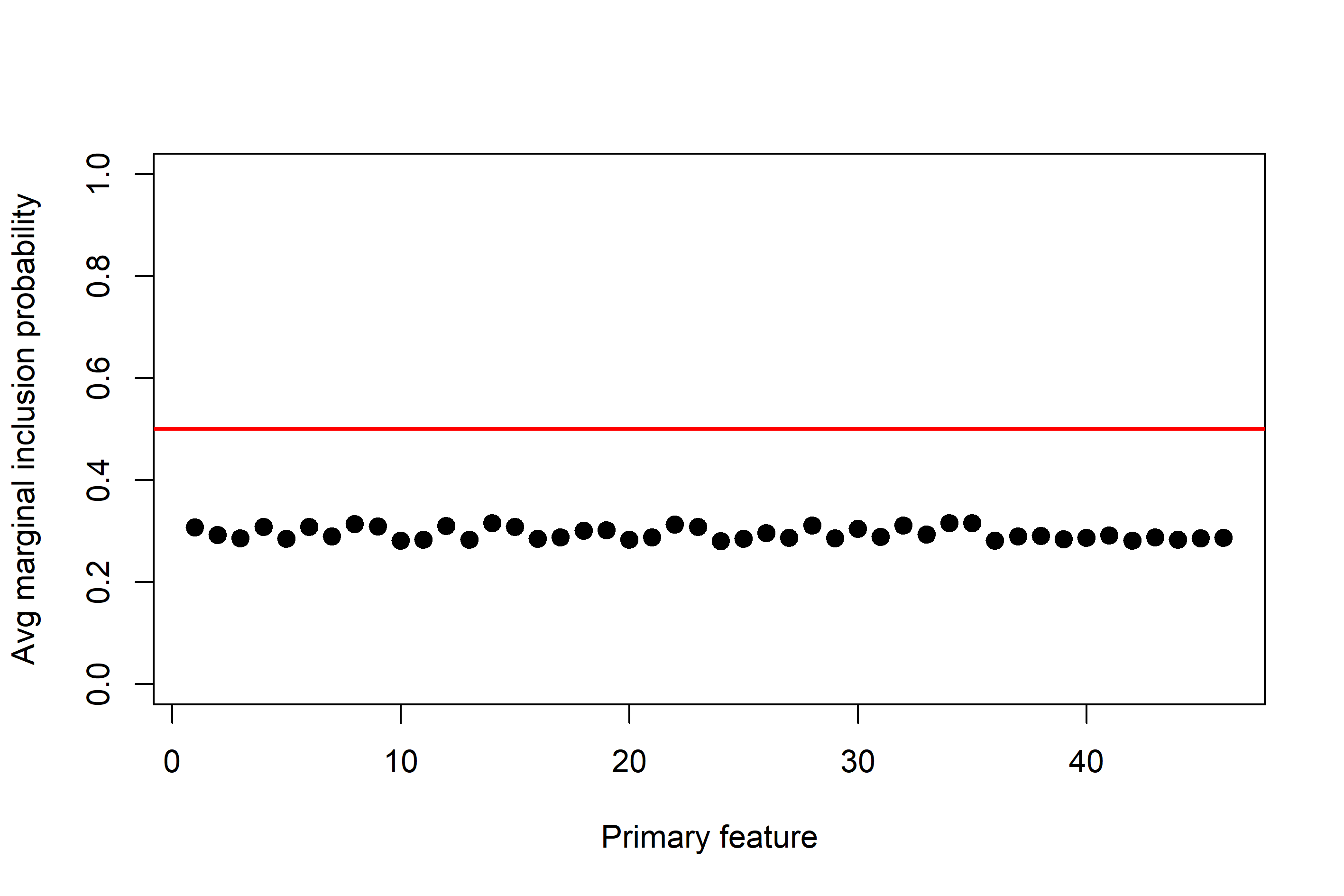}
    \caption{Average MIP of primary feature for iABC across 50 random partitions. Red solid line indicates MIP of 0.5.}
    \label{fig:supp4_abc_mip}
\end{figure}

During our analysis, we notice that both iABC and iNIS fail to select any descriptors at iteration 3. iABC did not select any descriptor because it did not select any primary feature $\bm{x}_1,...,\bm{x}_p$---as illustrated in Figure~\ref{fig:supp4_abc_mip}, no MIP is above the MPM selection threshold. Given this observation, we substitute the MPM selection criterion with the permutation test of BART-G.SE and use iABC$+$perm for further comparison. For iNIS, it only completed iteration 1 in 25/50 cases and failed to complete iteration 2 in all 50 cases due to numerical issues. Thus we exclude it from all analyses in this section.

\begin{figure}[h]
    \centering
    \includegraphics[width=\textwidth]{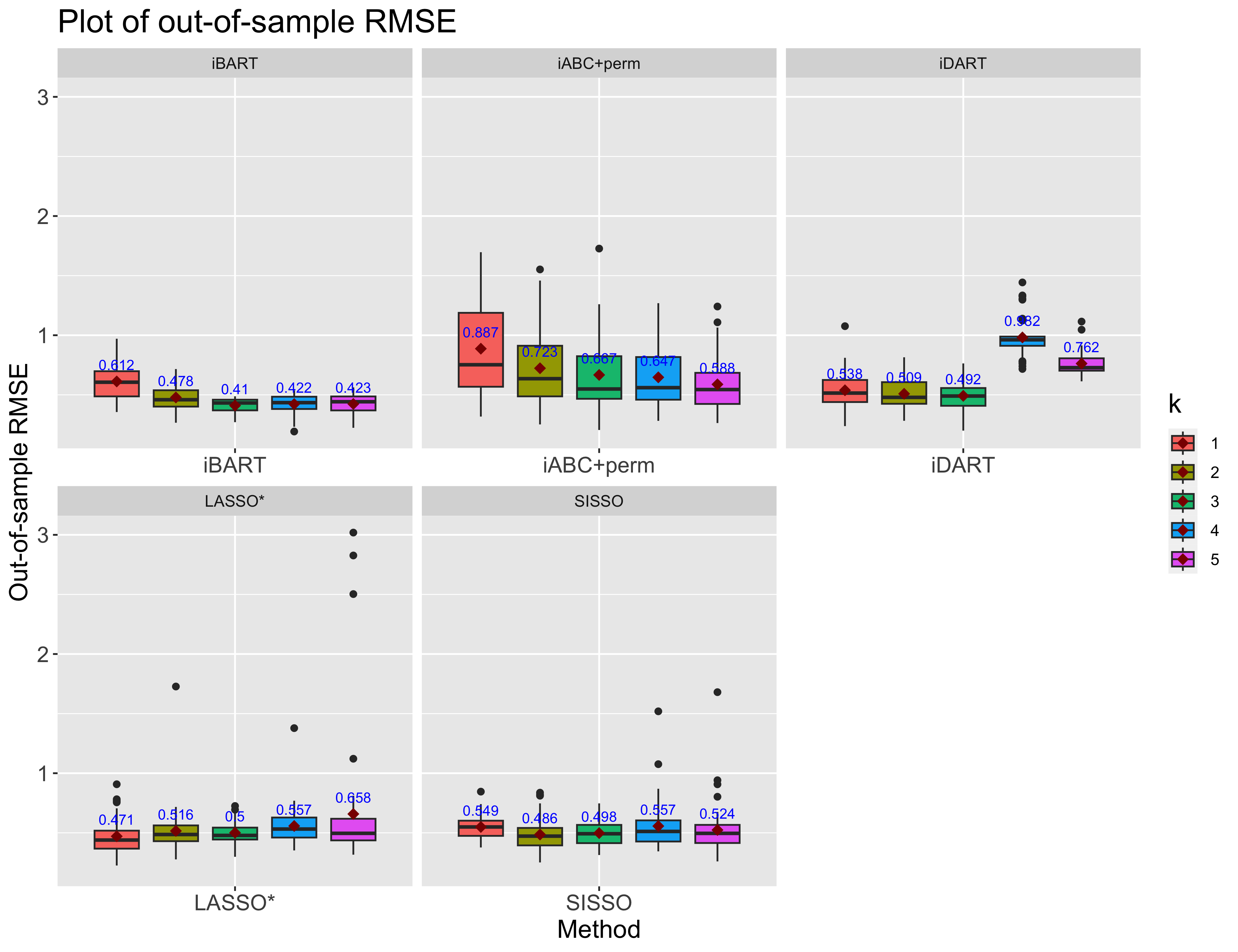}
    \caption{Boxplot of the out-of-sample RMSE for each method across 50 random partitions with $1 \leq k \leq 5$ (left to right in each plot). The blue numbers and the red rhombuses are the mean out-of-sample RMSE.}
    \label{fig:supp4_rmse_out}
\end{figure}
We use out-of-sample RMSE to evaluate the model accuracy of each method. For the sake of interpretability, we restrict our analyses to at most 5-descriptor models, as described in Section 4 of the main paper. iBART has the smallest average RMSE for all $k = 1, \ldots, 5$, with only one exception in the comparison to iDART when $k = 1$. For iDART, we observe a significant increase in its out-of-sample RMSE when $k \geq 4$, suggesting a tendency of overfitting. We note that iABC$+$perm is substantially slower than iBART or iDART; for example, it took 7.58 hours in one cross-validation on Intel Xeon Gold 6230, compared to several minutes required by iBART or iDART.

\section{Additional simulation results using correlated primary features} \label{sec:corr}
In this section, we investigate iBART's performance relative to LASSO and SISSO when primary features are correlated. We follow the same data-generating scheme as in Section 3.4 of the main paper but add an autoregressive covariance structure to the primary features $\bm{x}_1,...,\bm{x}_p$. In particular, we first generate latent primary features $\Tilde{\bm{x}}_1, \ldots, \Tilde{\bm{x}}_p$ from a multivariate Gaussian distribution, $\mathcal{N}_n(\bm{0}, \bm\Sigma)$, where $\bm\Sigma = (\Sigma_{ij})_{p \times p}$ is an autoregressive covariance matrix with $\Sigma_{ij} = \Sigma_{ji} = \rho^{|i - j|}$ for $1 \leq i, j \leq p$. We then transform the latent primary features as follows
\begin{equation}
    \bm{x}_i = 2\Phi(\Tilde{\bm{x}}_i) - 1, \quad 1 \leq i \leq p,
\end{equation}
so that the primary features, $\bm{x}_1,\ldots,\bm{x}_p$, follow a correlated Uniform distribution from $-1$ to $1$. We choose $\rho \in \{0, 0.2, 0.5\}$ to represent different correlation settings, where $\rho = 0$ is reported in Section 3.4 of the main paper. 

\begin{figure}[h]
    \centering
    \includegraphics[width = \linewidth]{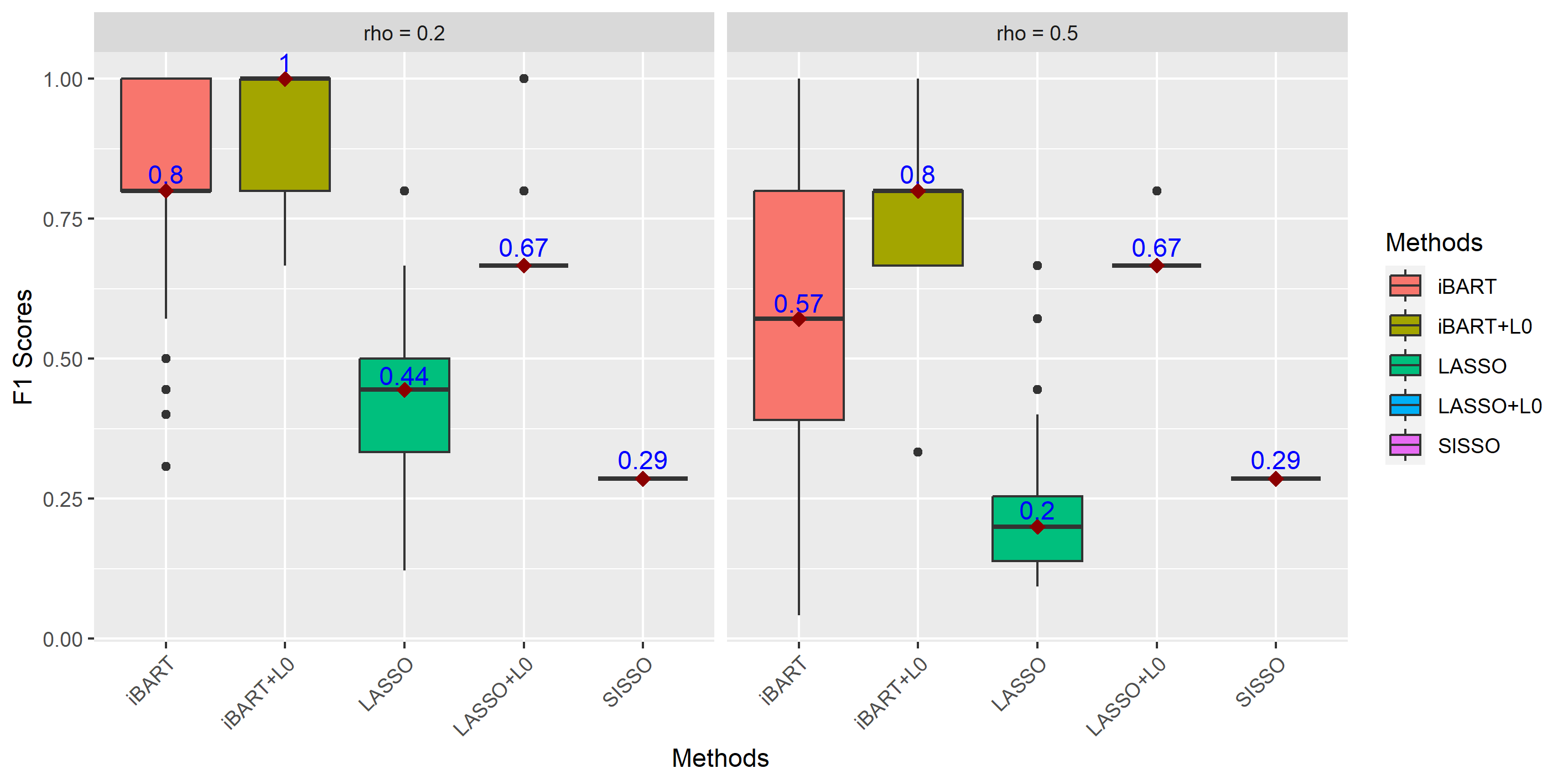}
    \caption{Boxplots of $F_1$ scores over 100 simulations for different methods with $\rho \in \{0.2, 0.5\}$.}
    \label{fig:Sec3.5_low_and_mid}
\end{figure}

As shown in the left panel of Figure~\ref{fig:Sec3.5_low_and_mid}, $F_1$ scores of most methods remain unchanged when $\rho = 0.2$. When $\rho = 0.5$, the performance of SISSO and LASSO$+\ell_0$ remains unchanged, suggesting robustness to collinearity, but it is still not comparable to that of iBART$+\ell_0$. In both correlation settings, iBART and iBART$+\ell_0$ maintain a TP rate of 100\% across 100 simulations. Thus, their degraded performances are due to higher FP selection as collinearity increases. LASSO and LASSO$+\ell_0$ also achieve a TP rate of 100\% across 100 simulations but they incur more FP, leading to lower $F_1$ scores in comparison to those of iBART and iBART$+\ell_0$. SISSO, on the other hand, only correctly identifies one descriptor out of two, which aligns with our observation reported in the main paper when the primary features are independent.

\section{Additional results in real data application without unit information} \label{sec:without_unit_info}
We replicate the experiments in Section 4 of the main paper without constraining the constructed descriptors to be \textit{unit consistent}, that is, we do not compare the unit of the primary features to enforce physical interpretability. This will lead to a larger search space and may lead to the selection of non-physical descriptors, e.g. a descriptor whose unit is $\textit{km/h} + \textit{lbs}$. All other settings mirror Section 4 of the main paper.  

iBART and SISSO are used to analyze this dataset. LASSO$^\star$ is excluded as the increased descriptor space exceeds the matrix size it can handle and also that it did not yield better performance than iBART and SISSO in the main paper. 

The performance of each method is gauged by the out-of-sample RMSE, runtime, and the number of generated descriptors. To calculate out-of-sample RMSE, we randomly partition the $n = 91$ observations into a 90\% training set (82 samples) and a 10\% testing set (9 samples) and repeat this process 50 times. For brevity, we herein refer to out-of-sample RMSE as RMSE.

\begin{figure}[h]
    \centering
    \includegraphics[width = 0.9\linewidth, trim = 0 0 250 0, clip = TRUE]{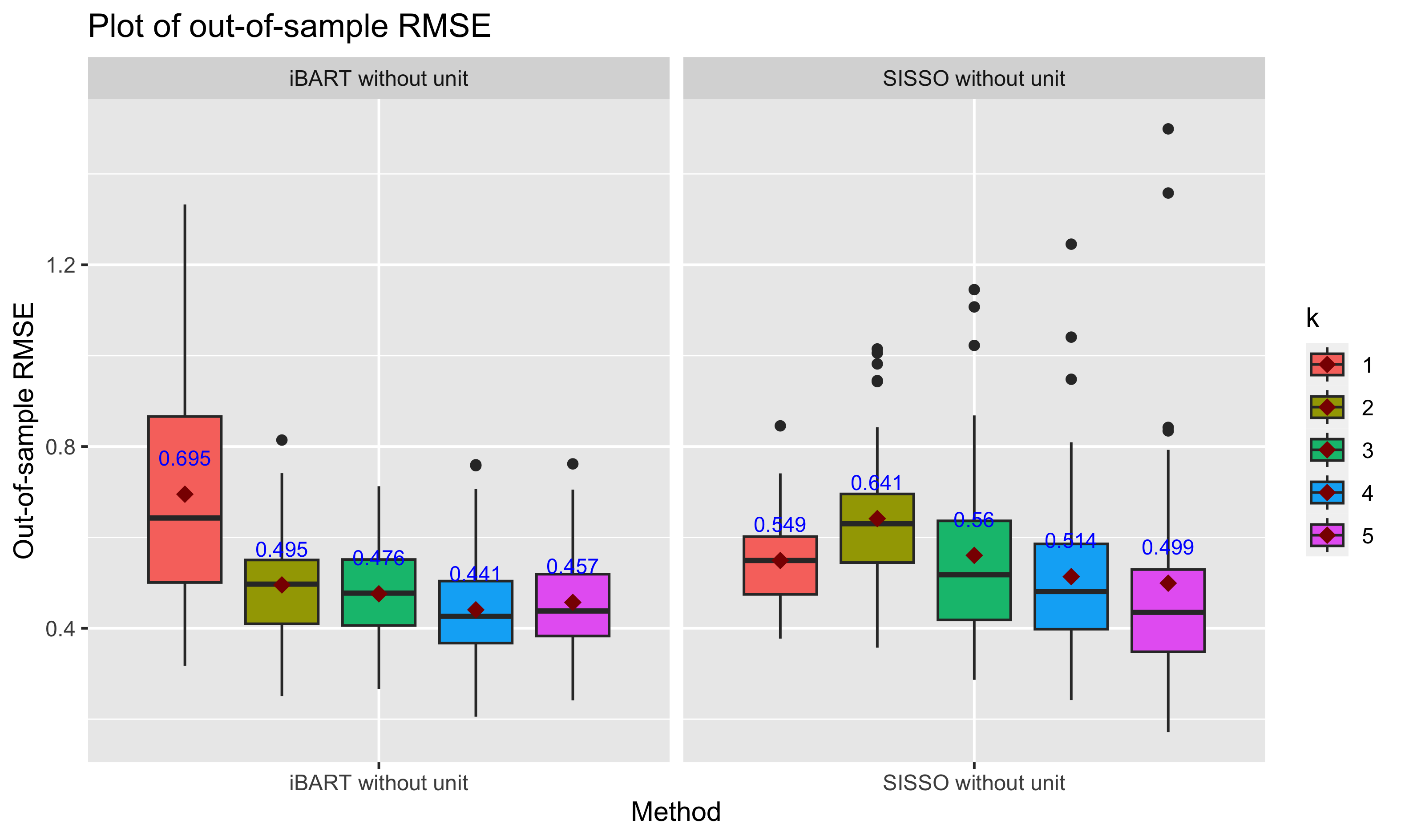}
    \caption{Boxplot of the out-of-sample RMSE for each method across 50 random partitions with $1 \leq k \leq 5$ (left to right in each plot). The blue numbers and the red rhombuses indicate the average out-of-sample RMSE.}
    \label{fig:Connor_RMSE_out_no_unit}
\end{figure}

From Figure \ref{fig:Connor_RMSE_out_no_unit}, the smallest average RMSE is attained at $k = 4$ for iBART (0.441) and $k = 5$ for SISSO (0.499), that is, iBART reduces the RMSE by 11\% relative to SISSO; also see the first row of Table~\ref{tab:runtime_no_unit}. 
In addition to the performance gain in RMSE, iBART also leads to a substantial reduction in computing time and memory usage. Table~\ref{tab:runtime_no_unit} shows that iBART using 1 CPU core is faster than SISSO using 40 CPU cores by about 15\%, tested on an Intel Xeon Gold 6230 CPU @ 2.10 GHz. The excellent scalability of iBART transforms into memory efficiency as the descriptor space in iBART is orders of magnitude smaller than that of competing methods. Table~\ref{tab:runtime_no_unit} shows that iBART generates a descriptor space of size $1166 < O(p^2)$ in the last iteration on average. Thanks to this significantly smaller descriptor space, we were able to run iBART on a laptop with only 10GB of memory; in contrast, SISSO failed at the descriptor generation step owing to the enormous descriptor space they try to generate, and require server-grade computing facilities.   

\begin{table}[ht]
    \caption{\label{tab:runtime_no_unit}Performance comparison of three methods: out-of-sample RMSE, runtime, and the number of generated descriptors, averaged over 50 cross validations.}
    \centering
    \fbox{
        \begin{tabular}[t]{lccc}
        & iBART (1 CPU core) & SISSO (40 CPU cores) \\
        \hline
        RMSE & 0.441 & 0.499 \\
        Runtime & 269.4 sec & 317.5 sec \\
        Number of generated descriptors & 1166 & $2.2 \times 10^8$\\
        \end{tabular}}
\end{table}

Overall, the comparison between iBART and SISSO is similar to Section 4 in the main paper. Incorporating the unit information reduces the size of descriptor space and reduces the runtime for both methods (compare Table~\ref{tab:runtime_no_unit} and Table 1 in the main paper). 

Table~\ref{tab:k_model_table} reports the selected descriptors by iBART with various $k$ using the full dataset, and Table~\ref{tab:primary_feature_names} reports the physical meanings of the selected primary features.

\begin{table}[ht]
    \caption{\label{tab:k_model_table}Selected linear models by iBART for $k \in \{1,2,3,4,5\}$.}
    \renewcommand\arraystretch{2}
    \centering
    \fbox{
        \begin{tabular}[t]{ll}
        $k$ & Selected descriptors\\
        \hline
        1 & $\text{Z}^\text{s} \cdot \text{IE}_4^\text{s} \cdot (\Delta H_{\text{sub}} - \Delta H_{\text{f,ox,bulk}})$ \\ 
        2 & $\text{Z}^\text{s} \cdot \text{IE}_4^\text{s} \cdot (\Delta H_{\text{sub}} - \Delta H_{\text{f,ox,bulk}})$, 
            $\left|\frac{(\eta^{1/3})^\text{m}/\Delta E_{\text{vac}}}{\text{N}_{\text{val}}^\text{m} - (\eta^{1/3})^\text{m}}\right|$\\
        3 & $\text{Z}^\text{s} \cdot \text{IE}_4^\text{s} \cdot (\Delta H_{\text{sub}} - \Delta H_{\text{f,ox,bulk}})$, 
            $\left|\frac{\text{IE}_4^{\text{s}} \cdot (\eta^{1/3})^\text{m} \cdot (\Delta H_{\text{sub}} - \Delta H_{\text{f,ox,bulk}})}{\text{N}_{\text{val}}^\text{m}} \right|$, 
            $\left|\frac{\text{N}_{\text{val}}^\text{m} - \text{CN}_{\text{surf}}^\text{m}}{\exp(\text{N}_{\text{val}}^\text{m} - \text{CN}_{\text{surf}}^\text{m})} \right|$ \\
        4 & $\text{Z}^\text{s} \cdot \text{IE}_4^\text{s} \cdot (\Delta H_{\text{sub}} - \Delta H_{\text{f,ox,bulk}})$, 
            $\left|\frac{\text{IE}_4^{\text{s}} \cdot (\eta^{1/3})^\text{m} \cdot (\Delta H_{\text{sub}} - \Delta H_{\text{f,ox,bulk}})}{\text{N}_{\text{val}}^\text{m}} \right|$, 
            $\left|\frac{\text{N}_{\text{val}}^\text{m} - \text{CN}_{\text{surf}}^\text{m}}{\exp(\text{N}_{\text{val}}^\text{m} - \text{CN}_{\text{surf}}^\text{m})} \right|$,
            $\left|\frac{\text{N}_{\text{val}}^\text{m} - (\eta^{1/3})^\text{m}}{\Delta H_{\text{f,ox,bulk}}}\right|$ \\
        5 & $\text{EA}^\text{s} \cdot \Delta H_{\text{f,ox,bulk}}$, 
            $\frac{(\eta^{1/3})^\text{m}}{\phi^{\text{s}}}$,
            $\left|\frac{\text{IE}_4^{\text{s}} \cdot (\eta^{1/3})^\text{m} \cdot (\Delta H_{\text{sub}} - \Delta H_{\text{f,ox,bulk}})}{\text{N}_{\text{val}}^\text{m}} \right|$, 
            $\left|\frac{\text{N}_{\text{val}}^\text{m} - \text{CN}_{\text{surf}}^\text{m}}{\exp(\text{N}_{\text{val}}^\text{m} - \text{CN}_{\text{surf}}^\text{m})} \right|$, 
            $\left|\frac{(\eta^{1/3})^\text{m}/\Delta E_{\text{vac}}}{\text{N}_{\text{val}}^\text{m} - (\eta^{1/3})^\text{m}}\right|$
        \end{tabular}}
\end{table}

\begin{table}[ht]
    \caption{\label{tab:primary_feature_names}Descriptions of the selected primary features by iBART.}
    \centering
    \fbox{
        \begin{tabular}{*{2}{l}}
        Primary feature               & Physical meaning\\
        \hline
        $\text{Z}^\text{s}$           & Atomic number of the surface material\\
        $\text{IE}_4^\text{s}$        & \makecell[l]{4th ionization energy of support with the bulk metal in\\ the $4^+$ oxidation state}\\
        $\Delta H_{\text{sub}}$       & Heat of sublimation\\
        $\Delta H_{\text{f,ox,bulk}}$ & Oxidation energy of the bulk metal\\
        $(\eta^{1/3})^\text{m}$       & Discontinuity in electron density of metal adatom\\
        $\Delta E_\text{vac}$         & Oxygen vacancy energy\\
        $\text{N}_{\text{val}}^\text{m}$            & Number of valence electrons in metal adatom\\
        $\text{CN}_{\text{surf}}^\text{m}$          & {Coordination number of the support metal in the surface phase}\\
        $\text{EA}^\text{s}$                        & Electron affinity of support\\
        $\phi^s$                      & Chemical potential of the electrons in support\\
        \end{tabular}}
\end{table}

\begin{figure}[h]
    \centering
    \includegraphics[width = 0.8\linewidth]{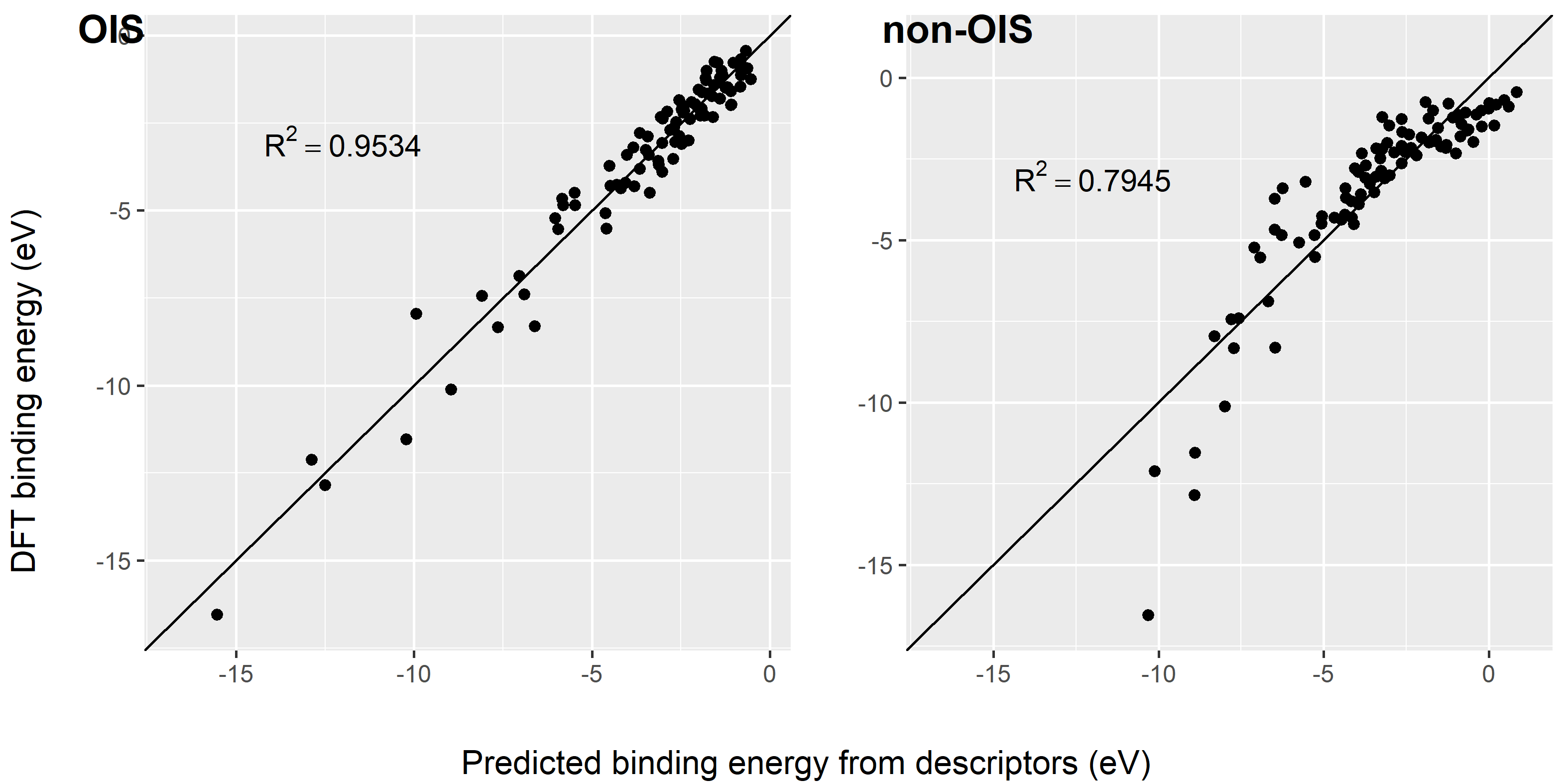}
    \caption{DFT binding energies versus predicted values using linear models with OIS and without OIS. The black line is $y = x$. Each model has $k=4$ descriptors.}
    \label{fig:in_sample_plot}
\end{figure}

Figure \ref{fig:in_sample_plot} reports the fitting performance of the OIS model and the non-OIS model. The OIS model is the iBART model with $k=4$, the optimum model suggested by RMSE. The non-OIS model is a simple least squares model with $\bm{X}$ as the design matrix and no feature engineering step. For ease of comparison, the non-OIS model also has $k=4$ predictors determined by best subset selection. The plot centers around the black line $y = x$, suggesting model adequacy. The OIS model yields an $R^2$ of 0.965, indicating high explanatory power. In contrast, the non-OIS model shows poor fitting performance and considerable deviation from the line $y = x$. Compared to the OIS model, this model gives an $R^2 = 0.8085$, leaving 15.65\% more of the variance in the response unexplained. The fitted OIS model is
\begin{align*}
    \hat{{y}}_{\text{OIS}} &= -3.435 - 1.180 \cdot \text{Z}^\text{s} \cdot \text{IE}_4^\text{s} \cdot (\Delta H_{\text{sub}} - \Delta H_{\text{f,ox,bulk}})\\
    &\qquad - 1.463 \cdot \left|\frac{\text{IE}_4^{\text{s}} \cdot (\eta^{1/3})^\text{m} \cdot (\Delta H_{\text{sub}} - \Delta H_{\text{f,ox,bulk}})}{\text{N}_{\text{val}}^\text{m}} \right|\\
    &\qquad -0.416 \cdot \left|\frac{\text{N}_{\text{val}}^\text{m} - \text{CN}_{\text{surf}}^\text{m}}{\exp(\text{N}_{\text{val}}^\text{m} - \text{CN}_{\text{surf}}^\text{m})} \right|\\
    &\qquad + 0.246 \cdot \left|\frac{\text{N}_{\text{val}}^\text{m} - (\eta^{1/3})^\text{m}}{\Delta H_{\text{f,ox,bulk}}}\right|,
\end{align*}
and the fitted non-OIS model is
$
    \hat{y}_{\text{non-OIS}} = -12.045 + 0.610 \cdot \text{IE}_2^\text{m} - 2.698 \cdot r^\text{s} + 2.153 \cdot \text{BV}_{\text{bulk}}^{\text{m}} - 0.257 \cdot \text{IE}_3^\text{m}.
$
Here $\text{Z}^\text{s}$, $\text{IE}_4^\text{s}$, $\Delta H_{\text{sub}}$, $\Delta H_{\text{f,ox,bulk}}$, $(\eta^{1/3})^\text{m}$, $\Delta E_\text{vac}$,  $\text{N}_{\text{val}}^\text{m}$, $\text{CN}_{\text{surf}}^\text{m}$, $\text{EA}^\text{s}$, and $\phi^s$ are physical properties that correlate with the binding energy of metal-support pair. 

While the identified model above without enforcing unit consistency continues to maintain a high explanatory power, unlike the model identified in Section 4 in the main paper, it becomes challenging to interpret. 
\color{black}

\section{Proof of Theorem 2.1} \label{sec:proof}
\begin{proof}
Let $f^{(M)}(\bm{X})$ be an $M$-composition descriptor in $\mathcal{O}^{(M)}(\bm{X})$. Let $\bm{o} = \{o_1, o_2, ..., o_M\}$ be the sequence of operators used to construct $f^{(M)}(\bm{X})$ and let $f^{(m)}(\bm{X})$ be an $m$-composition descriptor such that
$$f^{(M)}(\bm{X}) = o_M \circ o_{M-1} \circ \cdots \circ o_{m+1} \circ f^{(m)}(\bm{X}), \quad \forall\, 1 \leq m \leq M.$$
Note that $o_m$ is either in $\mathcal{O}_u$ or $\mathcal{O}_b$ for all $m$. We first show that there exists $M_u \geq M$ such that $\mathcal{O}_{A_u}^{(M_u)}(\bm{X}) \supseteq \mathcal{O}^{(M)}(\bm{X})$ using mathematical induction. 
    
If $o_1 \in \mathcal{O}_u$, then $f^{(1)}(\bm{X}) \in \mathcal{O}_{A_u}^{(1)}(\bm{X})$. Otherwise, $f^{(1)}(\bm{X}) = o_1 \circ I(\bm{X}) \in \mathcal{O}_{A_u}^{(2)}(\bm{X})$. Suppose $f^{(m)}(\bm{X}) \in \mathcal{O}_{A_u}^{(m')}(\bm{X})$ for some $m \in \mathbb{N}$ and some $m' \geq m$. If $o_{(m+1)} \in \mathcal{O}_u$ and $m'$ is even, then $f^{(m+1)}(\bm{X}) \in \mathcal{O}_{A_u}^{(m'+1)}(\bm{X})$. If $o_{(m+1)} \in \mathcal{O}_u$ and $m'$ is odd, then $f^{(m+1)}(\bm{X}) = o_{(m+1)} \circ \pi_1 \circ f^{(m)}(\bm{X}) \in \mathcal{O}_{A_u}^{(m'+2)}(\bm{X})$. If $o_{(m+1)} \in \mathcal{O}_b$ and $m'$ is odd, then $f^{(m+1)}(\bm{X}) \in \mathcal{O}_{A_u}^{(m'+1)}(\bm{X})$. Lastly, if $o_{(m+1)} \in \mathcal{O}_b$ and $m'$ is even, then $f^{(m+1)}(\bm{X}) = o_{(m+1)} \circ I \circ f^{(m)}(\bm{X}) \in \mathcal{O}_{A_u}^{(m'+2)}(\bm{X})$. As this holds for any arbitrary $m \in \mathbb{N}$, by mathematical induction, there exists $M_u \geq M$ such that $f^{(M)}(\bm{X}) \in \mathcal{O}_{A_u}^{(M_u)}$ for all $M \in \mathbb{N}$. Note that for each $f^{(M)}(\bm{X}) \in \mathcal{O}^{(M)}(\bm{X})$, there exists such an $M_u$. Now take the largest $M_u = \max(M_{u_1}, M_{u_2}, \ldots)$. Then for all $f^{(M)}(\bm{X}) \in \mathcal{O}^{(M)}$, we have $f^{(M)}(\bm{X}) \in \mathcal{O}_{A_u}^{(M_u)}$, i.e., $\mathcal{O}_{A_u}^{(M_u)}(\bm{X}) \supseteq \mathcal{O}^{(M)}(\bm{X})$. To see that there exists $M_b \geq M$ such that $\mathcal{O}_{A_u}^{(M_u)}(\bm{X}) \supseteq \mathcal{O}^{(M)}(\bm{X})$, we use a similar argument. If $o_1 \in O_b$, then $f^{(1)}(\bm{X}) \in \mathcal{O}_{A_b}^{(1)}(\bm{X})$. Otherwise, $f^{(1)}(\bm{X}) = o_1 \circ \pi_1(\bm{X}) \in \mathcal{O}_{A_b}^{(2)}(\bm{X})$. Suppose $f^{(m)}(\bm{X}) \in \mathcal{O}_{A_b}^{(m')}(\bm{X})$ for some $m \in \mathbb{N}$ and some $m' \geq m$. If $o_{(m+1)} \in \mathcal{O}_b$ and $m'$ is even, then $f^{(m+1)}(\bm{X}) \in \mathcal{O}_{A_b}^{(m'+1)}(\bm{X})$. If $o_{(m+1)} \in \mathcal{O}_b$ and $m'$ is odd, then $f^{(m+1)}(\bm{X}) = o_{(m+1)} \circ I \circ f^{(m)}(\bm{X}) \in \mathcal{O}_{A_b}^{(m'+2)}(\bm{X})$. If $o_{(m+1)} \in \mathcal{O}_u$ and $m'$ is odd, then $f^{(m+1)}(\bm{X}) \in \mathcal{O}_{A_b}^{(m'+1)}(\bm{X})$. Lastly, if $o_{(m+1)} \in \mathcal{O}_u$ and $m'$ is even, then $f^{(m+1)}(\bm{X}) = o_{(m+1)} \circ I \circ f^{(m+1)}(\bm{X}) \in \mathcal{O}_{A_b}^{(m'+2)}(\bm{X})$. By mathematical induction, there exists $M_b \geq M$ such that $f^{(M)}(\bm{X}) \in \mathcal{O}_{A_b}^{(M_b)}$ for all $M \in \mathbb{N}$. Following similar arguments in the unary case, we have $\mathcal{O}_{A_b}^{(M_b)}(\bm{X}) \supseteq \mathcal{O}^{(M)}(\bm{X})$. This completes the proof. 
\end{proof}

\bibliographystyle{jasa_acs}
\bibliography{ref}